\documentclass[sigconf, screen, nonacm]{acmart}

\usepackage[toc,page,titletoc]{appendix}

\input{settings}

\AtBeginDocument{%
  \providecommand\BibTeX{{%
    \normalfont B\kern-0.5em{\scshape i\kern-0.25em b}\kern-0.8em\TeX}}}

\newcommand{\bmhead}[1]{\bigskip\noindent\textbf{#1}.\quad}

\begin{document}


\title{Fourier-Based 3D Multistage Transformer for Aberration Correction in Multicellular Specimens}


\settopmatter{authorsperrow=1} 
\newcommand{\tsc}[1]{\textsuperscript{#1}} 

\author{
Thayer Alshaabi\tsc{1,2}\footnotemark[1], 
Daniel E. Milkie\tsc{1},
Gaoxiang Liu\tsc{2}, 
Cyna Shirazinejad\tsc{2}, 
Jason L. Hong\tsc{2},
Kemal Achour\tsc{2}, 
Frederik G\"orlitz\tsc{2},
Ana Milunovic-Jevtic\tsc{2},
Cat Simmons\tsc{2},
Ibrahim S. Abuzahriyeh\tsc{2},
Erin Hong\tsc{2},
Samara Erin Williams\tsc{2},
Nathanael Harrison\tsc{2},
Evan Huang\tsc{2},
Eun Seok Bae\tsc{2},
Alison N. Killilea\tsc{2},
David G. Drubin\tsc{2},
Ian A. Swinburne\tsc{2},
Srigokul Upadhyayula\tsc{2,3,4}\footnotemark[1],
Eric Betzig\tsc{1,2,5,6}\footnotemark[1]\bigskip
}
\affiliation{
  \institution{\tsc{1}Howard Hughes Medical Institute, Ashburn, VA}
  \institution{\tsc{2}Department of Molecular and Cell Biology, University of California Berkeley, Berkeley, CA}
  \institution{\tsc{3}Lawrence Berkeley National Laboratory, Berkeley, CA}
  \institution{\tsc{4}Chan Zuckerberg Biohub, San Francisco, CA}
  \institution{\tsc{5}Department of Physics, University of California, Berkeley, CA}
  \institution{\tsc{6}Helen Wills Neuroscience Institute, Berkeley, CA}
  \state{}
  \country{}
}

\thanks{\bigskip$*$~Corresponding authors: alshaabit@hhmi.org, sup@berkeley.edu, betzige@hhmi.org.}

\renewcommand{\shortauthors}{Thayer Alshaabi, et al.}

\begin{abstract}
\begin{raggedright}
\let\thefootnote\relax\footnotetext{
\noindent\textbf{Model code}.
Source code for training and evaluation (and all pretrained models) are available at \url{https://github.com/cell-observatory/aovift}.
\medskip

\noindent\textbf{Synthetic data generator}.
Code for simulating beads is available at \url{https://github.com/cell-observatory/beads_simulator}.
\medskip

\noindent\textbf{Package}.
Docker image is available at \url{https://github.com/cell-observatory/aovift/pkgs/container/aovift}.
\medskip}
\let\thefootnote\relax\footnotetext{\textbf{Contributions}. \noindent
T.A. designed models, developed training pipelines and evaluation benchmarks. 
D.E.M. designed the preprocessing algorithms and developed the microscope software for the imaging experiments.
T.A. and D.E.M. designed Fourier embedding, and developed the synthetic data generator for training and validation. 
C. Shirazinnejad, C. Simmons, I.S.A., S.E.W., N.H., E. Hong, E. Huang, E.S.B., A.N.K., D.G.D., and I.A.S. generated the zebrafish reagents. 
A.N.K., prepared the cultured SUM159 cells.
C. Shirazinnejad, J.L.H., K.A., and A.M.J. prepared samples.
G.L. and J.L.H. performed the imaging experiments with zebrafish.
J.L.H. and K.A. performed the imaging experiments with cells.
G.L., K.A. and F.G. performed the imaging experiments with beads. 
T.A., D.E.M., and S.U. performed analysis and prepared figures.
T.A. wrote the manuscript with input from all co-authors. 
D.E.M., E.B. and S.U. edited the manuscript. 
T.A., E.B. and S.U. supervised the project.

}
\end{raggedright}

High-resolution tissue imaging is often compromised by sample-induced optical aberrations that degrade resolution and contrast. While wavefront sensor-based adaptive optics (AO) can measure these aberrations, such hardware solutions are typically complex, expensive to implement, and slow when serially mapping spatially varying aberrations across large fields of view. Here, we introduce \model~(Adaptive Optical Vision Fourier Transformer)---a machine learning-based aberration sensing framework built around a 3D multistage Vision Transformer that operates on Fourier domain embeddings. \model~infers aberrations and restores diffraction-limited performance in puncta-labeled specimens with substantially reduced computational cost, training time, and memory footprint compared to conventional architectures or real-space networks. We validated \model~on live gene-edited zebrafish embryos, demonstrating its ability to correct spatially varying aberrations using either a deformable mirror or post-acquisition deconvolution. By eliminating the need for the guide star and wavefront sensing hardware and simplifying the experimental workflow, \model~lowers technical barriers for high-resolution volumetric microscopy across diverse biological samples.

\end{abstract}

\maketitle

\newpage

\section{Introduction}\label{sec:introduction}

\begin{figure*}[h!]
    \centering
    \includegraphics[width=\textwidth]{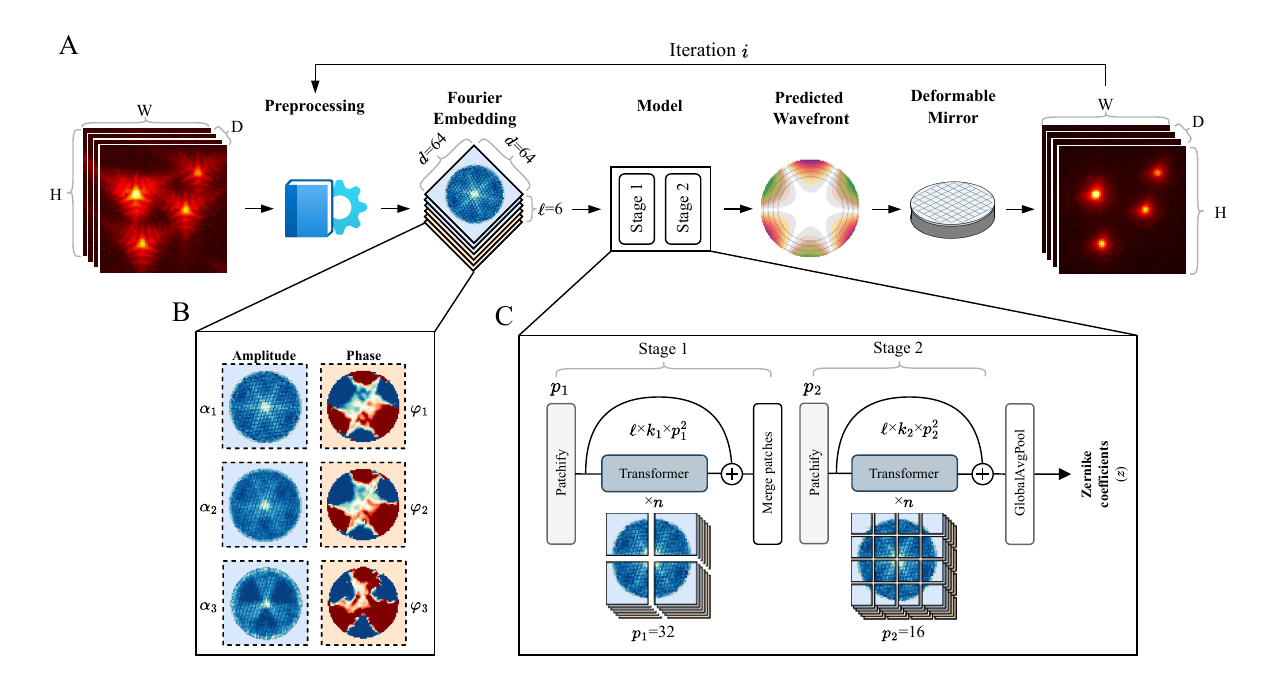}
    \caption{
        \textbf{\model~workflow}.
        \textbf{A}. \model~correction. An aberrated 3D volume is preprocessed and cast into a Fourier Embedding, which is passed to a 3D vision transformer model to predict the detection wavefront.  A deformable mirror (DM) compensates for this aberration, enabling acquisition of a corrected volume. 
        \textbf{B}. The Fourier embedding, $\mathcal{E}$. The Fourier Transform of the 3D volume is embedded into a lower space ($\mathcal{E} \in \mathbb{R}^{\ell \times d \times d}$), consisting of 3 amplitude planes ($\alpha_1,\alpha_2,\alpha_3$) and 3 phase planes ($\varphi_1,\varphi_2,\varphi_3$).
        \textbf{C}. \model~model. The Fourier embedding is input to a dual-stage 3D vision transformer model. 
        At each stage, the $\ell$ Fourier planes are tiled into $k$ patches (Patchify), applying a radially encoded positional embedding to each patch.  
        These patches are passed through $n$ Transformer layers.
        At the end of each stage, a residual connection is added, and the patches are merged back to the shape matching the stage input (Merge patches). 
        After all stages, the resulting patches are pooled (GlobalAvgPool) and connected with a dense layer to output the $z$ Zernike coefficients. 
    }
    \label{fig:pipeline-summary}
\end{figure*}

\begin{figure*}[!tp]
    \centering
    \includegraphics[width=\textwidth]{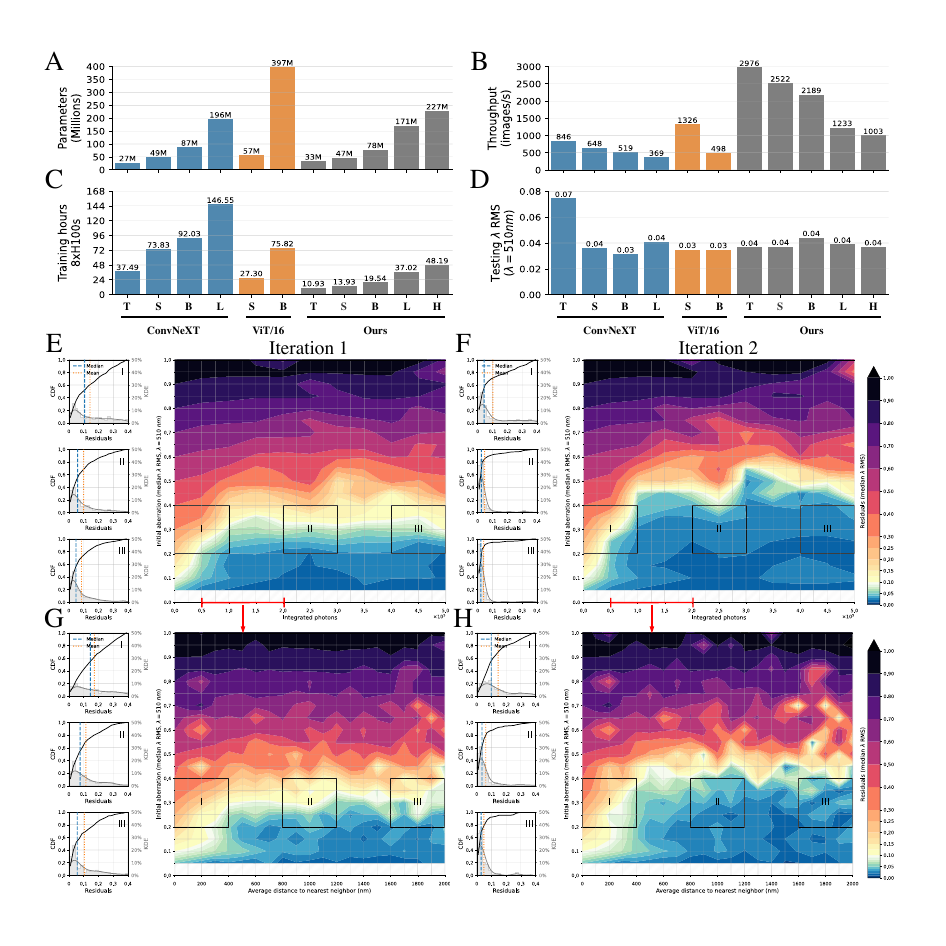}
    \caption{
        \textbf{Comparison of different state-of-the-art architectures when applied to 3D aberration sensing.}
        \textbf{A}. Total number of trainable parameters.
        \textbf{B}. Maximum predictions per second, using a batch size of 1024 on a single A100 GPU. Higher values are better.
        \textbf{C}. Training time on eight H100 GPUs. 
        \textbf{D}. Median $\lambda$ RMS residuals over 10K test samples after one correction, with aberrations ranging between $0.2\lambda$ to $0.4\lambda$, simulated with 50K to 200K integrated-photons.
        \textbf{E--F}. Median $\lambda$ RMS residuals using our Small model for a single bead over a wide range of SNR.
        \textbf{G--H}. Median $\lambda$ RMS residuals using our Small model for several beads (up to 150 beads), simulated at photon levels from 50K to 200K per bead.
        Lower values are better for all performance indicators listed here, except for \textbf{B}.
    }
    \label{fig:scaling-extended-summary-rms}
\end{figure*}

As we peer deeper into living organisms to reveal their inner workings, our view is increasingly compromised by sample-induced optical aberrations.  Numerous AO methods exist to compensate for these by using a wavefront shaping device that responds to a measurement of sample-induced aberration~\cite{zhang2023adaptive}.  
These methods differ in their complexity, generality, robustness, and practicality.  
In our lab, dependable success was had using a Shack-Hartmann (SH) sensor to measure the aberrations imparted on a guide star (GS) created by two-photon excitation (TPE) fluorescence within the specimen~\cite{wang2014rapid}, and we have used this approach extensively in adaptive optical lattice light sheet microscopy (AO-LLSM) to study four dimensional (4D) subcellular dynamic processes within the native environment of whole multicellular organisms~\cite{liu2018observing}.

Several recent approaches dispense with the cost and complexity of hardware-based wavefront measurement in favor of directly inferring aberrations from the microscope images themselves via machine learning (ML)~\cite{saha2020practical,rai2023deep,zhang2023deep,hu2023universal,kang2024coordinate} (Supplementary Table~\ref{tab:ao_comparisons}).  
Based on our experience with a variety of specimens, any ML-AO approach suitable for AO-LLSM must meet the following specifications:

\paragraph{Speed:}
To maximize the range of spatiotemporal events that can be visualized, the time for the ML model to infer the aberrations across any volume should be less than the time needed to image it---typically a few seconds in LLSM for a volume that encompasses a handful of cells.

\paragraph{Robustness:}
The model must accurately predict the vast majority of aberrations encountered in practice---for AO-LLSM in zebrafish embryos, typically up to $5\lambda$ peak-to-valley (P-V) in any combination of the first 15 Zernike modes ($Z_0^0$ through $Z_4^{\pm4}$ Supplementary Fig.~\ref{fig:zernike_pyramid}).

\paragraph{Accuracy:}
The method should be able to recover close to the theoretical 3D resolution limits of the microscope, regardless of the distribution of spatial frequencies within the specimen.

\paragraph{Noninvasiveness:}
The method should provide accurate correction without unduly depleting the fluorescence photon budget within the specimen or perturbing its native physiology.

As none of the aforementioned ML-AO methods meet all these specifications, we endeavored to create one better suited to the needs of AO-LLSM.
Our baseline model architecture, selected from an ablation study (Appendix~\ref{sup:ablation}, Supplementary Fig.~\ref{fig:model-transformer}--\ref{fig:embedding}), contains two transformer stages with patches of 32 and 16 pixels, respectively (Fig.~\ref{fig:pipeline-summary}c).  

Priors can greatly improve the performance of any ML approach.  
For our method, we depend on the prior that each isoplanatic subvolume (\ie having the same aberration) within the larger volume of interest contains one or more fluorescent puncta of true sub-diffractive size. 
Here, we introduce these by using genome-edited specimens expressing fluorescent protein-fused versions of AP2---an adaptor protein that targets clathrin-coated pits (CCPs) ubiquitously present at CCPs located on the plasma membrane of all cells (Methods).  
While this entails a one-time upfront cost for each specimen type,
it noninvasively produces a robust signal for AO correction
that does not preclude simultaneously imaging another subcellular target that occupies the same fluorescence channel, provided they are computationally separable~\cite{ashesh2025microsplit}.
\section{Results}
\label{sec:results}

\subsection{Benchmark comparisons of \model~to other architectures}

%
We created five variants of \model~by varying the numbers of layers and heads in each stage (Supplementary Table~\ref{tab:variants}) to explore the tradeoffs between model size (number of parameters and memory footprint), speed (floating-point operations (FLOPs) required, training time, and latency), and prediction accuracy (Supplementary Fig.~\ref{fig:scaling-multistage}).
To compare these to existing state-of-the-art architectures, we developed 3D versions of ViT and ConvNeXt for AO inference in three and four different size variants, respectively (Appendix~\ref{sup:benchmark}).  
We trained all models with the same set of $2 \times 10^6$ synthetic image volumes chosen to capture the full diversity of aberrations and imaging conditions likely to be encountered in AO-LLSM (Methods) and tested \model~on a separate set of $10^5$ image volumes created to find the limits of its accuracy when presented with an even larger range of aberration magnitudes, SNR, and number of fluorescent puncta (Appendix~\ref{sup:insilico}).  
We also tested the performance of all models and variants on $10^4$ image volumes from a test set that contained only a single punctum in each (Fig.~\ref{fig:scaling-extended-summary-rms}, Supplementary Fig.~\ref{fig:scaling-archs-rms}--\ref{fig:scaling-eval-rms}, and Supplementary Table~\ref{tab:benchmark}). 

While all models but the smallest variant of ConvNeXt were able to reduce the median residual error in a single iteration of aberration prediction to less than the diffraction limit (Fig.~\ref{fig:scaling-extended-summary-rms}d), 
\model~excelled in its parsimonious use of compute resources: 
training time using a node with eight NVIDIA H100 GPUs (Fig.~\ref{fig:scaling-extended-summary-rms}c); 
training FLOPs (Supplementary Fig.~\ref{fig:scaling-archs-rms}c); 
and memory footprint (Supplementary Fig.~\ref{fig:scaling-archs-rms}f).  
This reflects the benefits of our multistage architecture: faster convergence by learning features across different scales, accurate prediction even at comparatively modest model size (Fig.~\ref{fig:scaling-extended-summary-rms}a), highest inference rate among the models tested (Fig.~\ref{fig:scaling-extended-summary-rms}b), and fastest single-shot inference time (``latency'', Supplementary Fig.~\ref{fig:scaling-archs-rms}h).
Given its small size and low latency, we chose the Small variant of \model~as our primary model for evaluation.

\subsection{In silico evaluations of \model}

Diffraction-limited performance is conventionally defined by wavefront distortions below $\approx0.075\lambda$ RMS or $\lambda/4$ peak-to-valley, corresponding to a Strehl ratio of 0.8 under the Rayleigh quarter-wave criterion~\cite{mahajan1982strehl,bentley2012field}. 
In silico evaluation using the $10^4$ single punctum test images show \model~recovers diffraction-limited performance in a single iteration in nearly all trials where the initial aberration is $< 0.30\lambda~RMS$ and the integrated signal is $> 5 \times 10^4$ photons (Fig.~\ref{fig:scaling-extended-summary-rms}e).
The corrective range increases to 0.40, 0.50, 0.55, and $0.6\lambda~RMS$ for 2-5 iterations, respectively, although $\sim5 \times 10^4$ photons remains the floor of required signal (Fig.~\ref{fig:scaling-extended-summary-rms}f, Supplementary Fig.~\ref{fig:eval-snrheatmap}).  
This is comparable to the signal needed for SH wavefront sensing~\cite{wang2014rapid,liu2018observing} and at least 3x lower than that needed for PhaseRetrieval. 
In comparison, PhaseNet~\cite{saha2020practical} and PhaseRetrieval~\cite{hanser2004phase_retrieval} extend the diffraction limited range only slightly (initial aberration $< 0.15\lambda~RMS$) after a single iteration on the same test data (Supplementary Fig.~\ref{fig:comparisons-phase-retrieval-rms}a, b) and, in contrast to \model, do not appreciably increase this range after multiple iterations (Supplementary Fig.~\ref{fig:comparisons-phase-retrieval-psf-rms-iterations}a-f).  
PhaseRetrieval \textit{does} advantageously reduce residuals after a single iteration over a much broader range of initial aberration than \model, and this trend continues with further iteration, albeit never back to the diffraction limit (Supplementary Fig.~\ref{fig:comparisons-phase-retrieval-psf-rms-iterations}a-c). 
However, this advantage is lost if the fiducial bead is not centered in the field of view (FOV), and the predictive power of PhaseNet is lost completely under the same circumstances (Supplementary Fig.~\ref{fig:comparisons-phase-retrieval-rms}d, e), because the widefield 3D image of the bead is then clipped.  
Furthermore, PhaseRetrieval and PhaseNet assume \textit{a priori} the existence of only a single bead.  
\model~is trained on 1--5 puncta falling anywhere within the FOV but, thanks to the normalization step to eliminate phase fringes from multiple puncta (Supplementary Fig.~\ref{fig:embedding}d), produces inferences comparably accurate to a single punctum for up to 150 puncta, provided their mean nearest neighbor distance is $>$ 400nm (Fig.~\ref{fig:scaling-extended-summary-rms}g, h, Supplementary Fig.~\ref{fig:eval-densityheatmap}).
Indeed, \model~relies on the combined signal of multiple native but dim sub-diffractive biological assemblies such as CCPs to achieve accurate inferences.

\begin{figure*}[!tp]
    \centering
    \includegraphics[width=\textwidth]{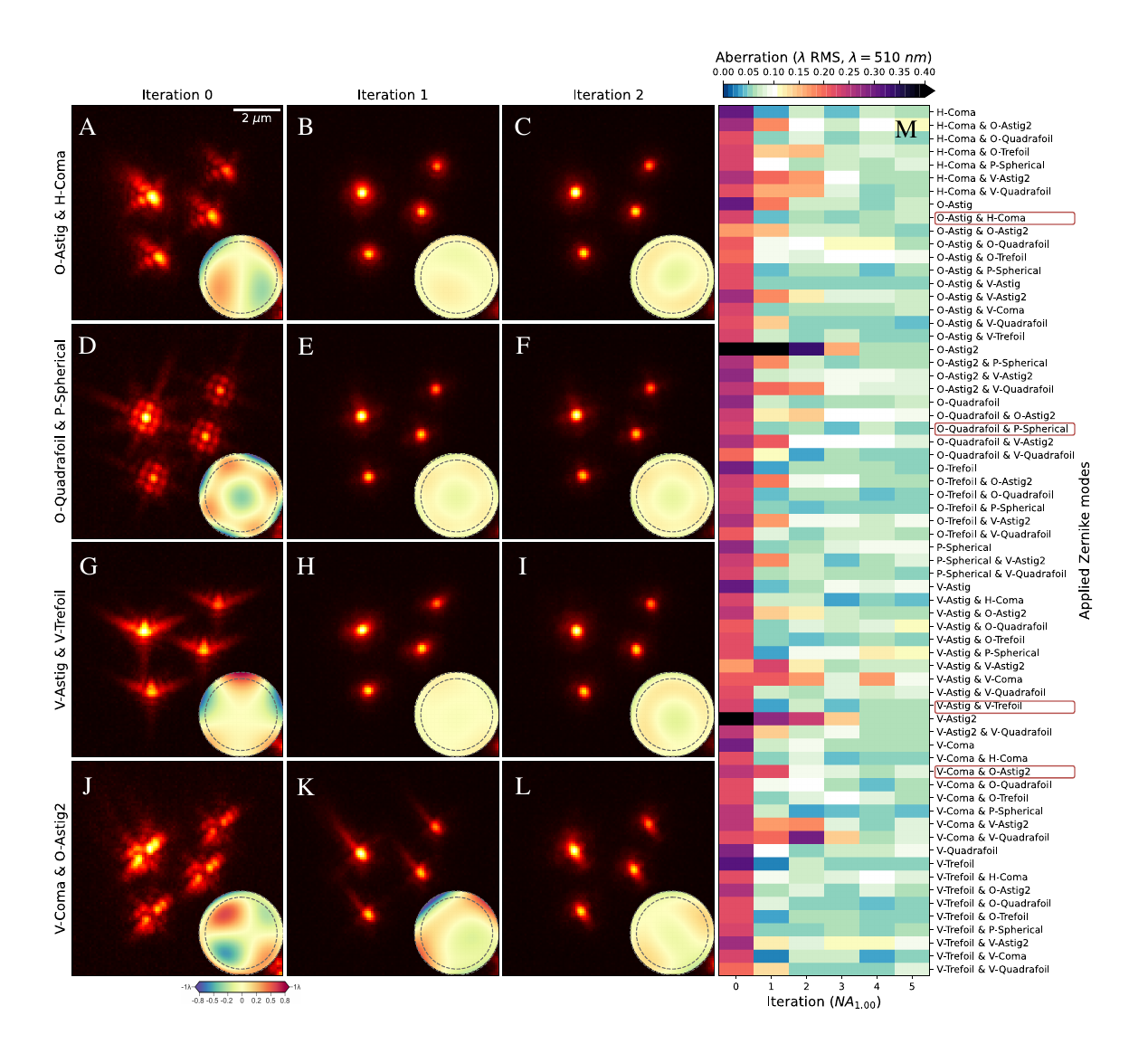}
    \caption{
    \textbf{Experimental correction of beads with initial artificial aberrations.} 
    \textbf{A--L}. Four examples, 
    O-Astig \& H-Coma $(Z^{m=\text{-}2}_{n=2} + Z^{m=1}_{n=3})$, 
    O-Quadrafoil \& P-Spherical $(Z^{m=\text{-}4}_{n=4} + Z^{m=0}_{n=4})$, 
    V-Astig \& V-Trefoil $(Z^{m=2}_{n=2} + Z^{m=\text{-}3}_{n=3})$, 
    V-Coma \& O-Astig2 $(Z^{m=\text{-}1}_{n=3} + Z^{m=\text{-}2}_{n=4})$, 
    where the initial aberration was artificially applied by the DM.
    \textit{Iteration 0} shows XY maximum projection of four beads with initial aberration imaged using LLS, providing initial conditions for \model~predictions. 
    \textit{Iteration 1} shows the resulting field of beads after applying \model~prediction to the DM.
    \textit{Iteration 2} shows the results after applying the \model~prediction measured from \textit{Iteration 1}. 
    Insets show the \model~predicted wavefront over the $\text{NA}=1.0$ pupil with a dashed line at $NA=0.85$
    \textbf{M}. Heatmap of the residual aberrations (measured via PhaseRetrieval on isolated bead) after applying \model~predictions, starting with a single Zernike mode up to Mode 14 ($Z^{m=4}_{n=4}$) across up to 5 iterations.
    }
    \label{fig:beads-main-rms}
\end{figure*}

\begin{figure*}[!tp]
    \centering
    \includegraphics[width=\textwidth]{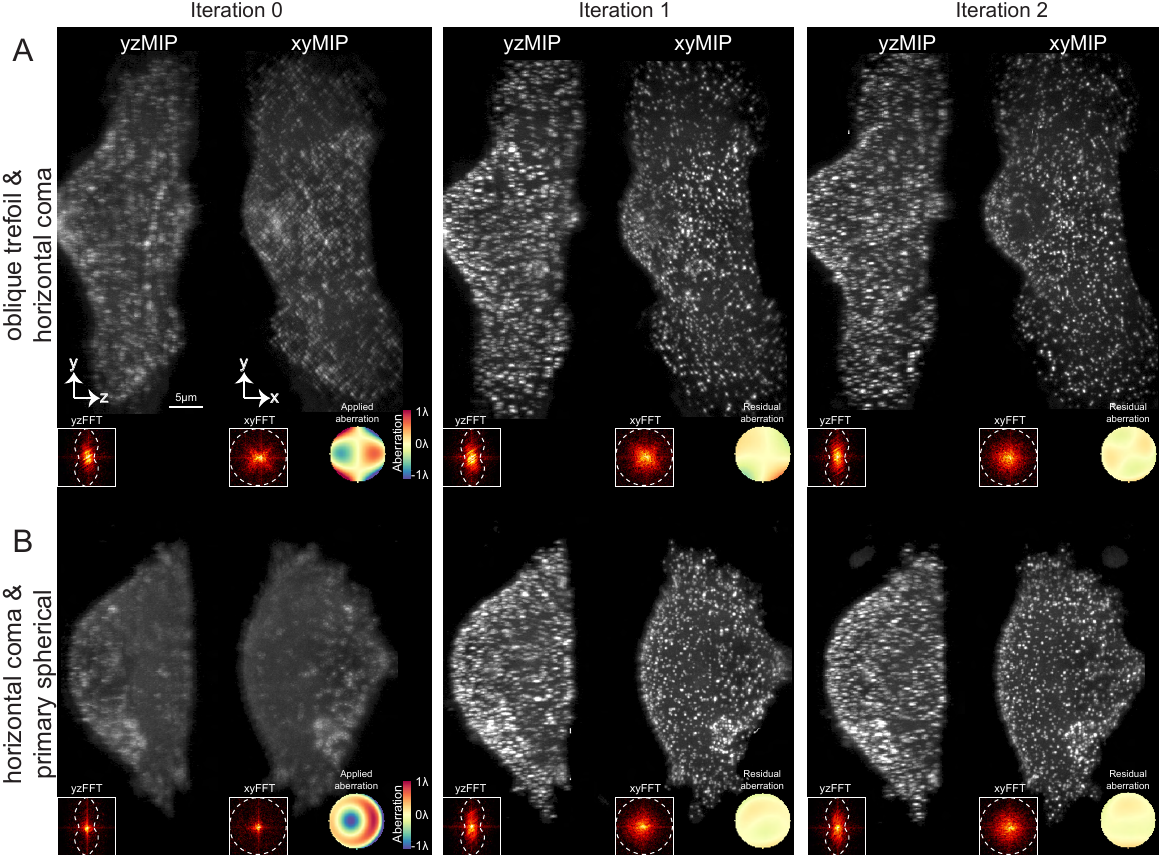}
    \caption{
    \textbf{Correction of aberrations in live SUM159-AP-2 cells expressing $\sigma$2-eGFP.} 
    \textbf{A.} 3D volume SUM159-AP2 cells represented as  XY and YZ MIPs covering a $15.7 \times 55.6 \times 25.6$ $\mu$m$^3$ FOV after applying a 2.9$\lambda$ peak-to-valley (P-V)  aberration to the DM. 
    This aberration combines horizontal coma $Z_3^1$ and oblique trefoil $Z_3^3$. 
    \textbf{B.} XY and YZ MIPs of a similar FOV with 3.1$\lambda$ P-V aberration composed of horizontal coma ($Z_3^1$) and primary spherical ($Z_4^0$).
    In both cases,  near diffraction-limited performance was recovered after two iterations.  
    The insets show FFTs and corresponding wavefronts for each iteration.  Scale bar, $5\mu$m
    }
    \label{fig:cells}
\end{figure*}

\begin{figure*}[!tp]
    \centering
    \includegraphics[width=.65\textwidth]{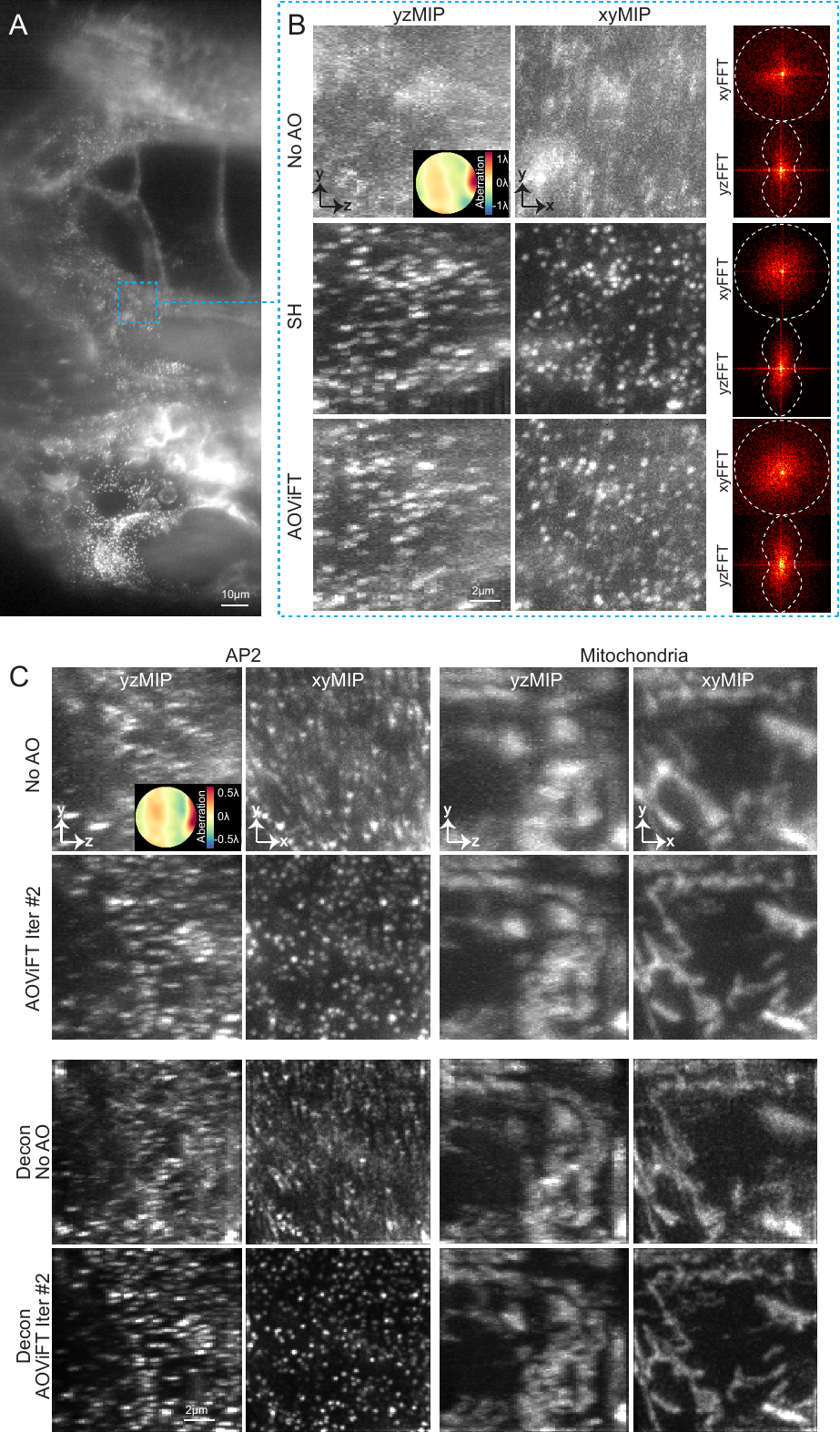}
    \caption{
    \textbf{In vivo, in situ correction of native aberrations in zebrafish embryos.} 
    \textbf{A.} XY MIP of a 72 hpf gene-edited zebrafish embryo expressing endogenous AP2-mNeonGreen, exhibiting native and spatially varying  aberrations near the notochord. Scale bar, $10\mu$m.
    \textbf{B.} Enlarged view of the dashed blue box in (A). The XY and YZ MIPs, along with the corresponding FFTs of a $12.5 \times 12.5 \times 12.8 \mu$m$^3$ FOV, show $\sim$2$\lambda$ P-V of sample-induced aberration without AO (top row),
        corrected by SH (middle row), and corrected by \model~(bottom row).
    The contrast for each volume was scaled to its 1st and 99.99th percentile intensity values. Scale bar, $2\mu$m.
    \textbf{C.} XY and YZ MIPs of a different gene-edited zebrafish embryo  expressing exogenous AP2-mNeonGreen and injected mRNA for mChilada-Cox8a (to visualize mitochondria). 
    The AP2 signal was used to infer the underlying aberration, and the same correction was applied to both channels. 
    The top row shows $\sim$1.5$\lambda$ P-V aberration; 
    the second row shows \model~correction after two iterations. 
    The third and fourth rows present the results of OTF masked Wiener (OMW) deconvolution without and with \model~corrected volumes, respectively.  Scale bar, $2\mu$m.
    }
    \label{fig:fish_online}
\end{figure*}

\begin{figure*}[!tp]
    \centering
    \includegraphics[width=\textwidth]{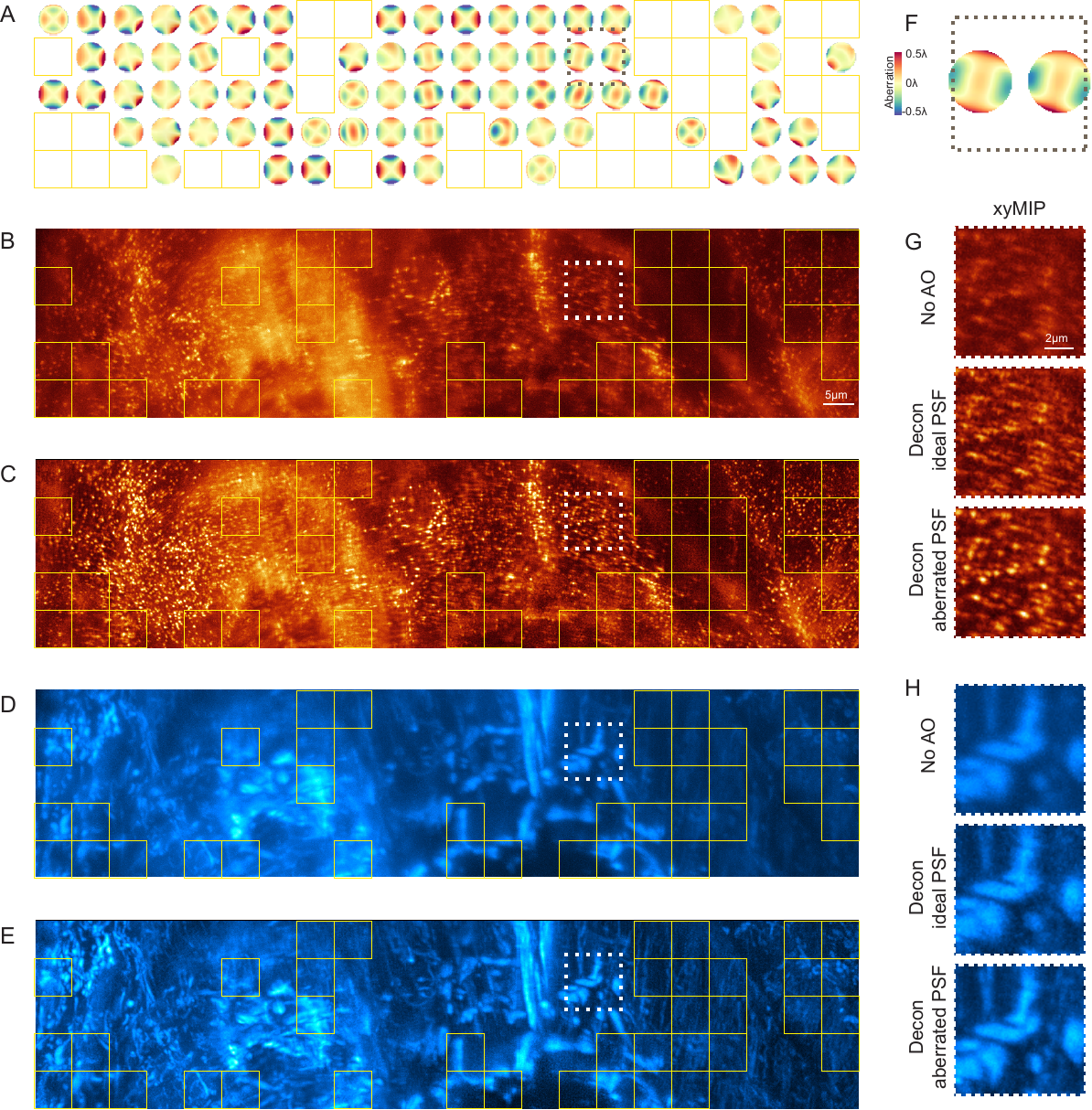}
    \caption{
    \textbf{Correcting aberrations post acquisition using spatially varying deconvolution in a zebrafish embryo.} 
    \textbf{A.} Isoplanatic patch map determined by \model~for 204 tiles ($6.3\times 6.3 \times 12.8 \mu$m$^3$ each), spanning $37\times 211 \times 12.8 \mu$m$^3$ FOV in a live, gene-edited zebrafish embryo expressing endogenous AP2-mNeonGreen. 
    The yellow box marks areas with insufficient spatial features to accurately determine aberrations; an ideal PSF was used for OMW deconvolution in these regions.
    \textbf{B.} XY MIP of the AP2 signal without AO. Scale bar, $5\mu$m. 
    \textbf{C.} XY MIP of each tile after deconvolution with spatially varying PSFs predicted by \model. 
    \textbf{D--E.} Raw (D), and deconvolved (E) XY MIPs of the mitochondria channel for the same region.
    \textbf{F.} Enlarged view of the wavefronts within the black dashed box in (A).
    \textbf{G--H.} Zoomed views of  AP2 (G), and mitochondria (H)  structures from (B--E), comparing No AO to OMW deconvolution using either an ideal PSF or spatially varying tile-specific aberrated PSFs predicted by \model. Scale bar, $2\mu$m.
    }
    \label{fig:fish_offline}
\end{figure*}

\subsection{Experimental characterization on fiducial beads}

We performed all experiments using the AO-LLSM microscope schematized in Supplementary Fig.~\ref{fig:schematic_lls}. 
For initial characterization of \model's ability to correct a wide range of possible aberrations, we performed 66 separate experiments wherein we:
\begin{enumerate}[label=\alph*)]
    \item introduced aberration by applying to the deformable mirror (DM) one of the 66 possible combinations of one or two Zernike modes (from the first 15, excluding piston, tip/tilt, and defocus), with each mode set to 0.2$\lambda~RMS$ amplitude;
    \item used AO-LLSM with the \textit{MBSq-35} LLS excitation profile of Supplementary Table~\ref{tab:lightsheets} to image a field of 100 nm diameter fluorescent beads with this aberration;
    \item used \model~to predict the aberration;
    \item applied the corrective pattern to the DM;
    \item repeated b-d for 5 iterations.
\end{enumerate}
In 45 cases, we recovered diffraction-limited performance in two iterations (Fig.~\ref{fig:beads-main-rms}), and in 5 iterations for 11 more (Supplementary Fig.~\ref{fig:beads-si}).  
In the remaining 10 cases, aberrations were reduced by at least 50\% after 5 iterations.

\subsection{Correction of aberrations on live cultured cells}

We next tested the ability of \model~to correct aberrations during live cell imaging under biologically relevant conditions of limited signal, dense puncta, and specimen motion.  
To this end, we applied aberrations to the DM and imaged cultured SUM 159 human breast cancer-derived cells gene edited to produce endogenous levels of the clathrin adapter protein AP-2 tagged with eGFP.  
This yielded numerous membrane-bound CCPs at various stages of maturation that were suitable for aberration measurement.  
In one example (Fig.~\ref{fig:cells}a), we applied a 2.9$\lambda$ peak-to-valley (P-V) aberration to the DM consisting of a mix of horizontal coma and oblique trefoil ($Z_3^1$ and $Z_3^3$), and recovered near diffraction-limited performance after two iterations (Supplementary Table~\ref{tab:cells_wavefronts}).  
Peak signal at the CCPs increased 2-3 fold post-correction, and the spatial frequency content as seen in orthoslices through the 3D FFT (insets at bottom) increased in every iteration. 
In another case (Fig.~\ref{fig:cells}b), we reduced a 3.1$\lambda$ P-V aberration composed of a combination of horizontal coma and primary spherical ($Z_3^1$  and $Z_4^0$) to 0.069$\lambda~RMS$ after two iterations, increasing CCP signal by 3-4 fold (Supplementary Table~\ref{tab:cells_wavefronts}).  
Four more examples of correction on cells and fiducial beads after applying single modes of 1$\lambda$ P-V aberration are given in Supplementary Fig. ~\ref{fig:Figure4_SI1}, and five more examples of two mode correction are shown in Supplementary Fig.~\ref{fig:Figure4_SI2}.

\subsection{In vivo correction of native aberration within a zebrafish embryo}

As a transparent vertebrate, zebrafish are a popular model organism for imaging studies.  
However, the spatially heterogenous refractive index within multicellular organisms and the discontinuity of refractive index at their surfaces with respect to the imaging medium result in aberrations that vary throughout their interiors (Fig.~\ref{fig:fish_online}a).  
We corrected $\sim$2$\lambda$ P-V aberration in one such region (Fig.~\ref{fig:fish_online}b, top) with \model~(Fig.~\ref{fig:fish_online}b, bottom) near the notochord of a transgenic zebrafish embryo 72 hours post fertilization expressing AP2-mNeonGreen in CCPs at the membranes of all cells (\textit{ap2s1}:\textit{ap2s1-mNeonGreenb$^{k800}$}, Methods) and recovered spatial frequencies across the corrected volume (FFTs at right) comparable to SH correction over the same region (Fig.~\ref{fig:fish_online}b, middle).  

In a second embryo expressing AP2-mNeonGreen in CCPs and mChilada-Cox8a in mitochondria (Fig.~\ref{fig:fish_online}c), we used the mNeonGreen signal to correct $\sim$1.5$\lambda$ P-V aberration (top row) in one region, which provided an aberration-corrected view of both CCPs and mitochondria (second row).  
Deconvolution of the aberrated images using an assumed ideal PSF only amplified high frequency artifacts (third row), but provided a more accurate representation of sample structure (bottom row) for the aberration-corrected ones by compensating for known attenuation of high spatial frequencies in the ideal optical transfer function (OTF).

\subsection{Correction of spatially varying aberrations in vivo}

With guide star illuminated SH sensors, aberration measurement is not accurate unless it is confined to a single isoplanatic region.  
However, these are often much smaller than the volume of interest, and their boundaries are not generally known \textit{a priori}.  
Consequently, microscopists are often forced to map aberrations by serial SH measurement over many small, tiled subregions whose dimensions are a matter of educated guesswork.  
On the other hand, with \model~we generated a complete map of 204 aberrations (Fig.~\ref{fig:fish_offline}a) at 6.3 $\mu$m intervals over 37$\times$211$\times$12.8 $\mu$m$^{3}$ in a live zebrafish embryo 48 hpf (Fig.~\ref{fig:fish_offline}b, d) in $\sim$1.5 minutes on a single node of four A100 GPUs.
Unfortunately, it is not possible to apply a corrective pattern to a single pupil conjugate DM and thereby correct this spatially varying aberration across the entire FOV.  
One option would be to apply each aberration in turn and image the tiles one by one, or together in groups of similar aberration.  
Although slow, this would recover the full information of which the microscope is capable.  
However, a much faster and simpler alternative is to deconvolve each raw image tile with its own unique aberrated PSF (Fig.~\ref{fig:fish_offline}c, e).  
This does not recover full diffraction-limited performance, but it does suppress aberration-induced artifacts and provides a more faithful representation of the underlying sample structure (Figs.~\ref{fig:fish_offline}f--h).

\section{Discussion}\label{sec:discussion}
\model~provides accurate mapping of spatially varying sample-induced aberrations in specimens having sub-diffractive puncta. Although \model~can be slower than using SH for a single ROI, it gains a substantial net speed advantage when mapping multiple ROIs across a large field of view (FOV) due to its parallelizable inference framework (Supplementary Table~\ref{tab:performance_SH_vs_model}).  Moreover, its throughput can be further accelerated via distributed GPU processing across multiple nodes and by compiling the model with TensorRT\footnote{\url{https://docs.nvidia.com/deeplearning/tensorrt/pdf/TensorRT-Developer-Guide.pdf}} for optimized inference.  
Unlike \model, SH measurement with a TPE guide star has the key advantage of being agnostic to the fluorescence distribution within each isoplanatic region, but requires additional hardware (TPE laser, galvos, SH sensor) and the TPE power level must be carefully monitored to minimize photodamage.  
In addition, since the isoplanatic regions are not known \textit{a priori}, the initial measurement grid for SH sensing must be very dense to accurately map aberrations and their rate of change across the FOV, requiring additional time at an additional cost to the photon budget.  
Conversely, \model~determines the aberration map from a single large 3D image volume, and can therefore iteratively adjust tile sizes or positions \textit{in silico} as needed until the map converges to an accurate solution.

Although trained for a specific LLS type (Supplementary Table~\ref{tab:lightsheets}), \model~retained predictive capability when tested \textit{in silico} with other light sheets as well (Supplementary Fig.~\ref{fig:eval-mb-rms}).  
While training specifically for such light sheets might increase the predictive range even further, a more fruitful path might be to augment the synthetic training data with light sheets axially offset from the detection focal plane to replace the closed-loop hardware based mitigation of such offsets needed now~\cite{liu2018observing}.  
Future models might leverage ubiquitous subcellular markers, such as plasma membranes or organelles, rather than genetically-expressed diffraction-limited puncta, provided these markers contain sufficient high spatial frequency content for accurate inference of aberrations. Finally, to enhance generalizability  of \model~and reduce overfitting to narrowly defined imaging scenarios, future models should incorporate a more diverse range of light-sheets, specimen types, and labeling strategies.

Development of \model~highlighted the challenges of constructing a 3D transformer-based architecture for AO correction. Each iteration of model design, training, and testing required specialized simulated data pipelines, large GPU resources, and extensive hyperparameter tuning---leading to lengthy model development cycles. A key bottleneck is the absence of universally applicable, large pre-trained models for volumetric imaging data---a limitation that extends beyond adaptive optics applications. 

Unlike the natural image domain, where ViT benefited from extensive training on standardized 2D datasets, a comparable ``foundation model'' for 3D microscopy is pending the collection of similar datasets.  This gap severely limits how far and how quickly new methods like \model~can be generalized. While our work highlights the feasibility of building a solution for a given task (\eg AO corrections under specific imaging conditions), adapting to new scenarios such as new sample types, microscope geometries, or aberration ranges typically requires substantial retraining and additional data curation.

These limitations highlight the need for pre-trained foundation models in volumetric microscopy.  We consider \model, a 3D vision transformer model, as a stepping stone towards the more ambitious goal of creating a 4D model pre-trained on massive volumetric microscopy datasets. Such a model could be fine-tuned for tasks across spatial scales (from molecules to organisms), and temporal scales (from stochastic molecular kinetics to embryonic development)~\cite{betzig2024cell}.  Realizing this vision would require petabytes of high-quality curated 4D datasets and significant computational resources. However, successful implementation would dramatically shorten development timelines, improve generalization, and reduce the overhead of custom training for varied experimental setups or microscope configurations.

\section{Methods}\label{sec:methods}

\subsection{AO-LLS microscope}\label{sec:microscope}

Imaging was performed using an adaptive optical lattice light-sheet (AO-LLS) microscope similar to one described previously~\cite{liu2018observing} (Supplementary Fig.~\ref{fig:schematic_lls}, Supplementary Table~\ref{tab:imaging_configuration}).  
Briefly, 488-nm and 560nm lasers (500 mW 2RU-VFL-P-500-488-B1R and 1000mW 2RU-VFL-P-1000-560-B1R, MPB Communications Inc.) were modulated via an acousto-optical tunable filter (AOTF; Quanta-Tech, AA OptoElectronic, AOTFnC-400.650-CPCh-TN) and shaped into a stripe by a Powell lens (Laserline Optics Canada, LOCP-8.9R20-2.0) and a pair of 50- and 250-mm cylindrical lenses (25 mm diameter; Thorlabs, ACY254-050, LJ1267RM-A). The stripe illuminated a reflective, phase-only, gray-scale spatial light modulator (SLM; Meadowlark Optics, AVR Optics, P1920-0635-HDMI, 1920 $\times$ 1152 pixels) located at a sample conjugate plane. 
An 8-bit phase pattern written to the SLM generated the desired light-sheet pattern in the sample, and an annular mask (Thorlabs Imaging) at a pupil conjugate plane blocked unwanted diffraction orders before the light passed through the excitation objective (Thorlabs, TL20X-MPL). 
A pair of pupil conjugate galvanometer mirrors (Cambridge Technology, Novanta Photonics, 6SD11226 and 6SD11587) scanned the light sheet at the sample plane.  The sample was positioned at the common foci of the excitation and detection objectives by a three axis XYZ stage (Smaract; MLS-3252-S, SLS-5252-S, SLS-5252-S). 
Fluorescence emission from the sample was collected by a detection objective (Zeiss, 20$\times$, 1.0 NA, 421452-9800-000), reflected off a pupil conjugate deformable mirror (DM; ALPAO, DM69) that applied aberration corrections, and then recorded on two sample conjugate cameras (Hamamatsu ORCA Fusion). 

Shack Hartmann measurements (Supplementary Fig.~\ref{fig:schematic_sh}) were performed on the same microscope by localizing the intensity maxima (on a Hamamatsu ORCA Fusion) formed by the emitted light after passage through a pupil conjugate lenslet array (Edmund Optics, 64-479).
The positional shifts of these maxima relative to those seen with no specimen present encode the pupil wavefront phase~\cite{wang2014rapid}, which can then be reconstructed.

\subsection{Integration with microscope}\label{subsec:mosaic}

\model~inference is routinely performed on the microscope acquisition PC (Intel Xeon, W5-3425, Windows 11, 512GB RAM, NVIDIA A6000 with 48GB VRAM). Inferences are made in an Ubuntu Docker container based on the TensorFlow NGC Container (24.02-tf2-py3) running in parallel with the microscope control software.  Data communication between \model~and the microscope control software is handled through the computer's file system.  Image files and command-line parameters are passed to the model, and an output text file reports the resultant DM actuator values (Fig.~\ref{fig:schematic_cycle}).  When a volume is large enough to require tiling and dozens of volumes need to be processed, model inferences are parallelized and run using a SLURM compute cluster consisting of 4 nodes, each node containing four NVIDIA A100 80GB.

\subsection{Fluorescent beads and cells expressing fluorescent endocytic adaptor AP2}\label{sec:cells}

The 25 mm coverslips (Thorlabs, CG15XH) used for imaging beads, cells and zebrafish embryos were first cleaned by sonication in 70\% ethanol followed by Milli-Q water, each for at least 30 minutes. They were then stored in Milli-Q water until use. Gene-edited SUM159 AP2-eGFP cells~\cite{aguet2016membrane} were grown in Dulbecco's modified Eagle's medium (DMEM)/F12 with GlutaMAX (Gibco, 10565018) supplemented with 5\% fetal bovine serum (FBS; Avantor Seradigm 89510-186), 10mM HEPES (Gibco 15630080), 1$\mu$g/ml hydrocortisone (Sigma H0888), 5$\mu$g/ml insulin (Sigma I9278). 
Fluorescent beads (0.2 $\mu$m diameter, Invitrogen FluoSpheres Carboxylate-Modified Microspheres, 505/515 nm, F8811 or 0.2 $\mu$m diameter Tetraspeck, Thermo Fisher Scientific Invitrogen T7280) alone or with cells at 30-50\% confluency were deposited onto plasma-treated and poly-D-lysine (Sigma-Aldrich, P0899) treated 25mm coverslips. Cells were cultured under standard conditions ($37^{\circ}$C, 5\% CO2, 100\% humidity) with twice weekly passaging. 
The SUM159 AP2-eGFP cells were imaged in Leibovitz's L-15 medium without phenol red (Gibco, 21083027), with 5\% FBS (American Type Culture Collection, SCRR-30-2020), 100 $\mu$M Trolox (Tocris, 6002), and 100 $\mu$g/ml Primocin (InvivoGen, ant-pm-1) at $37^{\circ}$C. 
Aberrations of approximately 1$\lambda$ P-V were induced using a deformable mirror in 10 configurations of Zernike modes ($Z_2^2$, $Z_3^{-3}$, $Z_3^{-1}$, $Z_4^0$, and their pairwise combinations). 
Widefield point spread functions were collected from 0.2 $\mu$m fluorescent beads to confirm the aberrations applied and residual aberrations after correction (Supplementary Table~\ref{tab:imaging_configuration}).

\subsection{Zebrafish embryos expressing fluorescent AP2 and mitochondria}\label{sec:fish}

Genome edited \textit{ap2s1} expressing zebrafish (Genome editing of \textit{ap2s1}, \textit{ap2s1}:\textit{ap2s1-mNeonGreen$^{bk800}$}, Appendix~\ref{sup:ap2s1}) were injected with cox8-mChilada mRNA for two color experiments.
The N-terminal 34 amino acids of Cox8a was cloned into a pMTB backbone with a linker and mChilada coding sequence on the c-terminus (unpublished, gift from Nathan Shaner).  
The plasmid was linearized, and mRNA was synthesized using a SP6 mMessage mMachine transcription kit (Thermo Fisher). 
RNA was purified using an RNeasy kit (Qiagen) and embryos were injected with 2 nl of 10 ng/$\mu$l Cox8a-mChilada, 100 mM KCl, 0.1\% phenol red, 0.1 mM EDTA, and 1 mM Tris, pH 7.5.  
Zebrafish embryos were first nanoinjected with 3 nL of a solution containing 0.86 ng/$\mu$L $\alpha$-bungarotoxin protein, 1.43$\times$ PBS, and 0.14\% phenol red. The injected embryos were mounted for imaging using a custom, volcano-shaped agarose mount. 
Each mount was constructed by solidifying a few drops of 1.2\% (wt/wt) high-melting agarose (Invitrogen UltraPure Agarose, 16500--100, in 1$\times$ Danieau buffer) between a 25 mm glass coverslip and a 3D-printed mold (Formlabs Form 3$+$, printed in clear v4 resin). This created ridges that formed a narrow groove. 
A hair-loop was used to orient the embryo within the agarose groove, positioning the left lateral side upward. Subsequently, 10–20 $\mu$L of 0.5\% (wt/wt) low-melt agarose (Invitrogen UltraPure LMP Agarose, 16520--100, in 1$\times$ Danieau buffer) preheated to $40^{\circ}$C, containing 0.2 $\mu$m Tetraspeck microspheres, was added on top of the embryo. This layer solidified around the embryo to secure it while providing fiducial beads for sample finding. 
Once the low-melt agarose solidified, the volcano-shaped mount was held by a custom sample holder for imaging. The embryo was oriented so that its anterior-posterior axis lay parallel to the sample x-axis, with the anterior end facing the excitation objective and the posterior end facing the detection objective. The microscope objectives and the sample was immersed in a $\sim$50 mL bath of 1$\times$ Danieau buffer were fully submerged, ensuring the embryo remained in buffered media.
Measurements for \model~and Shack Hartmann were done serially on the same FOV to compare the aberration corrections of both methods (Supplementary Table~\ref{tab:imaging_configuration}).

\subsection{Spatially varying deconvolution}\label{sec:SVdecon}

To compensate for sample-induced aberrations post-acquisition, we performed a tile-based spatially varying deconvolution on each 3D volume. Each volume was first subdivided into multiple 3D tiles approximating isoplanatic patches. A \model~predicted PSF (for compensation) or an ideal PSF (for no compensation) was assigned to each tile, and aberrations were corrected using OTF masked Wiener (OMW) deconvolution~\cite{ruan2024image}. To minimize boundary artifacts during deconvolution, the tile size was extended by half the PSF width at each boundary (32 pixels); after deconvolution, these overlaps were removed and the deconvolved core regions were stitched together to form the final corrected volume. All computations were done in MATLAB 2024a (Mathworks).

\subsection{Synthetic training/testing datasets}\label{sec:dataset}

To train a model for predicting optical aberrations from images of sub-diffractive objects in biological samples, we generated synthetic datasets encompassing a range of relevant variables (\eg~aberration modes and amplitudes, number and density of puncta, SNR). This synthetic dataset generation procedure is as follows:

For a single sub-diffractive punctum, the electric field in the rear pupil of the detection objective is given by:
\begin{align}\label{eq:system_pupil}
    E(k_x,k_y) = A(k_x,k_y) \exp(i\phi(k_x,k_y))
\end{align}
where $A(k_x,k_y)$  is the pupil amplitude and  $\phi(k_x,k_y)$ is the pupil phase. Under aberration-free conditions, $\phi(k_x,k_y)$ is a constant. 
We can empirically determine $A(k_x,k_y)$ by acquiring a widefield image of an isolated  sub-diffractive object (100 nm fluorescent bead), performing phase retrieval~\cite{hanser2004phase_retrieval,campbell2004generalized}, and applying  the opposite of the retrieved phase using a pupil conjugate DM so that $\phi(k_x,k_y)$ becomes a constant.

The electric field for the image of a single aberrated punctum is:
\begin{align}\label{eq:aberrated_pupil}
    E_{\text{abb}}(k_x,k_y) = A(k_x,k_y) \exp(i\phi_{\text{abb}}(k_x,k_y))
\end{align}
where the $\phi_{\text{abb}}(k_x,k_y)$ is described as a weighted sum of Zernike modes of unique amplitudes:
\begin{align}\label{eq:aberrated_phase}
    \phi_{\text{abb}}(k_x,k_y) = \sum_{m,n} \alpha_n^m Z_n^m(k_x,k_y)
\end{align}

Empirically, zebrafish induced aberrations for the microscopes used here are well described by combinations of 11 of the first 15 Zernike modes~\cite{lakshminarayanan2011zernike} (Supplementary Fig.~\ref{fig:zernike_pyramid}), for which $n \leq 4$, excluding  piston ($Z_0^0$), tip ($Z_1^{-1}$), tilt ($Z_1^1$), and defocus ($Z_2^0$) (as these represent phase offsets or sample translation). 
The distributions and amplitudes of the remainder are used to build the training set as discussed below.

The aberrated 3D detection PSF of a sub-diffractive punctum is  approximated by:
\begin{align}\label{eq:detection_psf}
    \text{PSF}_{\text{abb}}^{\text{det}}(x,y,z) = \left| \iint_{\text{pupil}} E_{\text{abb}}(k_x,k_y) \exp[i(k_x x + k_y y + k_z z)] \, \text{d}k_x \, \text{d}k_y \right|^2
\end{align}
where $k_z = \sqrt{\left( \frac{2\pi\eta}{\lambda} \right)^2 - k_x^2 - k_y^2}$,
 $\eta$ is the refractive index of the imaging medium, and $\lambda$ being the free-space wavelength of the fluorescence emission. 

For light sheet microscopy, the aberrated 3D overall PSF is:
\begin{align}\label{eq:overall_psf}
    \text{PSF}_{\text{abb}}^{\text{overall}}(x,y,z) = \text{PSF}^{\text{exc}}(z) \cdot \text{PSF}_{\text{abb}}^{\text{det}}(x,y,z)
\end{align}
where $\text{PSF}^{\text{exc}}(z)$ is given by the cross-section of the swept light sheet used for imaging. 
Examples of these PSFs are shown in Supplementary Fig.~\ref{fig:eval-mb-rms}I-V, with \textit{MBSq-35} in Supplementary Table~\ref{tab:lightsheets} used for training and imaging 
(see~\cite{liu2023characterization} for additional information on these light sheets).

Each synthetic training volume sample $V$ is 64$\times$64$\times$64 voxels in size spanning 8$\times$8$\times$12.8 $\mu$m$^{3}$ (with 125$\times$125$\times$200 nm$^{3}$ voxels) and containing between  $J = 1$ to 5 puncta chosen from a uniform distribution and located randomly at points $(x_j,y_j,z_j)$ within the volume. 
Each punctum is modeled as a Gaussian of full width at half maximum $w_j$ randomly chosen from the set $[100, 200, 300, 400]$ nm, allowing for slightly larger than the diffraction-limit features. 
The image of each punctum is generated by its convolution with the aberrated PSF:
\begin{align}\label{eq:bead}
    I_j^{\text{bead}}(x,y,z) = \text{PSF}_{\text{abb}}^{\text{overall}}(x,y,z) \otimes \exp\left[-4 \ln(2) \frac{x^2 + y^2 + z^2}{w_j^2}\right]
\end{align}

The integrated photons $N_o$ per punctum were selected from a uniform distribution of 1 to 200,000 photons. 
The total intensity distribution is:
\begin{align}\label{eq:intensity_distribution}
   I_{\text{photon}}(x,y,z) &= \Upsilon \cdot \sum_{j=1}^J I_j^{\text{bead}}(x - x_j, y - y_j, z - z_j)
 \end{align}  
where,
\begin{align}\label{eq:intensity_distribution_upsilon}
    \Upsilon &= \dfrac{N_o}{\iiint_{-\infty}^{\infty} I_j^{\text{bead}}(x,y,z) \, \text{d}x \, \text{d}y \, \text{d}z}
\end{align}

Since the signal from each aberrated punctum can exceed the boundary of $V$, total signal $S_V$ within $V$ is:
\begin{align}\label{eq:total_signal}
    S_V = \iiint_V I(x,y,z) \, \text{d}x \, \text{d}y \, \text{d}z \leq J N_o
\end{align}

After accounting for partial signal contributions ($S_V$) the photons per voxel were converted to camera counts by applying the quantum efficiency QE, Poisson shot noise $\eta$, and camera read noise $\epsilon$ to arrive at the final synthetic training set example:
\begin{align}\label{eq:synthetic_sample}
    I_{\text{camera}}(x,y,z) = QE \cdot I_{\text{photon}}(x,y,z) + \eta[QE \cdot I_{\text{photon}}(x,y,z)] + \epsilon
\end{align}

\subsubsection{Zernike distributions}\label{sec:zernike_distributions}
To ensure diversity in the training set to cover potential aberrations, each training example was chosen from the amplitudes of the eleven included aberration modes shown in color in Supplementary Fig.~\ref{fig:zernike_pyramid} with equal probability from one of four different distributions:

\paragraph{Single mode (Supplementary Fig.~\ref{fig:zernike_distribution}b)} 
One mode randomly chosen, with amplitude $\alpha$ randomly chosen from $0 \leq \alpha \leq 0.5 \lambda$ RMS.

\paragraph{Bimodal (Supplementary Fig.~\ref{fig:zernike_distribution}c)}
An initial target for the total amplitude $\alpha_t$ is randomly chosen from $0 \leq \alpha_t \leq 0.5 \lambda$ RMS. 
A second partitioning factor $\epsilon$ is randomly chosen from $0 \leq \epsilon \leq 1$. 
The amplitudes of the two modes are then $\alpha_1 = \epsilon \alpha_t$ and $\alpha_2 = (1 - \epsilon) \alpha_t$.

\paragraph{Powerlaw (Supplementary Fig.~\ref{fig:zernike_distribution}d)}
An initial target for the total amplitude $\alpha_t$ is randomly chosen from $0 \leq \alpha_t \leq 0.5 \lambda$ RMS. 
The initial partitioning factors $\epsilon_n$ for the modes are randomly chosen from a Lomax (\ie Pareto II) distribution~\cite{lomax1954business}:
\begin{align}\label{eq:powerlaw}
    \epsilon_n = \frac{\gamma}{(x_n + 1)^{\gamma + 1}} \quad \text{where} \quad \gamma = 0.75
\end{align}
where each $x_n$ is randomly chosen from $0 \leq x_n \leq 1$. 
They are then re-normalized:
\begin{align}\label{eq:powerlaw_norm}
    \epsilon_n' = \frac{\epsilon_n}{\sum_{n=1}^{11} \epsilon_n}
\end{align}
and the final amplitudes of the modes are $\alpha_n = \epsilon_n' \alpha_t$.

\paragraph{Dirichlet (Supplementary Fig.~\ref{fig:zernike_distribution}e)}
An initial target for the total amplitude $\alpha_t$ is randomly chosen from $0 \leq \alpha_t \leq 0.5 \lambda$ RMS. 
The initial partitioning factors $\epsilon_n$ for the modes are randomly chosen from $0 \leq \epsilon_n \leq 1$. They are then re-normalized:
\begin{align}\label{eq:dirichlet}
    \epsilon_n' = \frac{\epsilon_n}{\sum_{n=1}^{11} \epsilon_n}
\end{align}
and the final amplitudes of the modes are $\alpha_n = \epsilon_n' \alpha_t$.

Together, the training examples from these four distributions create a diverse set of overall aberration amplitudes and number of significant modes in the training data, with all eleven modes contributing equally across the dataset (Supplementary Fig.~\ref{fig:zernike_distribution}a).

\subsubsection{Training dataset}\label{sec:training_dataset}
For the model training, a dataset of 2 million synthetic 3D volumes was created, with aberration magnitude uniform sampled from  0.0 to 0.5~$\lambda$~RMS  (at wavelength $\lambda = 510$~nm), uniform distribution of the number of objects between 1 and 5, and photons ranging between 1 and 200,000 integrated photons per object.

\subsubsection{Test dataset}\label{sec:test_dataset}
To evaluate our models, we created a test dataset with 100,000 3D volumes.
The parameter distribution was the same as training, but extended the aberration magnitude up to 1.0~$\lambda$~RMS, and up to 500,000 integrated photons.
To test the operational limit of our models, this test dataset included up to 150 objects in any given volume.

\subsection{Fourier embedding}\label{subec:fourier_embedding}

Most ML vision models operate on real-space representations of the data, which lack clearly defined limits on image size or feature descriptors of their content.  
Instead, we used Fourier domain embeddings (Supplementary Fig.~\ref{fig:embedding}). These are bound by the microscope's OTF.
Aberrations within an isoplanatic patch globally effect all photons within that patch, producing a unique, learnable ``fingerprint'' pattern in the FFT amplitude and phase (Appendix~\ref{sup:fourier_embedding_design}, Supplementary Figs.~\ref{fig:modes-embedding-fingerprints}--\ref{fig:modes-embedding-planes}).

\subsubsection{Preprocessing}\label{subec:preprocessing}
To create Fourier embeddings (Fig.~\ref{fig:pipeline-summary}b) for our model, we preprocess the input 3D image stack $W$ of CCPs within an isoplanatic region to suppress noise and edge artifacts (Fig.~\ref{fig:pipeline-summary}a),  
\begin{align} \label{eq:preprocessing}
V &= \Upsilon(W).
\end{align}  
The preprocessing module ($\Upsilon$) begins with a set of filters to extract sharp-edged objects which reveal the aberration signatures:
a Gaussian high-pass filter to remove inhomogeneous background and a low-pass filter via a Fourier frequency filter, with cutoff set at the detection NA limit ($\sigma$=3 voxels). 
A Tukey window (Tukey cosine fraction=0.5, in \textit{$\hat{x}\hat{y}$} only) is applied to remove FFT edge artifacts from the volume borders.  
No windowing is applied along the axial direction, \textit{$\hat{z}$}, because embeddings are constructed near $k_z=0$ where aberration information is maximized.  

\subsubsection{Embedding}\label{subec:embedding}
Once preprocessed, a ratio of the resultant 3D FFT amplitude, to the 3D FFT amplitude of the ideal PSF (undergoing identical preprocessing steps) is used to generate the amplitude embedding, $\alpha(k_z)$ at each $k_z$ plane:
\begin{align} \label{eq:alpha}
V_{ideal} &= \Upsilon(PSF_{ideal}) \\
\alpha &= \dfrac{|\mathcal{F}(V)|}{|\mathcal{F}(V_{ideal})|}
\end{align}  
where $\mathcal{F}$ denotes the 3D Fourier transform. 
The most useful information content is located at $k_z = 0$, the principal plane located at the midpoint of the \textit{$\hat{k_z}$}-axis.  
Three 2D planes from  $\alpha_1$, $\alpha_2$, and $\alpha_3$  along \textit{$\hat{k_z}$}-axis as are necessary to extract axial information for inputs to the model as follows:
\begin{align} \label{eq:alpha_planes}
\alpha_1 &= \alpha_{k_{z=0}} \\
\alpha_2 &= \frac{1}{5}\sum_{i=0}^{4}{\alpha_{k_{z=i}}}  \\
\alpha_3 &= \frac{1}{5}\sum_{i=5}^{9}{\alpha_{k_{z=i}}} 
\end{align} 
where $\alpha_1$ is the principal plane along the $k_x$-axis and $k_y$-axis, 
$\alpha_2$ is the mean of five consecutive 2D planes starting from the principal plane,
and $\alpha_3$ is the mean of five consecutive 2D planes starting from the $k_z=5$ plane (Supplementary Fig.~\ref{fig:embedding}a,c and Supplementary Fig.~\ref{fig:modes-embedding-planes}).

For the phase embedding, $\varphi$, we first remove interference from multiple puncta in the FOV that may obscure the aberration signature in the phase image.
The interference patterns are removed using: Peak local maxima (\textit{PLM})\footnote{\url{https://scikit-image.org/docs/stable/auto\_examples/segmentation/plot\_peak\_local\_max.html}} for peak detection in real space using Normalized Cross-Correlation (\textit{NCC})\footnote{\url{https://scikit-image.org/docs/stable/auto\_examples/registration/plot\_masked\_register\_translation.html}} with a kernel cropped from the highest peak in \textit{V}.
The neighboring voxels around the detected puncta peaks are masked off, creating a volume, $\mathcal{S}$. 
The OTF with interference removed, $\tau'$, can now be obtained as well as a real space reconstructed volume, $V'$, via inverse FFT,
\begin{align} \label{eq:varphi}
M &= \text{PLM}(\text{NCC}(V)) \\
\mathcal{S} &= V \times M \\
\tau &= \dfrac{\mathcal{F}(V)}{\mathcal{F}(\mathcal{S})} \\
V' &= \mathcal{F}^{\text{-}1}(\tau)
\end{align} 

The phase $\varphi(k_z)$ at each $k_z$ plane is then given by the unwrapped phase of $\tau$ at that plane (Supplementary Fig.~\ref{fig:embedding}b,d). 
We calculate the three phase embeddings in the same manner as our amplitude embedding such that:
\begin{align} \label{eq:phi_planes}
\varphi_1 &= \varphi_{k_{z=0}} \\
\varphi_2 &= \frac{1}{5}\sum_{i=0}^{4}{\varphi_{k_{z=i}}}  \\
\varphi_3 &= \frac{1}{5}\sum_{i=5}^{9}{\varphi_{k_{z=i}}} 
\end{align} 

Combining the six planes together, we define the input to the model as a Fourier embedding,
\begin{align} \label{eq:embedding}
\mathcal{E} &= \{\alpha_1, \alpha_2, \alpha_3, \varphi_1, \varphi_2, \varphi_3\}
\end{align} 

A notable advantage of this approach is that, although the signal from each individual CCP is weak, those in the same isoplanatic region contain near-identical spatial frequency distributions which add together to yield Fourier embeddings of high signal-to-noise ratio (SNR) suitable for accurate inference of the underlying aberration (Supplementary Fig.~\ref{fig:modes-embedding-objects}).

\subsection{\model: Adaptive Optical Vision Fourier Transformer}\label{sec:model_design}

Below, we outline the key components of \model,  which uses a 3D multistage vision transformer architecture. 
This model efficiently captures Fourier-domain features at multiple spatial scales, enabling robust aberration prediction.

\subsubsection{Multistage}
Recent advances in attention-based transformers have demonstrated scalability, generalizability, and multi-modality for a range of computer vision applications~\cite{dosovitskiy2020vit,carion2020end,arnab2021vivit,cheng2022masked,he2022masked}.

Multiscale (or hierarchical) vision transformers such as Swin~\cite{liu2021swin}, and MViT~\cite{fan2021multiscale} 
are designed with specialized modules (\eg shifted-window partitioning~\cite{liu2021swin}, and hybrid window attention~\cite{li2022mvitv2}) to excel at a variety of detection tasks for 2D natural images using supervised training on ImageNet~\cite{deng2009imagenet}. 
Although these variants are more efficient than their ViT counterparts in terms of FLOPs and number of parameters, they often incorporate specialized modules as noted above.
Hiera~\cite{ryali2023hiera} showed that these designs can be streamlined without performance loss by leveraging large-scale self-supervised pretraining.  

Current multi-scale architectures use a Feature Pyramid Network scheme~\cite{lin2017feature}---downsampling the spatial resolution of the image for each stage while expanding the embedding size for deeper layers. 
Instead, in our work, we use $\Omega$ stages and do not downsample during any of the stages, but rather select different patch sizes for each stage (Fig.~\ref{fig:pipeline-summary}). 
This allows the embedding dimension within each stage to be fixed to the number of voxels in the patch of that stage, rather than expanding with increasing depth as in some hierarchical models. 

\subsubsection{Patch encoding}
The input to the model is the Fourier Embedding, a 3D tensor $\mathcal{E} \in \mathbb{R}^{\ell \times d \times d}$,
where $\ell$=6 is the number of 2D planes each with a height and width of $d$. 
For each model stage, $i$,
patchifying begins by dividing the input tensor $\mathcal{E}$ into non-overlapping 2D tiles (each  $p_i \times p_i$) that are each flattened into a 1D patch for a total of $k_i$ patches in a plane. 
After patchifying, the input tensor is transformed into $x_p \in \mathbb{R}^{\ell \times k_i \times p_i^2}$ (Fig.~\ref{fig:pipeline-summary}b).

The initial ViT model uses a set of consecutive transformer layers with a fixed patch size for all transformers, 
where each transformer layer can capture local and global dependencies between patches via self-attention~\cite{dosovitskiy2020vit}. 
The computation needed for the self-attention layers scales quadratically \wrt the number of patches (\ie sequence length). 
While using a smaller patch size could be useful to capture visual patterns at a finer resolution, using a large patch size is computationally cheaper.  

Our baseline model uses a two-stage design with patch sizes of 32 and 16 pixels, respectively (Fig.~\ref{fig:pipeline-summary}c).  
Appendix~\ref{sup:ablation} shows an ablation study using several stages with patch sizes ranging between 8 and 32 pixels.

\subsubsection{Positional encoding}
Rather than adopting the Cartesian positional encoding of ViT~\cite{dosovitskiy2020vit}, we use a polar coordinate system ($r,\theta$)  to encode the position of each patch. This choice is motivated by the radial symmetries of the Zernike polynomials and the efficiencies gained in NeRF~\cite{mildenhall2022nerf}, coordinate-based MLPs~\cite{tancik2020fourier}, and RoFormer~\cite{su2024roformer}.
For a given plane in $\mathcal{E}$ (Eq.~\ref{eq:embedding}), the radial positional encoding vector (\textit{RPE}) is calculated for every patch,
\begin{align} \label{eq:rpe}
\text{RPE}(r,\theta) &= [r, \sin \theta, \cos \theta, \dots, \sin m \theta, \cos m \theta]  
\end{align}
where $(r, \theta)$ are the polar coordinates for the center of each patch, and $m=16$. 
All patches and their positional encoding are then mapped into a sequence of learnable linear projections $\zeta \in \mathbb{R}^{\ell \times k_i \times p_i^2}$ that we use as our input to the transformer layers in the model.  

\subsubsection{Transformer building blocks}
Each stage has $n$ transformer layers, where each layer has $h$ multi-head attention (MHA) layers that map the inter-dependencies between patches, 
followed by a multi-layer perceptron block (MLP) that learns the relationship between pixels within a patch. 
The stage's embedding size, $\epsilon_i = p_i^2$, is set to match the number of voxels in a patch for that stage.
The MLP block is four times wider than the embedding size (Fig.~\ref{fig:model-transformer}c).
Layer normalization (LN)~\cite{ba2016layernorm} is applied before each step, and a skip/residual connection~\cite{he2016deep} is added after each step
\begin{align} \label{eq:model}
\zeta_1 &= \text{LN}(\text{MHA}(\zeta)) + \zeta \\
\zeta_2 &= \text{LN}(\text{MLP}(\zeta_1)) + \zeta_1
\end{align} 

In addition to the skip connections in each transformer layer, we also add a skip connection between the input and output of each stage. 
We use a dropout rate of 0.1 for each dense layer~\cite{srivastava2014dropout}, and stochastic depth rate of 0.1~\cite{huang2016deep}.
The patches from the final stage are pooled using a global average along the last dimension and passed to a fully connected layer to output $z$ Zernike coefficients.

\subsubsection{Attention modules}
We use self-attention~\cite{vaswani2017attention} as our default attention module for all transformer layers in our model.
Complementary to our approach, recent studies have looked into alternative attention methods to reduce the quadratic scaling of self-attention~\cite{cheng2022masked,dao2022flashattention,yang2022focal}.
Our architecture is compatible with these attention mechanisms, which would further improve our model's efficiency.

\subsection{In-silico evaluations}\label{sec:insilico}

Appendix~\ref{sup:ablation} shows an ablation study of our
synthetic data simulator (Appendix~\ref{sup:synthetic_data}), 
our multi-stage design (Appendix~\ref{sup:architecture}),
our training dataset size (Appendix~\ref{sup:datasetsize}, Supplementary Fig.~\ref{fig:scaling-dataset-summary-rms}),
and details of our training hyperparameters (Appendix~\ref{sup:params}, Supplementary Table~\ref{tab:variants}, Supplementary Table~\ref{tab:training_config}).
We also introduce a novel way of measuring prediction confidence of our model using digital rotations in Appendix~\ref{sup:confidence} (Supplementary Fig.~\ref{fig:pipeline-confidence}).

We present a detailed cost analysis benchmark comparing our architecture with other widely used models like ConvNeXt~\cite{liu2022convnet} and ViT~\cite{dosovitskiy2020vit} in Appendix~\ref{sup:benchmark}.
To further diagnose our model's performance, we carried out a series of experiments to understand our model's 
sensitivity to SNR (Appendix~\ref{sup:snr}, Supplementary Fig.~\ref{fig:modes-eval-single-mode-5-7}--\ref{fig:modes-eval-single-mode-10-12}), 
generalizability to other light-sheets (Appendix~\ref{sup:lightsheets}), 
number of objects in the FOV (Appendix~\ref{sup:objectdensity}), 
and object size (Appendix~\ref{sup:objectsize}, Supplementary Fig.~\ref{fig:modes-eval-bead-size-single-mode-5-7-rms}).


\section*{Acknowledgments}

We thank 
X. Ruan,
M. Mueller, P. Zwart,
and H. York for helpful discussions and comments. 
SUM159 cells used in this study were a gift from Kirchhausen Lab.   
We thank N. Shaner for providing the mChilada fluorescent protein plasmid to I.A.S., which was used to generate reagents for this study.
We gratefully acknowledge the support of this work by the Laboratory Directed Research and Development (LDRD) Program of Lawrence Berkeley National Laboratory under US Department of Energy contract No. DE-AC02-05CH11231. 
We thank J. White for managing our computing cluster. 
T.A., G.L., F.G., J.L.H. and S.U. are partially funded by the Philomathia Foundation (awarded to E.B. and S.U.). 
T.A. and G.L. are partially funded by the Chan Zuckerberg Initiative (awarded to S.U.).
T.A. and S.U. are partially supported by Lawrence Berkeley National Laboratory's LDRD program 7647437 and 7721359 (awarded to S.U.). 
T.A., D.E.M., and E.B. are funded by HHMI (awarded to E.B.). 
C. Shirazinnejad. and D.G.D. are partially supported by NIH Grant R35GM118149 (awarded to D.G.D.).
Cat Simmons, I.S.A., and I.A.S. are partially supported by NIH Grant 1R01DC021710 (awarded to I.A.S.).
F.G. is partially funded by the Feodor Lynen Research Fellowship, Humboldt Foundation. 
S.U. is funded by the Chan Zuckerberg Initiative Imaging Scientist program 2019-198142 and 2021-244163. 
E.B. is an HHMI Investigator. 
S.U. is a Chan Zuckerberg Biohub – San Francisco Investigator.

\section*{Declarations}

\bmhead{Author contribution}
\noindent
T.A. designed models, developed training pipelines and evaluation benchmarks. 
D.E.M. designed the preprocessing algorithms and developed the microscope software for the imaging experiments.
T.A. and D.E.M. designed Fourier embedding, and developed the synthetic data generator for training and validation. 
C. Shirazinnejad, C. Simmons, I.S.A., S.E.W., N.H., E. Hong, E. Huang, E.S.B., A.N.K., D.G.D., and I.A.S. generated the zebrafish reagents. 
A.N.K., prepared the cultured SUM159 cells.
C. Shirazinnejad, J.L.H., K.A., and A.M.J. prepared samples.
G.L. and J.L.H. performed the imaging experiments with zebrafish.
J.L.H. and K.A. performed the imaging experiments with cells.
G.L., K.A. and F.G. performed the imaging experiments with beads. 
T.A., D.E.M., and S.U. performed analysis and prepared figures.
T.A. wrote the manuscript with input from all co-authors. 
D.E.M., E.B. and S.U. edited the manuscript. 
T.A., E.B. and S.U. supervised the project.

\bmhead{Data availability}
Data for demos is available on our Github repository at \url{https://github.com/cell-observatory/aovift}. 
The full datasets for training and testing are too large to be hosted on public repositories, they can be shared upon reasonable request.

\bmhead{Code availability}
Source code for training and evaluation (and all pretrained models) are available at \url{https://github.com/cell-observatory/aovift}.
Docker image is available at \url{https://github.com/cell-observatory/aovift/pkgs/container/aovift}.
Deconvolution was performed using PetaKit5D(\url{https://github.com/abcucberkeley/PetaKit5D}).

\bmhead{Ethics approval and consent to participate}
All experiments with zebrafish were done in accordance with protocols approved by the University of California, Berkeley's Animal Care and Use Committee and following standard protocols (animal use protocol number AUP-2019-09-12560-1) . 
All zebrafish used in this study were embryos younger than 72hpf. 
Sex determination was not a factor in our experiments.
All husbandry and experiments with zebrafish were done in accordance with protocols approved by the University of California, Berkeley's Animal Care and Use Committee and following standard protocols (animal use protocol numbers AUP-2019-09-12560-1 (Upadhyayula Lab), AUP-2020-10-13737-1 (Swinburne Lab), and AUP-2021-05-14347-1 (Zebrafish Facility Core Protocol)).

\bmhead{Conflict of interest/Competing interests}
None.

\bibliographystyle{unsrt}
\bibliography{acm}
\clearpage


\renewcommand{\appendixname}{}


\renewcommand{\thefigure}{S\arabic{figure}}
\renewcommand{\thetable}{S\arabic{table}}
\setcounter{section}{0}
\setcounter{figure}{0}
\setcounter{table}{0}
\setcounter{footnote}{0}

\begin{appendices}

\startcontents[si]
\renewcommand\contentsname{Appendices}
\printcontents[si]{l}{1}[3]{\section*{\contentsname}}

\section{Ablation study}
\label{sup:ablation}

\subsection{Fourier embedding design}
\label{sup:fourier_embedding_design}

Most ML approaches to AO rely on real-space representations of training data~\cite{saha2020practical,rai2023deep,zhang2023deep,hu2023universal,kang2024coordinate}. However, open-source AO training datasets remain scarce, and acquiring large, annotated 3D volumes that capture diverse biological features over a broad aberration space is prohibitively expensive.  Training on small datasets leads to overfitting and poor generalization; scaling existing 2D models to 3D also comes at significant computational cost.

To address these challenges, we use synthetic data transformed to the Fourier domain.
We embed diffraction-limited puncta (such as signal from sub-diffractive beads or fluorescent protein labeled AP2) into Fourier space using a pre-processing step
(see Sec.~\ref{subec:fourier_embedding}).
Within an isoplanatic patch, the aberration affects signal from all emitters,
yielding a consistent Fourier `fingerprint'.
As shown in Supplementary Fig.~\ref{fig:embedding}a,b, and demonstrated in noise-free Fourier principal planes
(Supplementary Fig.~\ref{fig:modes-embedding-fingerprints}), the amplitude and phase embedding planes contain unique signatures for each Zernike mode.
To illustrate the similarity of aberration-specific patterns across a variety of real-space samples,
we apply 0.1 $\mu$m~RMS of coma $Z_3^{-1}$ to three different examples:
a single bead (Supplementary Fig.~\ref{fig:modes-embedding-objects}a),
a volume with five synthetic beads resembling training data (spanning $9.3 \times 9.3 \times 12.8$ $\mu$m$^3$, Supplementary Fig.~\ref{fig:modes-embedding-objects}b),
and a volume of 100 synthetic beads resembling an AP2-labeled zebrafish embryo (corresponding to density beyond the parameters of the training data, Supplementary Fig.~\ref{fig:modes-embedding-objects}c).
Despite drastic differences in real-space appearance, the Fourier embeddings ($\alpha_1$, $\alpha_2$, $\alpha_3$, $\varphi_1$, $\varphi_2$, $\varphi_3$) remain consistent.

In contrast, real-space models trained on limited data (\eg a few synthetic beads) did not generalize on dense puncta, where isolating individual features is difficult and aberrations are spatially variable. By moving to the Fourier domain, we leverage the OTF, which bounds the frequency space. Despite never seeing more than five beads during training, our model generalizes to volumes with 100+ resolvable puncta because each Zernike pattern persists in Fourier space (Appendix~\ref{sup:objectdensity}) regardless of the real-space density.

Beyond generalization, Fourier embeddings reduces the inference cost by compressing the original isoplanatic patch from D$\times$H$\times$W to 6$\times$64$\times$64 voxels. A real-space model operating on $64^3$ voxels requires more than ten times the computation. Further downsampling (to 6$\times$32$\times$32) accelerates inference but loses sensitivity to discriminate subtle changes in aberrations. Conversely, scaling up to 6$\times$128$\times$128 yields minimal gains for the first 15 Zernike modes, although may be necessary for higher-order modes.

During preprocessing, we also normalize phase embeddings to remove interference patterns introduced by sample structures (Supplementary Fig.~\ref{fig:modes-embedding-planes}d,e). Without phase information, the model cannot recover aberration signs, since amplitude alone lacks sign encoding (Supplementary Fig.~\ref{fig:modes-embedding-planes}f,g).
While the principle amplitude ($\alpha_1$)
and phase ($\varphi_1$) planes capture lateral aberrations,
 additional planes ($\alpha_2$, $\alpha_3$ \& $\varphi_2$, $\varphi_3$) are necessary for axial distortions
(\eg spherical aberrations $Z_3^{-1}$, Supplementary Fig.~\ref{fig:modes-embedding-planes}f).
To capture such aberrations, we average a small set of planes along \textit{$\hat{k_z}$}-axis,
rather than using the orthogonal principal planes, to improve the model generalizability. A single model trained on the principal plane plus additional  \textit{$\hat{k_z}$}-axis planes then generalized across diverse lightsheet configurations with distinct axial support
(Appendix~\ref{sup:lightsheets}), enabling robust aberration prediction under varied imaging setups.

\subsection{Synthetic training and validation data}
\label{sup:synthetic_data}

\begin{figure*}[!tp]
    \centering
    \includegraphics[width=.9\textwidth]{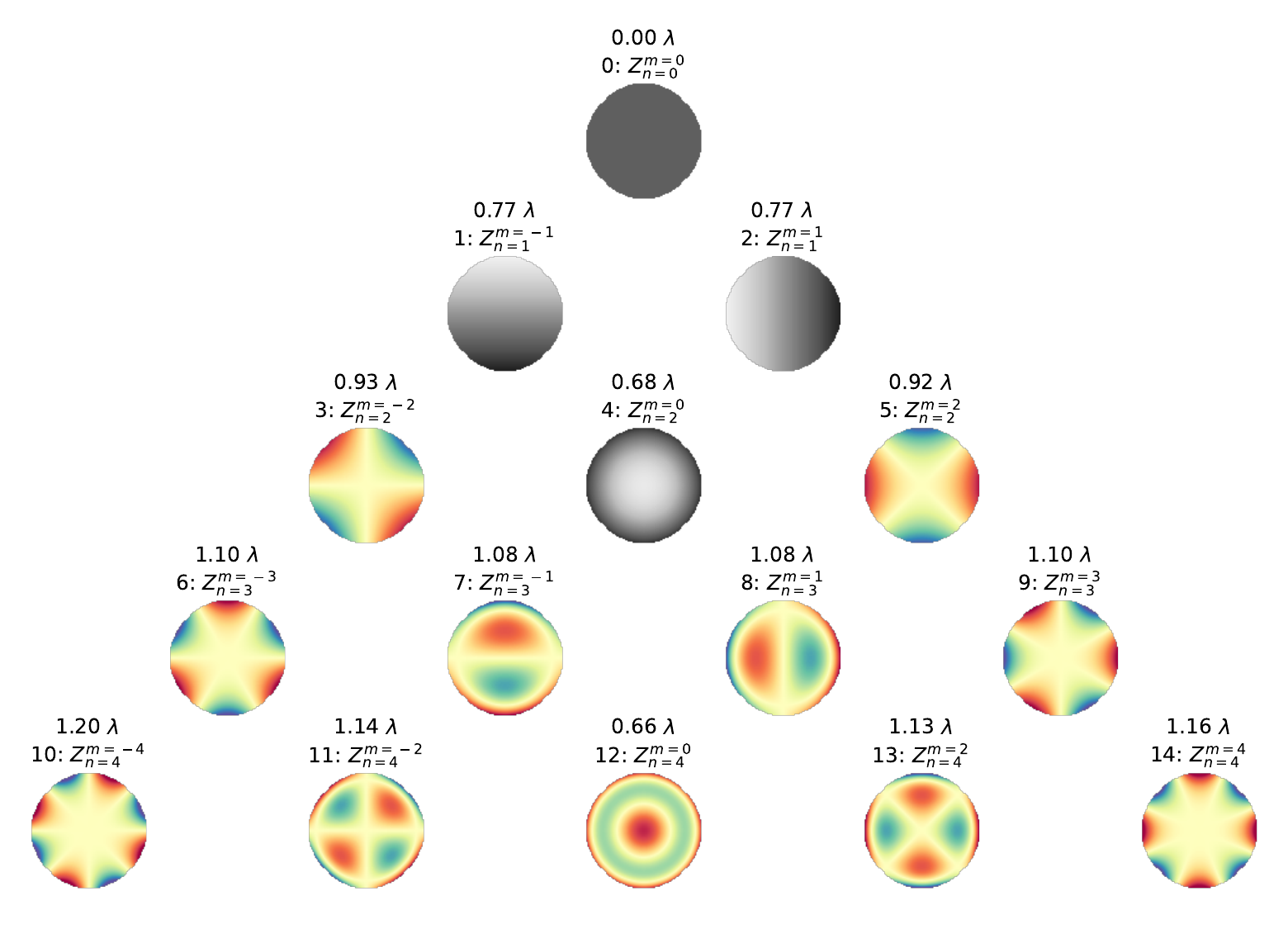}
    \caption{
        \textbf{Zernike modes (OSA/ANSI standard indexing)}. 
        Wavefronts for the first 15 Zernike modes ($Z_0^0$ through $Z_4^{\pm4}$) with 0.1 $\mu$m~RMS applied to each mode. 
        Undetectable modes (bias, tip, tilt, and defocus) are greyed out, and the measured peak-to-valley for each mode is reported in waves for a wavelength $\lambda = 510nm$.  
    }
    \label{fig:zernike_pyramid}
\end{figure*}

We used the synthetic data generator described in Sec.~\ref{sec:dataset} to create training and testing data for the first 15 Zernike modes, $Z_0^0$ through $Z_4^{\pm4}$ (Supplementary Fig.~\ref{fig:zernike_pyramid}).
We show the $\lambda$~RMS magnitude range for our training (blue) and testing (orange) distributions in Supplementary Fig.~\ref{fig:zernike_distribution}a.
The panels show an example wavefront and a breakdown of each distribution:
single (Supplementary Fig.~\ref{fig:zernike_distribution}b),
bimodal (Supplementary Fig.~\ref{fig:zernike_distribution}c),
Powerlaw---a heavy-tailed distribution with a few dominating modes (Supplementary Fig.~\ref{fig:zernike_distribution}d),
and Dirichlet---uniform weighting across all modes (Supplementary Fig.~\ref{fig:zernike_distribution}e).

To train our models,
we created two millions synthetic 3D volumes with aberrations ranging between 0.0 to 0.5~$\lambda$~RMS.
Each volume can have up to five beads with a uniform distribution of photons ranging between 1 and 200,000 integrated photons per bead.

We created a separate dataset with 100,000 volumes for testing.
This smaller dataset extended the aberration magnitude up to 1.0~$\lambda$~RMS, and the SNR range to 500,000 integrated photons.
We also simulated samples with much higher object density (up to 150 objects in any given volume).
We sample from this dataset to carry out individual test experiments described in Appendix~\ref{sup:insilico} to stress test our models.

\subsection{Multistage design}
\label{sup:architecture}

\begin{figure*}[!tp]
    \centering
    \includegraphics[width=\textwidth]{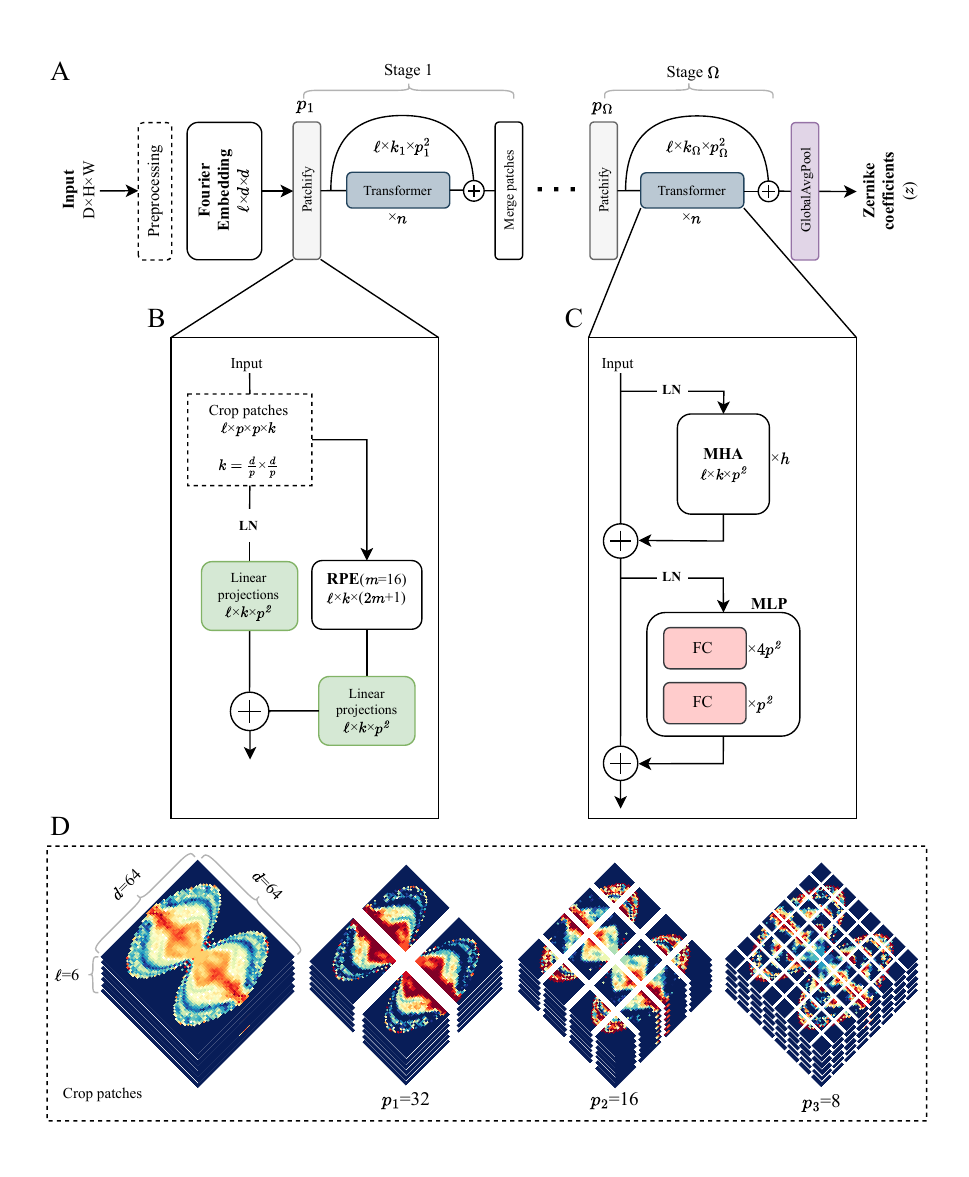}
    \caption{
    \textbf{Multistage architecture}.
    \textbf{A}. A schematic of the general form of our model with $\Omega$ stages. 
    \textbf{B}. A breakdown of the patchify layer. 
    \textbf{C}. A breakdown of the transformer layer. 
    \textbf{D}. An example output of the crop patches block using a patch size of 32, 16, and 8, respectively.   
    }
    \label{fig:model-transformer}
\end{figure*}

We used Fourier embedding ($\mathcal{E} \in \mathbb{R}^{\ell \times d \times d}$) as input to a vision transformer model with $\Omega$ stages (Supplementary Fig.~\ref{fig:model-transformer}a).
For each stage, the $\ell$ Fourier planes are tiled into $k$ patches (Supplementary Fig.~\ref{fig:model-transformer}b), applying the radial encoded Positional Embedding to each patch.  
These are passed through a sequence of $n$ Transformer (Supplementary Fig.~\ref{fig:model-transformer}c) layers, each consisting of $h$ parallel multi-head attention layers (MHA), followed by a multi-layer perceptron block (MLP).   
At each stage end, a residual connection is added, and the patches are merged back into the shape matching the stage input (Merge patches). 
After all stages, the resulting patches are pooled (GlobalAvgPool) and connected with a dense layer to output the $z$ Zernike coefficients. 

To determine the optimal number of stages for our architecture, 
we designed several models by changing the number of transformer layers $n$ used for each patch size $p$, 
while keeping the total number of transformer layers used in each model to 8, and choosing a patch size from $p_1=32$, $p_2=16$, or $p_3=8$.

\begin{figure*}[!tp]
    \centering
    \includegraphics[width=.75\textwidth]{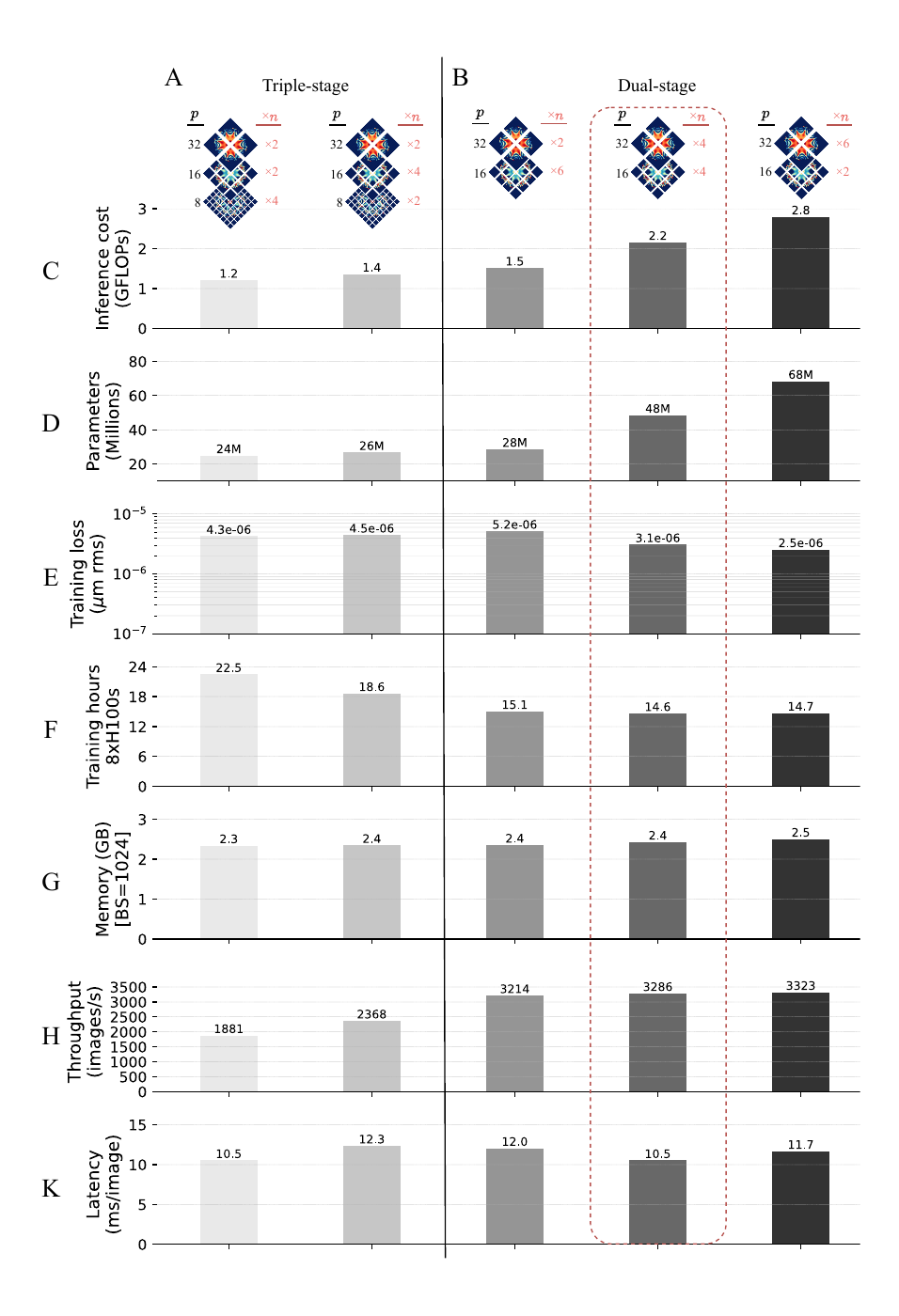}
    \caption{
    \textbf{Performance analysis of our multistage design using a total of $n=8$ transformer layers, and several patch sizes $p_1=32$, $p_2=16$, and $p_3=8$}.
    \textbf{A}. Triple-stage design with a patch size of 32, 16, and 8 respectively. 
    \textbf{B}. Dual-stage design with a patch size of 32 and 16. 
    \textbf{C}. Inference cost per image measured in GFLOPs ($10^{9}$ FLOPs).
    \textbf{D}. Total number of trainable parameters.
    \textbf{E}. Training loss using a dataset of 2M synthetically generated samples.
    \textbf{F}. Training time using a single node with eight H100 GPUs. 
    \textbf{G}. Memory footprint of each model with a batch of 1024 images using 16-bit floating point precision.
    \textbf{H}. Throughput (average number of predictions/images per second) using a batch size of 1024 on a single A100 GPU.
    \textbf{K}. Latency (average inference time per image) measured in milliseconds for a batch of 1024 examples using a single A100 GPU.
    }
    \label{fig:scaling-multistage}
\end{figure*}

Supplementary Fig.~\ref{fig:scaling-multistage} shows our performance analysis of each model trained on the same dataset of 2 million synthetically generated samples.
We created two models using a triple-stage design in Supplementary Fig.~\ref{fig:scaling-multistage}a, and three models using a dual-stage design in Supplementary Fig.~\ref{fig:scaling-multistage}b.
We used 8 transformers for each model, but we vary the number of transformers used for each patch size. 

The cost of running our triple-stage models is lower than our dual-stage models in terms of GFLOPs (Supplementary Fig.~\ref{fig:scaling-multistage}c), 
and both of our triple-stage models are smaller than the rest of the models we tested (Supplementary Fig.~\ref{fig:scaling-multistage}d).  
However, our dual-stage models converged to a better training loss, 
and were substantially faster in terms of training (Supplementary Fig.~\ref{fig:scaling-multistage}f), and inference (Supplementary Fig.~\ref{fig:scaling-multistage}h--k).

In general, using smaller patch sizes is more expensive, as the compute for self-attention layers scales quadratically \wrt the number of patches.
Our analysis suggests that the smallest patch size $p_3=8$ is not necessary for our application, 
which is expected given that our starting input images are relatively small.
Furthermore, the models that use more transformers for the biggest patch size ($p_1=32$) are faster, and perform better in terms of training loss.
Given that most aberrations extend beyond ($16 \times 16$) pixels in Fourier space,
we chose a dual-stage layout with an equal number of transformers for each patch size ($p_1=32, p_2=16$) to balance between accuracy and training/inference time.

In all, we derived five variants---scaling the model size from 34 million up to 228 million parameters.
Supplementary Table~\ref{tab:variants} shows the number of transformer layers $n$, number of heads $h$, embedding size $\epsilon$, and MLP size $x$, highlighting the hyperparameters used to create each of our variants.
Further evaluation for all variants of our model can be found in Appendix~\ref{sup:benchmark}---where they are also compared to other prominent architectures.

\subsection{Training dataset size}
\label{sup:datasetsize}

In this section, we present a small case study to understand the scaling of our models in terms of model size, and training dataset size. 
For simplicity, we chose two models for this analysis: Tiny (T) and Small (S).
We trained each model on a synthetic dataset that has up to eight million samples.

Supplementary Fig.~\ref{fig:scaling-dataset-summary-rms} shows the cumulative distribution functions for our residuals after a single correction over 10,000 test samples with initial aberrations ranging from 0.2$\lambda$~RMS up to 0.4$\lambda$~RMS, simulated with 50,000 up to 200,000 integrated photons using a single bead. 
The colors indicate the number of training samples used for each model:
500K (blue), 1M (orange), 2M (green), 4M (red), 6M (purple), and 8M (brown). 

The performance of our Tiny model starts to plateau after four million training samples---indicating it saturated its learning capacity given the limited number of trainable parameters that it can adjust during training (Supplementary Fig.~\ref{fig:scaling-dataset-summary-rms}a).
On the other hand, our Small model continued to improve with more training data (Supplementary Fig.~\ref{fig:scaling-dataset-summary-rms}b). 
We show an additional analysis, scaling our model size up to 227 million trainable parameters in Appendix~\ref{sup:benchmark}.

\subsection{Training hyperparameters}
\label{sup:params}

We trained all models on a two million sample dataset of synthetically aberrated beads, described in Sec.~\ref{sec:training_dataset}.
Training uses Adam~\cite{kingma2015adam}, augmented with decoupled weight decay regularization~\cite{loshchilov2018decoupled} with weight decay of 0.001, $\beta_1=0.9$, $\beta_2=0.99$, and a layerwise adaptive rate scaling~\cite{you2020large} for faster training using large batch sizes.

We trained all models using a node with eight NVIDIA H100 GPUs for a total of 500 epochs with a two-phase learning rate scheduler---linearly increasing our learning rate from 0 to $1e^{-3}$ over 25 epochs (warmup phase), followed by a cosine decay~\cite{loshchilov2016sgdr} for the remaining epochs (decay phase). 
We used mean squared error (MSE) as our loss function with a batch size of 4096 unless otherwise noted. 
Supplementary Table~\ref{tab:training_config} shows the breakdown of our training configuration.

We applied dropout~\cite{srivastava2014dropout,gal2016dropout} with probability 0.1 after all Dense layers except the patch encoder layer~\cite{dosovitskiy2020vit}. 
We also used stochastic depth regularization with linearly increasing dropout rates and a max dropout rate of 0.1~\cite{huang2016deep}. 

We used Tensorflow~\cite{abadi2016tensorflow} to implement our models. 
We derived five variants---scaling the model size from 34 million up to 228 million parameters---to compare our model design with the architectures described below.
Supplementary Table~\ref{tab:variants} highlights the hyperparameters used to create each of our variants.

\subsection{Measuring prediction confidence}
\label{sup:confidence}

To validate a single \model~prediction, we perform inferences on 361 digitally rotated copies of $\mathcal{E}$ through a range of 0–-360 degrees and check if these predictions are self-consistent.
For each Zernike mode $Z^{m\neq0}_{n}$, there is a known rotational period, $\frac{2\pi}{m}$. 
Confidence in the prediction of each mode is established when the rotated inferences match the known Zernike modes' rotational periods, (Supplementary Fig.~\ref{fig:pipeline-confidence}) within a desired tolerance. 
For rotationally invariant modes (\eg spherical aberration), the variance of the Zernike amplitude must remain within a tolerance.  
When the model predictions exceed tolerance, the typical cause is there is not enough information in the data to make an accurate prediction for that mode.  
This will occur when the underlying sample structure does not have visible puncta (\eg high spatial frequency content) or the acquisition has too much noise.

Supplementary Fig.~\ref{fig:pipeline-confidence}a shows an example of our pipeline for measuring prediction confidence. 
The subpanels show the predicted amplitudes for each digital rotation, and a regression fit between the digitally rotated angle and the predicted twin angle to evaluate the variance in the predictions for 360 digital rotations.

Particularly, we show the predicted amplitudes for:
\begin{itemize}
    \item Supplementary Fig.~\ref{fig:pipeline-confidence}b: Oblique astigmatism ($Z^{m=\text{-}2}_{n=2}$, blue), vertical astigmatism ($Z^{m=2}_{n=2}$, orange), and the magnitude for both twin modes in black.
    \item Supplementary Fig.~\ref{fig:pipeline-confidence}d: Vertical trefoil ($Z^{m=\text{-}3}_{n=3}$, blue), oblique trefoil ($Z^{m=3}_{n=3}$, orange), and the magnitude for both twin modes in black. 
    \item Supplementary Fig.~\ref{fig:pipeline-confidence}f: Vertical coma ($Z^{m=\text{-}1}_{n=3}$, blue), horizontal coma ($Z^{m=1}_{n=3}$, orange), and the magnitude for both twin modes in black. 
    \item Supplementary Fig.~\ref{fig:pipeline-confidence}h: Oblique quadrafoil ($Z^{m=\text{-}4}_{n=4}$, blue), vertical quadrafoil ($Z^{m=4}_{n=4}$, orange), and the magnitude for both twin modes in black. 
    \item Supplementary Fig.~\ref{fig:pipeline-confidence}j: Oblique secondary astigmatism ($Z^{m=\text{-}2}_{n=4}$, blue), vertical secondary astigmatism ($Z^{m=2}_{n=4}$, orange), and the magnitude for both twin modes in black. 
    \item Supplementary Fig.~\ref{fig:pipeline-confidence}l: Predicted amplitudes of primary spherical for each digital rotation ($Z^{m=0}_{n=4}$, blue). 
\end{itemize}

Supplementary Fig.~\ref{fig:pipeline-confidence}[c, e, g, i, and k] show the result of our regression fits. 
If the mean residual error (MSE) of these predicted twin angles exceeds 700 and the magnitude of the mode is above 0.05 $\mu~RMS$, 
then we flag that prediction as  notconfident (\eg, Supplementary Fig.~\ref{fig:pipeline-confidence}c, highlighted in red).
We highlight confident predictions in green (\eg Supplementary Fig.~\ref{fig:pipeline-confidence}e, and Supplementary Fig.~\ref{fig:pipeline-confidence}k).
If our MSE for the digital rotations is high but the magnitude of the mode is below 0.05 $\mu~RMS$, then we mark these modes in blue as confident zero predictions (\eg Supplementary Fig.~\ref{fig:pipeline-confidence}g and Supplementary Fig.~\ref{fig:pipeline-confidence}i).
We show the final prediction for each Zernike mode in Supplementary Fig.~\ref{fig:pipeline-confidence}m, and the corresponding wavefront in Supplementary Fig.~\ref{fig:pipeline-confidence}n.

\newpage
\section{Architecture benchmark}
\label{sup:benchmark}

\subsection{Baseline models}

We compared our model architecture with two widely used backbones, namely transformer-based ViT~\cite{dosovitskiy2020vit}, and convolution-based ConvNeXt~\cite{liu2022convnet} as shown in Fig~\ref{fig:scaling-archs-rms}.

ViT~\cite{dosovitskiy2020vit} originally was used for 2D RGB images and later adopted to 2D RGB videos~\cite{arnab2021vivit}.
To compare with our model, we extended the ViT architecture to our use case (3 spatial dimensions) by replacing each transformer layer with a new building block to process 3D volumes. 
Following the configurations proposed in~\cite{zhai2022scaling}, we created the variants: 

\begin{itemize}
    \item ViT-Small (layers=12, heads=6, EMB=384, MLP=1536), 
    \item ViT-Base  (layers=12, heads=12, EMB=768, MLP=3072), 
    \item ViT-Large (layers=24, heads=16, EMB=1024, MLP=4096), 
\end{itemize}
trying both 16 and 32 as the fixed patch size for all transformer layers in the variant.
We set the dropout rate to 0.1 for all variants~\cite{srivastava2014dropout, huang2016deep}.

ConvNeXt~\cite{liu2022convnet,woo2023convnext} was also designed for 2D RGB images and videos. 
To adopt the model backbone to 3D spatial data, we replaced each 2D convolutional layer with a 3D convolution, without any other modifications to the rest of the architecture. 
We used a kernel size of (1, 7, 7), a downscale ratio of (1, 2, 2) for each stage, and stochastic depth rate of 0.1~\cite{huang2016deep}.
As described in~\cite{liu2022convnet}, we create four variants of the model:
\begin{itemize}
    \item ConvNeXt-Tiny (C=[96, 192, 384, 768], B=[3, 3, 9, 3]),
    \item ConvNeXt-Small (C=[96, 192, 384, 768], B=[3, 3, 27, 3]), 
    \item ConvNeXt-Base (C=[128, 256, 512, 1024], B=[3, 3, 27, 3]), 
    \item ConvNeXt-Large (C=[192, 384, 768, 1536], B=[3, 3, 27, 3]),
\end{itemize}
where $C$ is the number of channels, and $B$ is the number of inverted bottleneck blocks~\cite{sandler2018mobilenetv2} for each stage in the model.

\subsection{Cost analysis}

To evaluate the performance of these models, we used a subsample of our testing dataset with single beads only (Sec.~\ref{sec:test_dataset}). 
Particularly, we use 10,000 samples with aberrations ranging between $0.1\lambda$ to $0.2\lambda$~RMS---simulated with 50,000 to 200,000 photons.
For each sample, we made three iterative corrections, measuring the RMS residuals of the aberration after each iteration. 
We show the median RMS residuals for all samples after two corrections in Supplementary Fig.~\ref{fig:scaling-archs-rms}a.
All models tested here, except for ConvNeXt-T, were able to reduce the residuals down to the diffraction-limit (\ie below $\approx0.075\lambda$ RMS or $\lambda/4$ peak-to-valley~\cite{mahajan1982strehl,bentley2012field}) using two corrections only. 
Supplementary Fig.~\ref{fig:scaling-eval-rms} shows the results for all three corrections. 

While equivalent accuracy is found between model variants, it is critical to choose the model with the optimal set of trade-offs, rather than using the model with marginally better loss (Supplementary Fig.~\ref{fig:scaling-archs-rms}B). 
Since all models are capable of reducing the aberration down to diffraction-limit with a reasonable number of iterations, we compare these models across three salient factors:
floating point operations per second (FLOPs);
model size (number of parameters and memory footprint);
speed (training time, throughput, and latency).

While our models' FLOPs for training (Supplementary Fig.~\ref{fig:scaling-archs-rms}c) and inference (Supplementary Fig.~\ref{fig:scaling-archs-rms}i) are slightly higher than ViT/32 because of the overhead of our multistage design,
FLOPs of our models are still better than ViT/16 and ConvNeXt. 
A model with lower FLOPs does not necessarily mean a better model overall,    
because the measurement of FLOPs does not factor in relevant details encoded into the architecture like parallelizable operations and the cost of any individual operation for a given hardware~\cite{dehghani2022efficiency}.
Since all models are trained on the same hardware, we can look at training time to measure the efficiency of these architectures.  
Our models' training hours (Supplementary Fig.~\ref{fig:scaling-archs-rms}d) are better because our design enables us to use a large batch size for training (Supplementary Fig.~\ref{fig:scaling-archs-rms}e), while maintaining a small memory footprint (Supplementary Fig.~\ref{fig:scaling-archs-rms}f), and a lower number of trainable parameters (Supplementary Fig.~\ref{fig:scaling-archs-rms}j).

ViT uses a fixed patch size for all of its transformer layers. 
Thus, using a smaller patch size (\eg ViT/16) will increase the sequence length (\ie number of patches) being processed by each layer, scaling the computational cost for self-attention quadratically for all transformer layers in the model. 
Our multistage architecture allows our models to converge faster by learning Fourier patterns from several scales, and also reduces the cost of our models by leveraging cheaper transformer layers with bigger patch sizes.    

Unlike ViT, our embedding size $\epsilon_i$ changes based on the number of voxels per patch in each stage $i$. 
Our initial stage has a patch size $p_1=32$, resulting in an embedding size $\epsilon_1=1024$. 
While our second stage uses a smaller patch size $p_2=16$, our embedding size $\epsilon_2=256$ is significantly smaller, forcing the model to compartmentalize and compress the learnable parameters of deeper layers.  
Therefore, our transformer layers with smaller patch sizes have a smaller embedding size, instead of keeping the embedding size fixed throughout the model. 
This design scales particularly well as we increase the model size up by adding transformer layers (Supplementary Fig.~\ref{fig:scaling-archs-rms}k), and increasing the number of transformer heads (Supplementary Fig.~\ref{fig:scaling-archs-rms}L) for each layer.

In terms of inference speed, we see an advantage of using our models over the other models tested here.
Our model variants perform exceptionally well \wrt throughput---maximum number of predictions per second using a batch size of 1024 (Supplementary Fig.~\ref{fig:scaling-archs-rms}g), and latency---average inference time (milliseconds) per image using a single NVIDIA A100 GPU (Supplementary Fig.~\ref{fig:scaling-archs-rms}h).
Table.~\ref{tab:benchmark} shows a breakdown of these cost indicators for all models tested here. 
Based on this analysis, we select our small model for use on experimental data.

\newpage
\section{In-silico evaluations}
\label{sup:insilico}

\subsection{Sensitivity to SNR}\label{sup:snr}

To test our model's sensitivity to SNR, we evaluated our Small model on 10,000 test samples using a single bead with a mixed distribution of aberrations of magnitudes up to 1~$\lambda$~RMS.
Supplementary Fig.~\ref{fig:eval-snrheatmap}a shows the initial aberrations in the test dataset as a function of integrated photons. 
Supplementary Fig.~\ref{fig:eval-snrheatmap}b--f shows five rounds of corrections. 

We also show first round corrections for specific single modes of aberrations under low SNR conditions (\ie $\leq$ 100,000 integrated photons) in Supplementary Fig.~\ref{fig:modes-eval-single-mode-5-7}---\ref{fig:modes-eval-single-mode-10-12}.

Supplementary Fig.~\ref{fig:modes-eval-single-mode-5-7}a shows the residual $\lambda$~RMS after a single correction for vertical astigmatism $Z^{m=2}_{n=2}$, 
along with XY MIPs of the residual aberration for various aberration amplitudes and SNR (Supplementary Fig.~\ref{fig:modes-eval-single-mode-5-7}b). 
Max counts is highlighted above each PSF. 

For single mode aberrations of $\leq 0.2~\lambda$~RMS amplitude, the model can reach diffraction-limited performance for as low as 20,000 photons.
However, it is exceptionally difficult to get good corrections for large aberrations at this SNR.  
We see similar trends for vertical trefoil $Z^{m=\text{-}3}_{n=3}$ (Supplementary Fig.~\ref{fig:modes-eval-single-mode-5-7}c),
vertical coma $Z^{m=\text{-}1}_{n=3}$ (Supplementary Fig.~\ref{fig:modes-eval-single-mode-5-7}e),
oblique quadrafoil $Z^{m=\text{-}4}_{n=4}$ (Supplementary Fig.~\ref{fig:modes-eval-single-mode-10-12}a),
oblique secondary astigmatism $Z^{m=\text{-}2}_{n=4}$ (Supplementary Fig.~\ref{fig:modes-eval-single-mode-10-12}c),
and primary spherical $Z^{m=0}_{n=4}$ (Supplementary Fig.~\ref{fig:modes-eval-single-mode-10-12}e).

\subsection{Generalization to other light sheets}\label{sup:lightsheets}

To evaluate our model's generalizability to other light-sheets, 
we created four synthetic datasets with 10,000 samples each having aberrations up to 1~$\lambda$~RMS---using a single bead with up to 500,000 integrated photons. 

Supplementary Fig.~\ref{fig:eval-mb-rms}a--c shows the residuals for three rounds of corrections using the same multi-Bessel LLS as used for generating the training data. 
The next row (Supplementary Fig.~\ref{fig:eval-mb-rms}d--f) shows the results for a LLS of the same type but at higher NA. 
It achieves similar performance for the same training set, despite the higher NA. 

However, the model's performance degrades when applied to Sinc (Supplementary Fig.~\ref{fig:eval-mb-rms}j--l), and Gaussian (Supplementary Fig.~\ref{fig:eval-mb-rms}m--o) light sheets, as these have notably different cross-sectional profiles. 
Supplementary Fig.~\ref{fig:eval-mb-rms}I--V show the excitation profile used to create each LLS \wrt the length of each LLS in microns.
Supplementary Table~\ref{tab:lightsheets} shows the LLS specifications we used for this test.

\subsection{Sensitivity to the density of objects}\label{sup:objectdensity}

Our model was trained with up to five simulated beads in any given FOV. 
However, most experimental data will have dozens if not hundreds of objects in a single volume. 
Therefore, we evaluated our model on a dataset of 10,000 samples simulated with up to 150 beads to understand its sensitivity to the density of objects ($\bar{\delta}$nm).

Supplementary Fig.~\ref{fig:eval-densityheatmap} shows the residual $\lambda$~RMS for five rounds of corrections as a function of the average distance to the nearest bead in that given FOV. 
The results suggest the model generalizes to a much larger number of objects than it was trained on. 
The heatmaps (Supplementary Fig.~\ref{fig:eval-densityheatmap}b--f) show continuous improvement over 5 iterations provided there is not much overlap among the objects ($\bar{\delta} \geq 200$nm).

\subsection{Sensitivity to object size}\label{sup:objectsize}

As mentioned in Sec.~\ref{sec:dataset}, we simulated individual objects (beads) as Gaussian kernels. 
We uniformly chose the kernel's full width at half max (FWHM) for each object between 100nm and 400nm to create our training dataset. 

Here, we show a test of the extent to which our model can reliably correct individual aberration modes of  0.3 $\lambda$~RMS amplitude \wrt object size.  

Supplementary Fig.~\ref{fig:modes-eval-bead-size-single-mode-5-7-rms}a--b shows the residual $\lambda$~RMS after a single correction using our Small model (S) for vertical astigmatism $Z^{m=2}_{n=2}$ simulated with 50,000 and 100,000 integrated photons, respectively. 
 
We also show representative XY MIPs of these aberrations using bead kernels of FWHM  100nm, 200nm, 300nm, and 400nm, respectively (Supplementary Fig.~\ref{fig:modes-eval-bead-size-single-mode-5-7-rms}I--IV). 
We find that as the objects get larger, the unique signature of these aberrations in real space and Fourier space is degraded. 
Better SNR helps improves correction by detecting smaller fringes in the Fourier embedded input images, but the ability to correct for non diffraction-limited objects still  depends on the nature of the aberrations. 
As a rule of thumb, our model can detect aberrations as long as they are reasonably apparent in the real space images. 
Supplementary Fig.~\ref{fig:modes-eval-bead-size-single-mode-5-7-rms}c--f show results for vertical trefoil $Z^{m=\text{-}3}_{n=3}$, and vertical coma $Z^{m=\text{-}1}_{n=3}$, respectively. 

\newpage
\section{Genome editing of ap2s1 for Zebrafish}
\label{sup:ap2s1}

Genome editing of \textit{ap2s1}, \textit{ap2s1:ap2s1-mNeonGreen$^{bk800}$} in zebrafish was performed using a combination of previously described techniques~\cite{hoshijima2016precisegenome,hoshijima2016preciseediting,gutierrez2018efficient}. 
First, a 2.4 kb DNA fragment that spanned the stop codon of \textit{ap2s1} was cloned from AB wild type genomic DNA (genomic DNA from the AB population at UC Berkeley). 
The 2.4 kb PCR fragment was designed to include 1.2 kb of the end of \textit{ap2s1} and 1.2 kb downstream of the stop codon that were the homology arms for homology dependent repair. 
The fragment was sequenced to ensure that the selected guide RNA target was present and lacked any SNPs. 
The target sequence of the guide RNA used to cut \textit{ap2s1} was ACTGATTGACAGTTTACTCC (identified using the CRISPscan~\cite{moreno2015crisprscan} software, guide RNA synthesized by IDT). The donor did not require any mutations to prevent cutting because the guide target spanned the stop codon of \textit{ap2s1}. 
The mNeonGreen used was designed and synthesized to minimize predicted RNA splicing donor and acceptor sequences without changing the protein~\cite{wang2006characterization}. 
Isothermal assembly was used to clone a linker and mNeonGreen green sequence before the stop codon of \textit{ap2s1}~\cite{rabe2020simple}. 
The backbone we cloned the genomic fragment into and engineered the donor within contains a set of flanking sequences that we used for both isothermal assembly and PCR of the genome editing donor.  
These sequences come from the inDrops single-cell transcriptomics library preparation protocol and worked well for efficient PCR (GCATACGAGATCTCTTTCCCTACACG, CACGGTCTCGGCATTCCTGCTGAAC)~\cite{klein2015droplet}. 
For making an HDR donor for genome editing, oligos with biotin added to the 5' ends were used to prevent oligomerization upon injection~\cite{gutierrez2018efficient}.   
For embryo injections, 2 nl was injected into the 1-4 cell embryo of our injection recipe: 5$\mu$M guide RNA (purchased from IDT), 50 ng/$\mu$l clean cap Cas9 mRNA (from Trilink), 100 mM KCl, 0.1\% phenol red, 0.1 mM EDTA, and 1 mM Tris, pH 7.5. 
Approximately 3000 embryos were injected by a group of 3-5 lab members. 
Injected embryos were screened and selected for health, and successful injections as indicated by broad fluorescence signal were raised to adulthood. 
After 2-3 months, zebrafish were screened for potential founders through germ-line transmission. 
Embryos with positive signal were established as a new allele once we confirmed by PCR that mNeonGreen had been inserted cleanly at the targeted site in \textit{ap2s1}. 
Founders were out-crossed with Casper (mitfa$^{w2/w2}$; mpv17$^{a9/a9}$)~\cite{white2008transparent} mutants, and their offspring were crossed with Roy Nacre mutants again to establish the heterozygous \textit{ap2s1-mNeonGreen} and homozygous Casper alleles. 
These fish were in-crossed and screened for homozygous \textit{ap2s1-mNeonGreen$^{bk800/bk800}$} embryos for imaging.

\bigskip\noindent\textbf{DNA Sequences used for zebrafish genome editing and labeling mitochondria}

\begin{itemize}
    \item \colorbox{yellow}{GCATACGAGATCTCTTTCCCTACACG}---forward biotinylated primer for generating HDR template by PCR
    \item \textcolor{cyan}{tgctaaagttgactg\dots}---left homology arm
    \item \colorbox{gray!50}{gggcGGatccggtggatccggtggatct}---GGS linker between Ap2s and mNG polypeptides
    \item \textcolor{green!70}{gtTagCaagggcg\dots}---mNeonGreen through to stop codon
    \item \textcolor{magenta}{actgtcaatcagtc\dots}---right homology arm
    \item \colorbox{green}{GTTCAGCAGGAATGCCGAGACCGTG}---reverse biotinylated primer for generating HDR template by PCR
\end{itemize}

\noindent\colorbox{yellow}{GCATACGAGATCTCTTTCCCTACACG}%
\noindent\textcolor{cyan}{\seqsplit{tgctaaagttgactgttgagttttacagtgtagttagtgttttacactatcccaaaataagtctgatcattgttttactaaggcatgtgaaatatttattatcgactaaaatattgttattgttgtcaatatgtgagctcatgtttagtatgccattcattttgtcaaaaataaatgaataaaaaaaaacatgtttagagagtcagcaagaatatactatgaaacttattagaaggcagttaaaaagatcaattcaatatataaaagcaaaaacatatacggttgaagtcagaattattagccccctttgaatttttttttctttttttaaatatttcccaaatgatgtttaacagagcaaggaaactttcacaatatgtctgattatattttttcttctggaaaaggtctttcttgttttatttcggctagaataaaagccatttttaatttttaaacaccattttaaggacaaaattattagccccttaaaggtaatttttttttttttacgatagtctacagaaccattgttatacaataacttggctaattaccctaacctgcctagttaacctaattaacatacacgcatacataccttttaacacgcatacaaacacacacaatttttattttcacaaaataatgtttttgaaattcccagagatgggttgcggccggaagggcatctggtgtgtaaaaatgtgctggataagttggcggttcattccgctgtggcgaccccagataaataaagggactaagccgacaagaaaattaatgaatgaacgaatgaatgtttttgaagtttaagtttgtgtagaactgcgttatagtcaagaaatatcccaaaaattgagtcatgttgacaaacatacattgattaagttaacttaattgtttttgcaaattaaagtggattggaactaaagcaattaatttttctctcaaaaaacttaagaattgatttagctcaactaacaaaaatatttttgagtgcaaacacatgtaaaaaatgcacgaagcagcaagtttttagttgaaaacagcggccctatatgtgaactgctattaagaagtattgtggatgtgtttcaggtgtatacggtggtggacgagatgtttctagcaggagagatcagagaaaccagccagacgaaggtgctcaagcagctcctcatgctgcagtccctgga}}%
\noindent\colorbox{gray!50}{gggcGGatccggtggatccggtggatct}%
\textcolor{green!70}{\seqsplit{gtTagCaagggcgaggagGACAAtATGGCCTCtCTGCCCGCAACACACGAGCTGCATATTTTCGGAAGCATCAAcGGcGTGGATTTCGAtATGGTtGGgCAaGGaACTGGAAACCCAAATGAcGGaTACGAGGAACTGAATCTGAAGTCAACCAAAGGCGACCTCCAaTTCtcaCCTTGGATTCTcGTtCCCCAtATTGGCTATGGaTTTCATCAaTATCTGCCaTAtCCTGAtGGAATGTCACCATTTCAaGCcGCTATGGTGGATGGATCTGGCTACCAaGTCCACcgcACCATGCAaTTTGAGGACGGcGCCtccCTGACTGTGAACTACCGCTATACCTACGAGGGATCtCATATCAAGGGCGAAGCACAaGTtAAAGGaACAGGATTCCCAGCTGAcGGCCCCGTCATGACAAACTCTCTGACCGCCGCCGACTGGAGCCGGTCCAAGAAAACTTACCCTAACGATAAGACCATCATCTCTACCTTCAAaTGGAGTTATACCACcGGCAACGGaAAGCGcTACAGAAGCACAGCCCGAACTACCTATACTTTTGCTAAGCCtATGGCTGCAAACTATCTGAAAAATCAGCCTATGTAtGTcTTCcGAAAaACcGAattgAAGCACTCCAAAACAGAACTGAATTTCAAGGAgTGGCAGAAGGCTTTTACCGATGTtATGGGcatggacgagctgtacaaAtaa}}%
\noindent\textcolor{magenta}{\seqsplit{actgtcaatcagtcaaccaaccaatcaaacaaccaccattgtgtcaggctccgcccaccccgctgcagccaagccatatgatttgtgtgtatgagtgttgtcatgctgaaaatcatcctctaacatttcacctttttctttacactcgcatccacctcacaggatgttgccgttctgtcctgcaacttcatgaattttcgaggcagataagaattcgtaaaacttcctcggttatgtgacctgattatatcttggcatagccagaattgaattcttcagcgattttatagcacactagacttcagtgtttccgactggaagcctctccaagcctgttcattaacactaacatagataatagcgtcatgtgggccaaaaaactgccatggcacgaaagtgacagtcataattgcaagattccggggctgcactgcaatatcgcaatgagaaaaacttgatttgatgtccatttaaggaaattcccatatatgtactattaaaaaaagaatcaaatgtttaatatgctgcactgcaaataataaatttgttactgctttgtcttgttttctggtacaagtatccatattgttaaaactccaaatacgtaaaggtgatcttcctgttggtgactatatctttttttatacaattaagatgttcataatggtgtcacaaaaagggtttcgcaacaactgttggcacttttgagagtaaaaaaaaaacatacaggcaaaaccaaattaactgaccgtttcagtatggagatgtcattaatccatgaacagttcctgcttgtcttaactatttctgaattgtaagaggcgattcggttgtaactaatgttaacatggctgctaaaaaaagtgtcaatatggacctttctggattctaataagatttggaaatgaagatataaaacagcaaatacataaataccattgttattatttacatccctcaaaatggtctacccactactgattcggtgaggtcctaagaagtgaaattaaaatttatgtcaaacaaaatacattattttagatccacagctttggccaattgtattctttttgttttgcctattattattattttttttttttcactctaaaaagtaataatcgggttaatcagcctaatggtgaatttaaagaaaaatgaaggagaaaacaaaatatttccttgccatactcttac}}%
\noindent\colorbox{green}{GTTCAGCAGGAATGCCGAGACCGTG}

\bigskip\noindent\textbf{COX8a-mChilada (mitochondria)}

\begin{itemize}
    \item \textcolor{violet}{ATGTCTGGACT\dots}---cDNA of N-terminal 34 amino acids of \textit{Cox8a}
    \item \textcolor{red}{GACAATATGG\dots}---mChilada 
    \item \colorbox{gray!50}{ggcGGatccggtgg}---linker
\end{itemize}

\noindent\textcolor{violet}{\seqsplit{ATGTCTGGACTTCTGAGGGGACTAGCTCGCGTCCGCGCCGCTCCGGTTCTGCGGGGATCCACGATCACCCAGCGAGCCAACCTCGTTACGCGACCCGCGAAG}}%
\noindent\colorbox{gray!50}{ggcGGatccggtggatccggtgga} 
\noindent\colorbox{gray!50}{AGCgtTagcaagggcgaggag}%
\noindent\textcolor{red}{\seqsplit{GACAATATGGCTATTATCAAAGAATATATGCGTTTTAAaGTtCATATGGAGGGCAGCGTCAATGGACATGAATTTGAAATTGAGGGcGAGGGAGAGGGCAGGCCATTCGAAGGGACTCAGACAGCGAAGTTGAAAGTTACCAAGGGGGGACCATTGCCATTTGCATGGCACATCCTCCCGCCCCAaTTTCAATATGGTTCTAAGGCATATGTTAAGCATCCTGCTGATATACCCGATTACTTCAAaCTgTCaTTTCCtGAgGGTTTTACCTGGGAACGCGAAATGAATTTTGAAGACGGCGGCGTtGTAACAGTCACACAGGACTCCAGTCTGCAaGACGGCGAGTTTATCTACAAAGTtAAACTTCGTGGAACCAACTTTCCCTCCGACGGGCCTGTCATGCAAAAGAAGACAATGGGGAACACCGCTTCAACcGAGCGtATGTACCCGGAAGATGGAGCTCTTAAgGGcGAAACGAAGTGGCGTCTTAAGTTGAAAGATGGTGGTCACTACGAGGCTGAAGTGAAGACCACGTATAAGGCGAAAAAGCCCGTACAGCTGCCTGGAGCGTACAATGTGGATCGTAAATTGAAGATAACCTACCATAATGAGGATTACACCATCGTGGAGCAGTATGAACGAGCCGAGGCTCGGCACTCAACGGGTggcatggacgagctgtacaagtaA}}

\clearpage

\clearpage        

\clearpage        

\clearpage
\begin{figure*}[!tp]
    \centering
    \includegraphics[width=\textwidth]{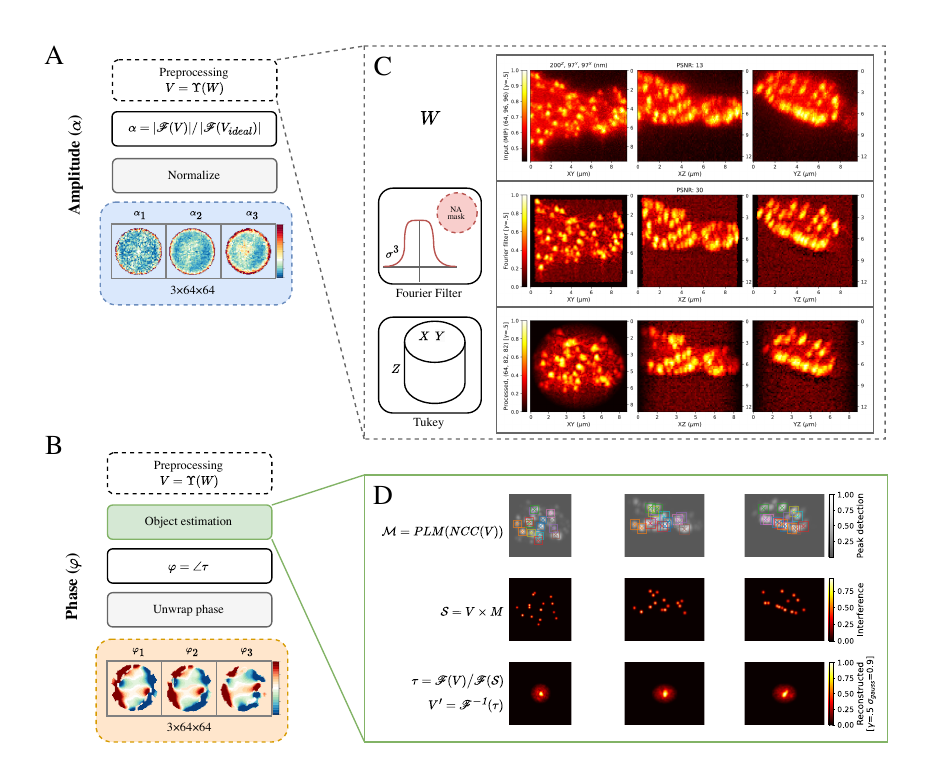}
    \caption{
        \textbf{Conversion of raw image volumes to Fourier embeddings}.
        The raw volume is first frequency filtered (Fourier Filter) and windowed (Tukey window, $\hat{x}\hat{y}$ only). The 3D FFT of the result is input into the Amplitude ($\alpha$) and Phase ($\varphi$) Fourier Embedding calculations.
        \textbf{A.} A breakdown of the steps needed to create the amplitude embedding ($\alpha$).
        \textbf{B.} A breakdown of the steps needed to create the phase embedding ($\varphi$).
        \textbf{C.} MIPs from a zebrafish embryo covering $9.3 \times 9.3 \times 12.8$ $\mu$m$^3$ FOV to illustrate the preprocessing modules we use to reduce noise and suppress edge artifacts.
        \textbf{D.} MIPs of the same FOV showing the intermediate steps to remove the interference patters from the given volume.
    }
    \label{fig:embedding}
\end{figure*}

\clearpage
\begin{figure*}[!tp]
    \centering
    \includegraphics[width=\textwidth]{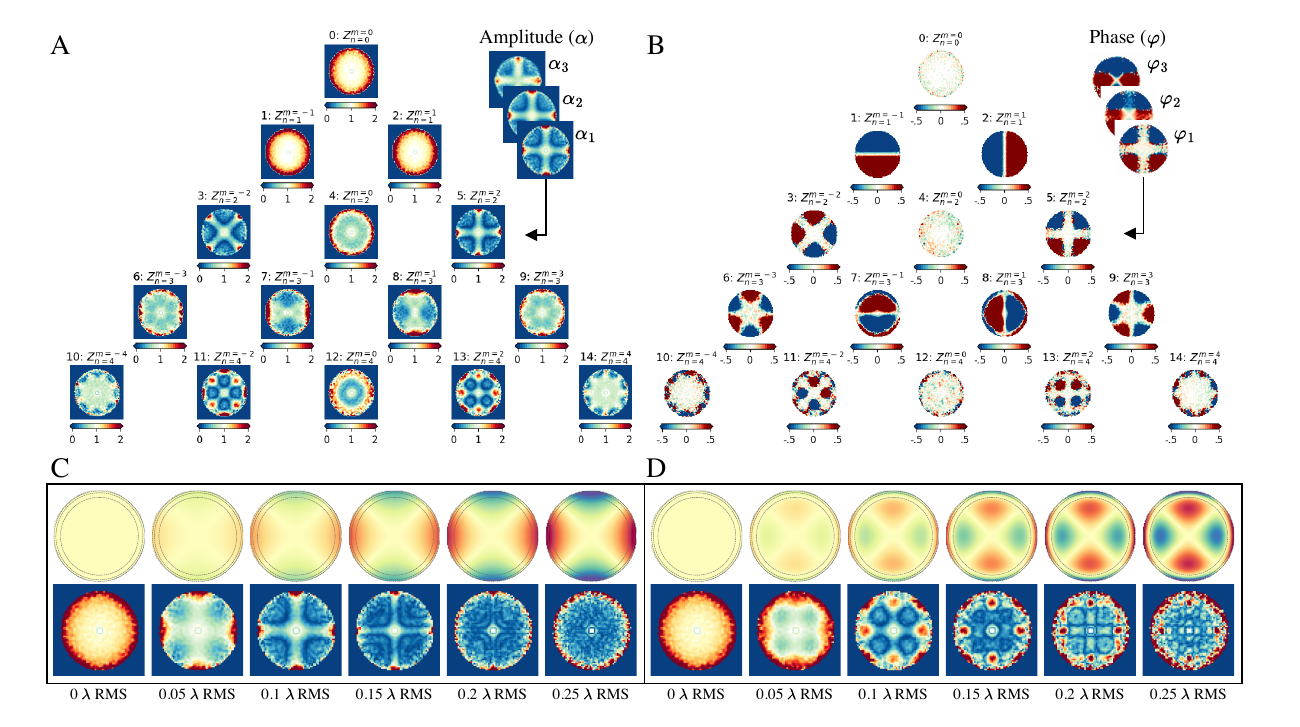}
    \caption{
        \textbf{Aberration fingerprints}.
   The principal Fourier plane of  \textbf{(A)} amplitude ($\alpha_1$) and \textbf{(B)} phase ($\varphi_1$) embeddings for the first 15 Zernike modes ($Z_0^0$ through $Z_4^{\pm4}$) each at 0.1$\lambda$~RMS, shown for an ideal noise-free volume containing a single point emitter.
     The wavefronts and corresponding amplitude embeddings ($\alpha_1$) for \textbf{(C)} $Z_2^{2}$ and \textbf{(D)}  $Z_4^{2}$ as aberration magnitudes increase  from 0 to 0.25$\lambda$~RMS.
    }
    \label{fig:modes-embedding-fingerprints}
\end{figure*}

\clearpage
\begin{figure*}[!tp]
    \centering
    \includegraphics[width=\textwidth]{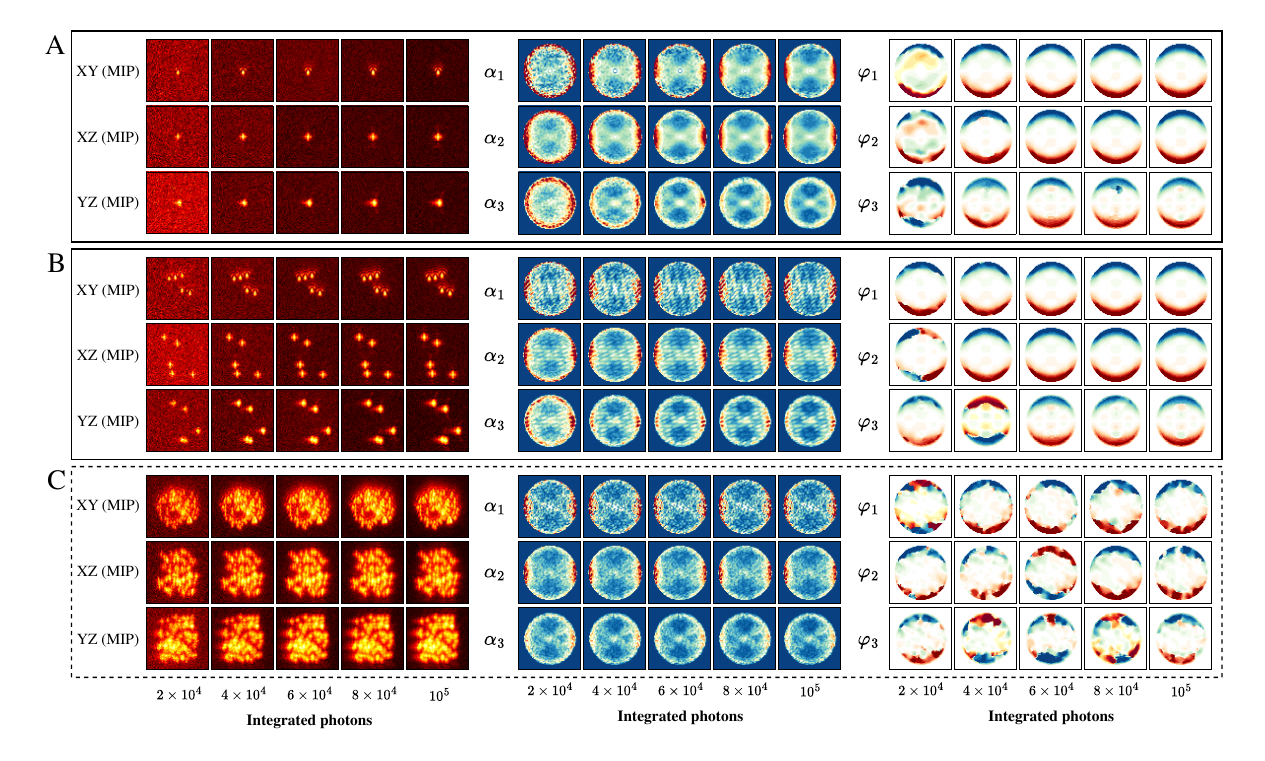}
    \caption{
        \textbf{Generalized representations via Fourier embedding}.
        Real-space (left) versus Fourier amplitude (middle) and phase (right) embeddings ($\alpha_1$, $\alpha_2$, $\alpha_3$, $\varphi_1$, $\varphi_2$, $\varphi_3$) for a $9.3 \times 9.3 \times 12.8$ $\mu$m$^3$
       volume. An aberration of 0.1$\lambda$~RMS in mode $Z_3^{-1}$ is applied to volumes containing \textbf{(A)} a single bead, \textbf{(B)} five beads, and \textbf{(C)}  one hundred beads.
    }
    \label{fig:modes-embedding-objects}
\end{figure*}

\clearpage
\begin{figure*}[!tp]
    \centering
    \includegraphics[width=\textwidth]{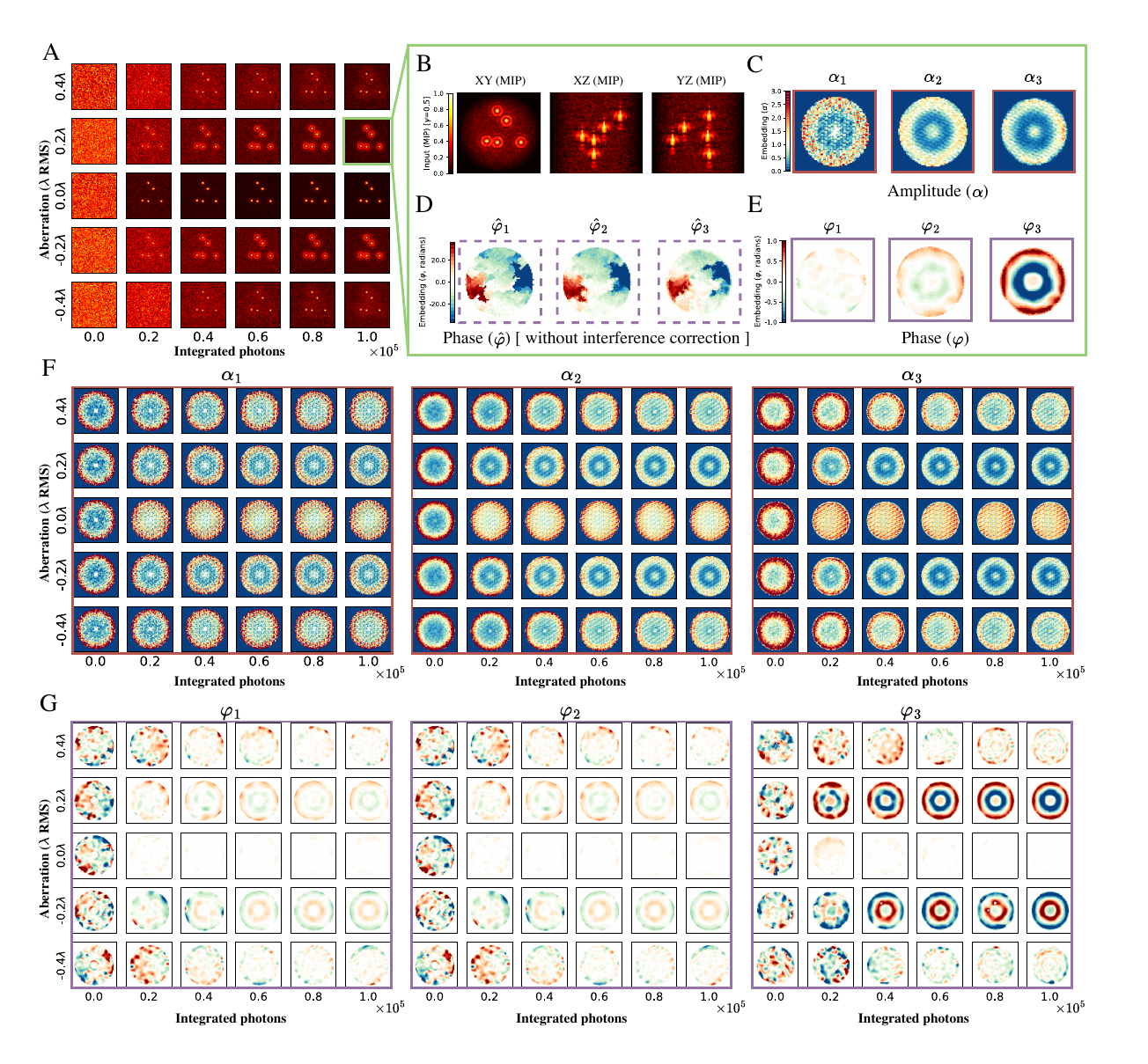}
    \caption{
        \textbf{Fourier embedding composition}.
        \textbf{A.} XY MIPs of five synthetic beads, each with an initial  aberration ranging from -0.2$\lambda$~RMS to 0.2$\lambda$~RMS in mode [$Z^{m=0}_{n=4}$] at different SNR levels.
        \textbf{B.} XY, XZ, and YZ MIPs of a single FOV simulated with $10^5$ photons.
        \textbf{C.} Amplitude embedding ($\alpha_1$, $\alpha_2$, $\alpha_3$) for the FOV in \textbf{B}.
        \textbf{D.} Phase embedding ($\varphi_1$, $\varphi_2$, $\varphi_3$) for the same FOV in \textbf{B}, shown without interference correction.
        \textbf{E.} Phase embedding after after removing bead-induced interference patterns.
        \textbf{F.} Amplitude embeddings for all examples shown in \textbf{A}.
        \textbf{G.} Phase embeddings for all examples shown in \textbf{A}.
    }
    \label{fig:modes-embedding-planes}
\end{figure*}

\clearpage        
\begin{figure*}[!tp]
    \centering
    \includegraphics[width=.85\textwidth]{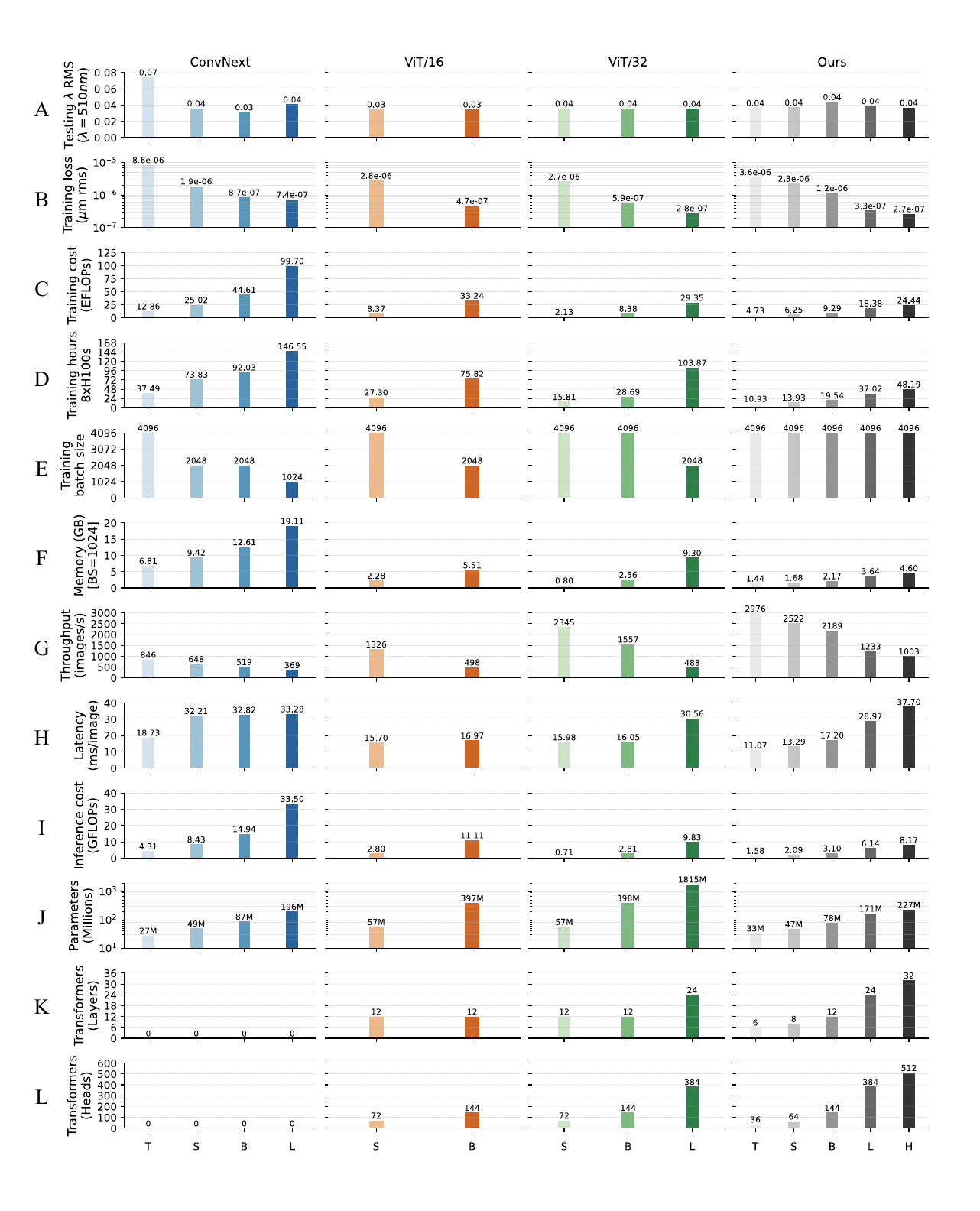}
    \caption{
        \textbf{Comparison of the current state-of-the-art architectures applied to 3D aberration sensing across cost indicators of model efficiency.}
        \textbf{A.} Median $\lambda$ RMS residuals over 10K test samples after one correction for aberrations ranging between $0.1\lambda$ to $0.2\lambda$ simulated with 50K and up to 200K integrated-photons.
        \textbf{B.} Training loss over a dataset consisting of 2M synthetically generated samples.
        \textbf{C.} Training cost measured in exaFLOPS ($10^{18}$ FLOPs) for 500 epochs of training.
        \textbf{D.} Training time using eight H100 GPUs. 
        \textbf{E.} Optimal batch size used for training based on memory usage for each model. 
        \textbf{F.} Memory footprint of each model with a batch of 1024 images using 16-bit floating point precision.
        \textbf{G.} Max number of predictions per second using a batch size of 1024 on a single A100 GPU.
        \textbf{H.} Average inference time (milliseconds) per image for a batch of 1024 examples using a single A100 GPU.
        \textbf{I.} Inference cost per image measured in gigaFLOPs ($10^{9}$ FLOPs).
        \textbf{J.} Total number of trainable parameters.
        \textbf{K.} Total number of transformer layers (not applicable for ConvNeXt models). 
        \textbf{L.} Total number of heads in all transformer layers (not applicable for ConvNeXt models). 
        Lower numbers are better for all cost indicators, except for \textbf{E}, \textbf{G}, \textbf{K}, and \textbf{L}.
    }
    \label{fig:scaling-archs-rms}
\end{figure*}

\clearpage        
\begin{figure*}[!tp]
    \centering
    \includegraphics[width=\textwidth]{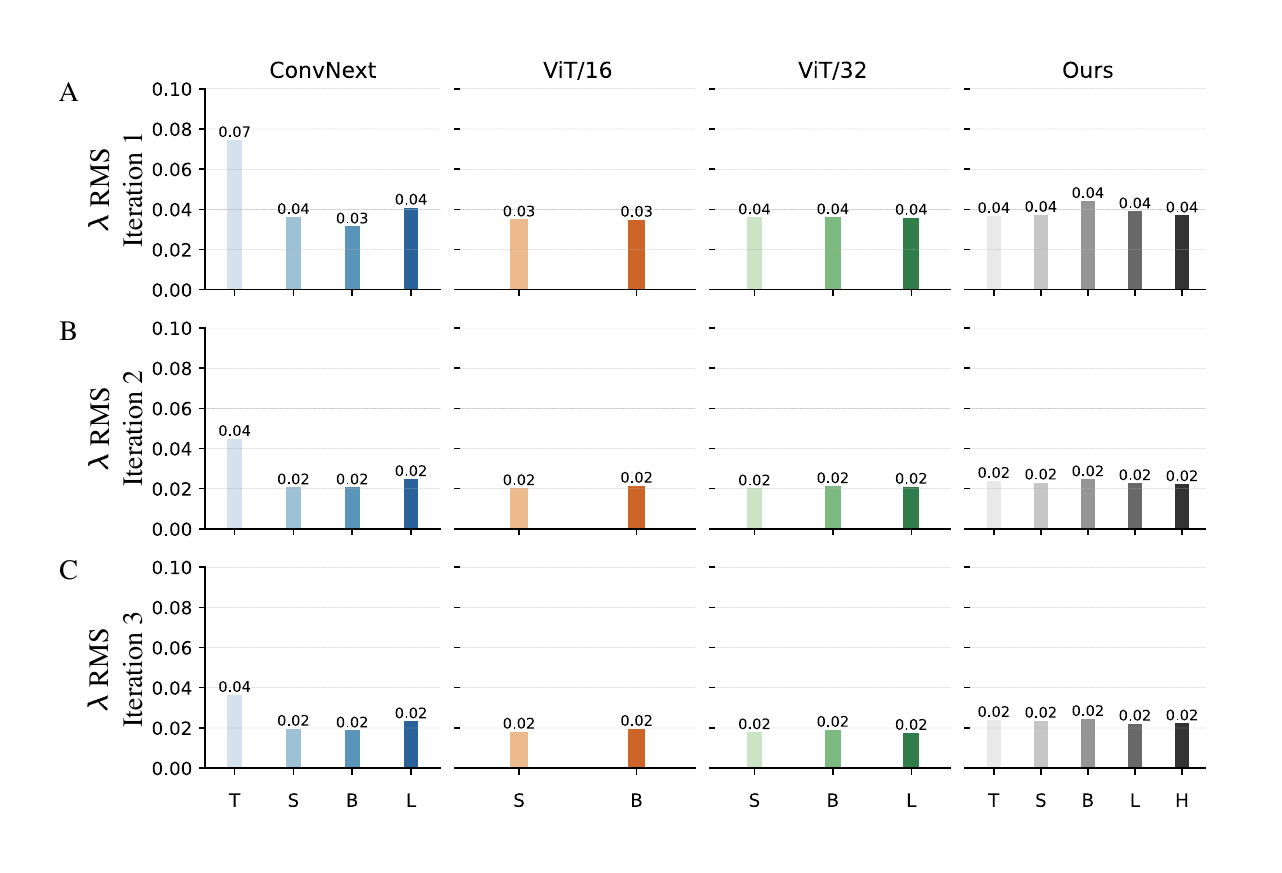}
    \caption{
    \textbf{Median $\lambda$ RMS residuals over 10K test samples with aberrations ranging between $0.2\lambda$ RMS to $0.4\lambda$ RMS, simulated with 50K and up to 200K integrated-photons}.
    \textbf{A}. First iteration.
    \textbf{B}. Second iteration.
    \textbf{C}. Third iteration.
    }
    \label{fig:scaling-eval-rms}
\end{figure*}

\clearpage        
\begin{figure*}[!tp]
    \centering
    \includegraphics[width=\textwidth]{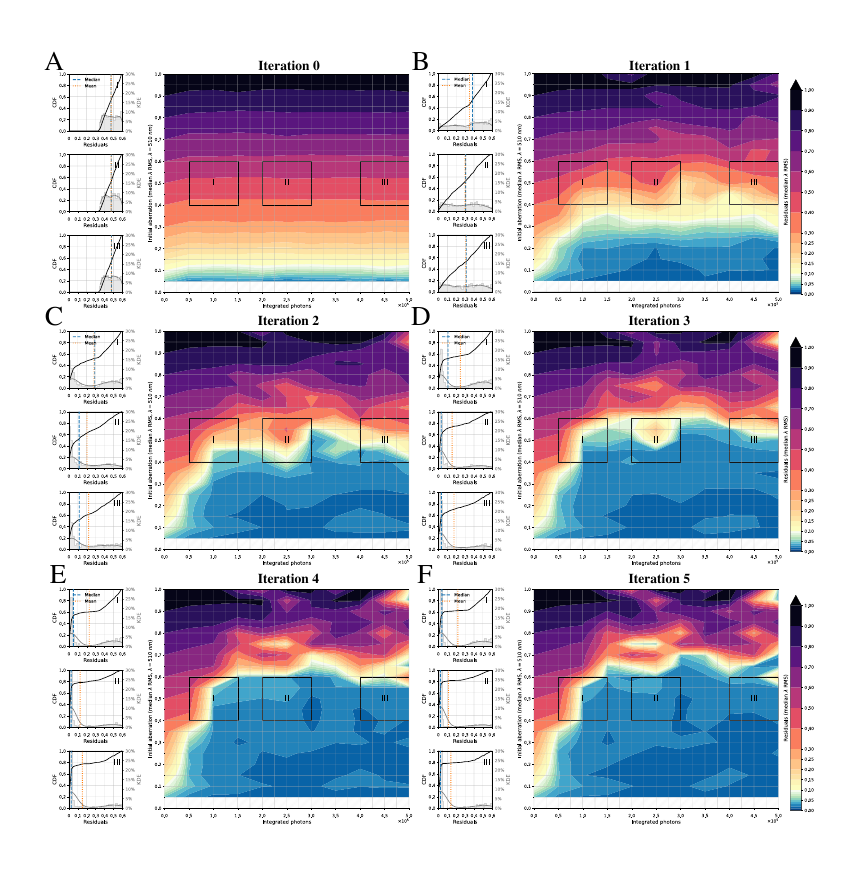}
    \caption{\textbf{Sensitivity to SNR.}
    Evaluation of our Small model (S) using simulated data of a single bead.
    \textbf{A}. The initial RMS distribution of test dataset as a function of SNR without AO.
    \textbf{B--F}. Residual $\lambda$~RMS with five rounds of AO corrections. 
    }
    \label{fig:eval-snrheatmap}
\end{figure*}

\clearpage        
\begin{figure*}[!tp]
    \centering
    \includegraphics[width=\textwidth]{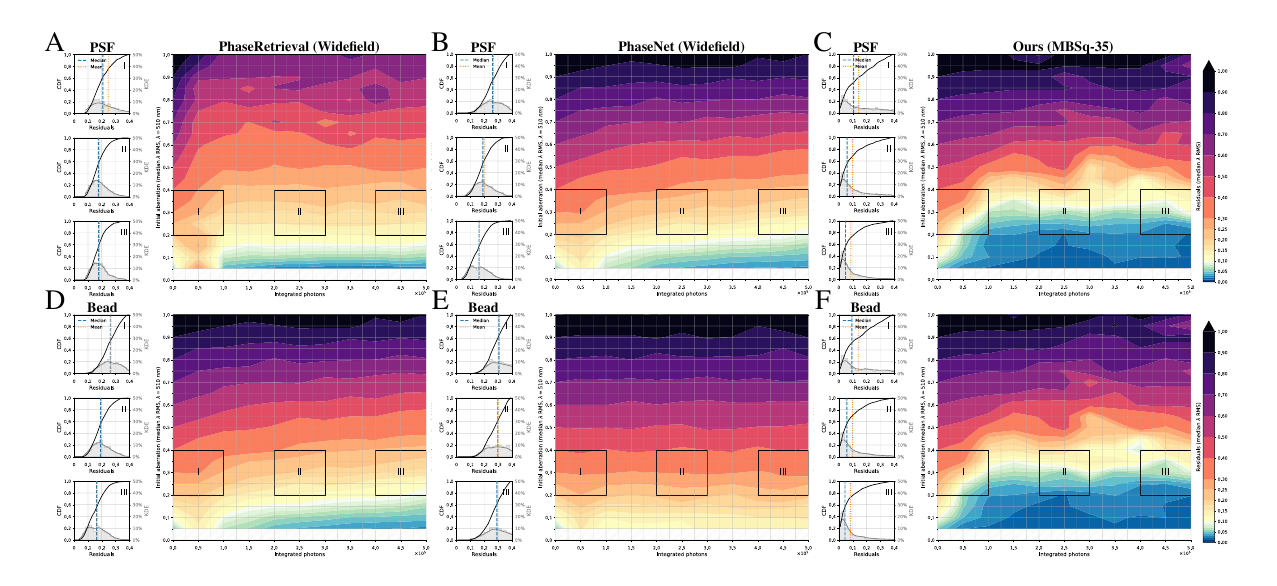}
    \caption{
    \textbf{Singleshot evaluation of phase retrieval using 10K synthetic PSFs}.
    \text{A--C}. Residual $\lambda$~RMS for centered bead using PhaseRetrieval, PhaseNet, and our model, respectively.
    \text{D--F}. Residual $\lambda$~RMS for a non-centered bead using the three methods mentioned above.
    }
    \label{fig:comparisons-phase-retrieval-rms}
\end{figure*}

\clearpage        
\begin{figure*}[!tp]
    \centering
    \includegraphics[width=\textwidth]{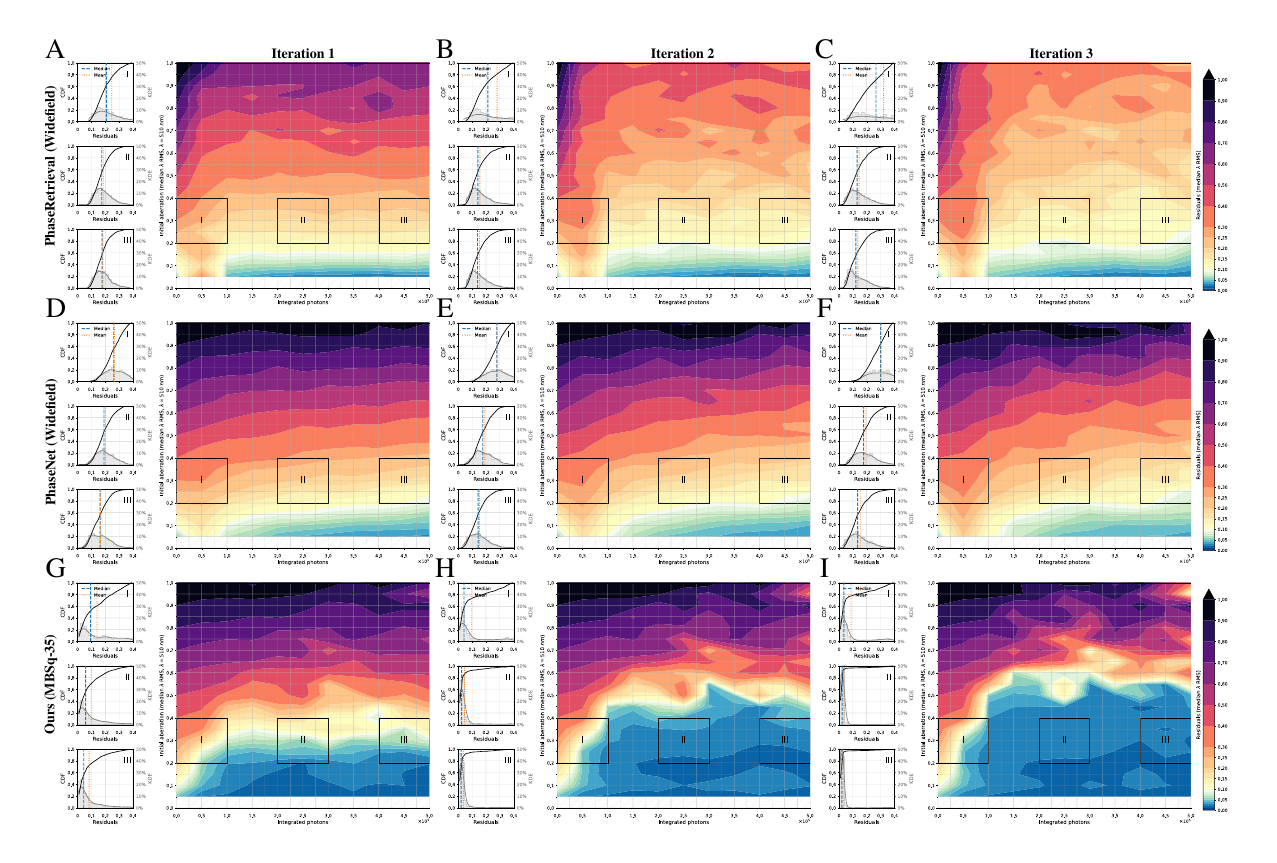}
    \caption{
    \textbf{Iterative evaluation of phase retrieval using 10K synthetic PSFs}.
    \text{A--C}. Residual $\lambda$~RMS forwidefield PSFs using PhaseRetrieval.
    \text{D--F}. Residual $\lambda$~RMS for widefield PSFs using PhaseNet.
    \text{G--I}. Residual $\lambda$~RMS for YuMB LLS PSFs using our Small model (S).
    }
    \label{fig:comparisons-phase-retrieval-psf-rms-iterations}
\end{figure*}

\clearpage        
\begin{figure*}[!tp]
    \centering
    \includegraphics[width=\textwidth]{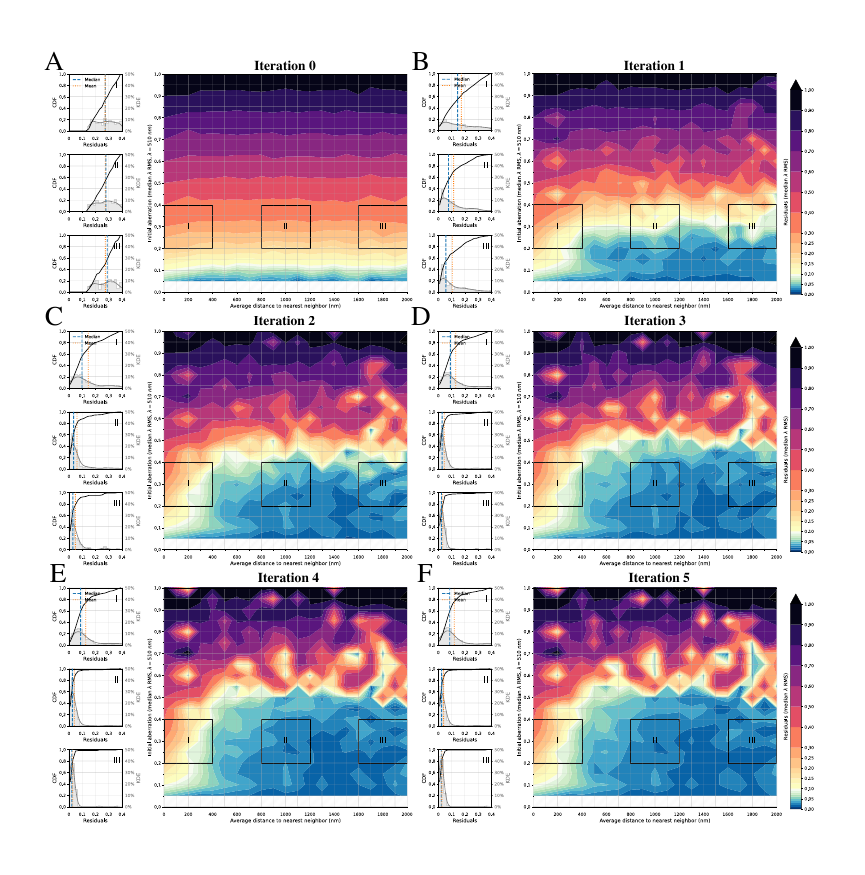}
    \caption{\textbf{Sensitivity to the density of objects}.
    Evaluation of our Small model (S) using simulated data with up to 150 beads in any given test sample.
    \textbf{A}. The initial RMS distribution of test dataset as a function of the average distance to the nearest neighbor (bead) without AO.
    \textbf{B--F}. Residual $\lambda$~RMS with five rounds of AO corrections. 
    }
    \label{fig:eval-densityheatmap}
\end{figure*}

\clearpage        
\begin{figure*}[!tp]
    \centering
    \includegraphics[width=0.95\textwidth]{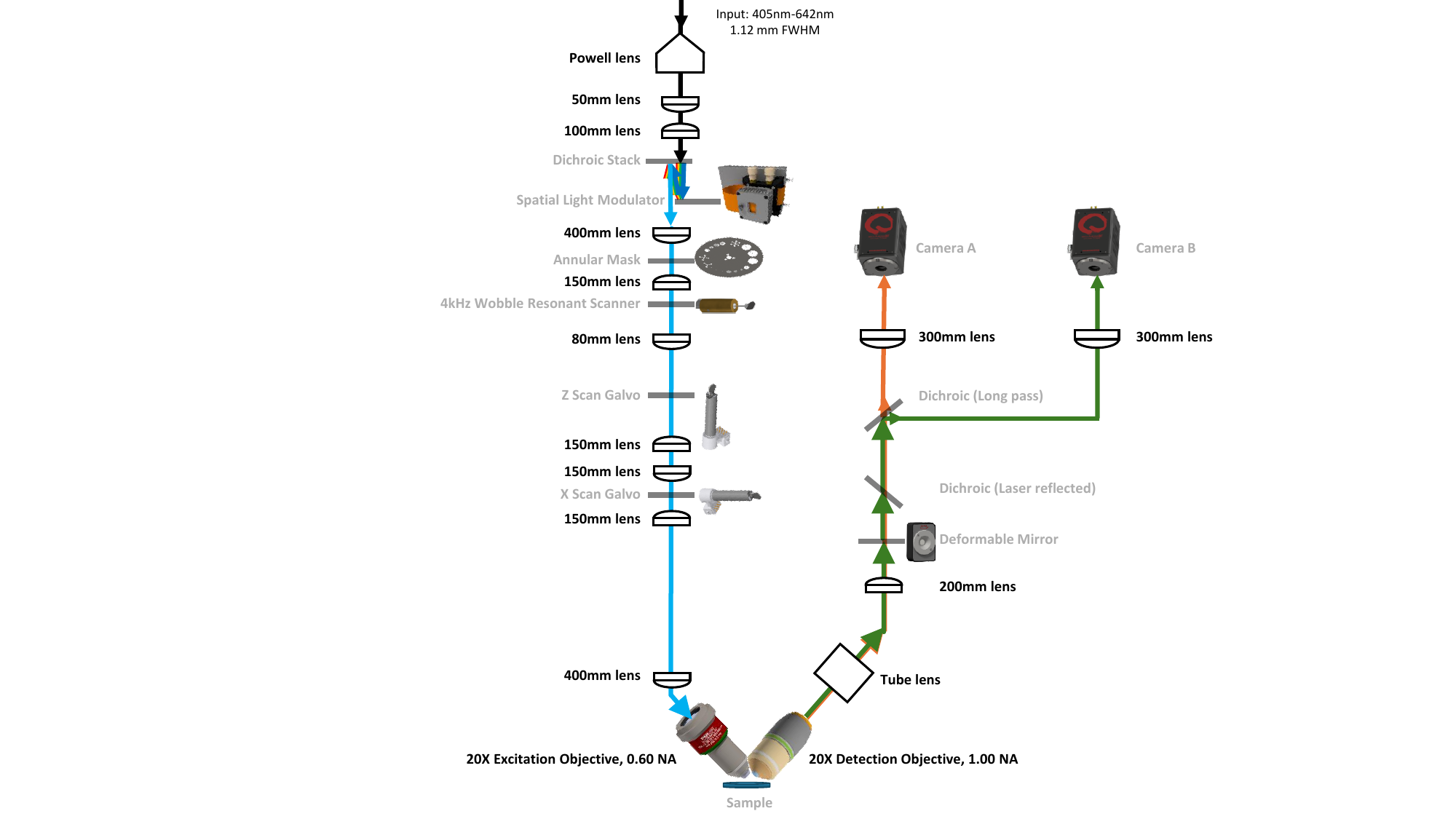}
    \caption{
        \textbf{Schematic of AO-LLS microscope used for all experiments}.    
    }
    \label{fig:schematic_lls}
\end{figure*}

\clearpage        
\begin{figure*}[!tp]
    \centering
    \includegraphics[width=\textwidth]{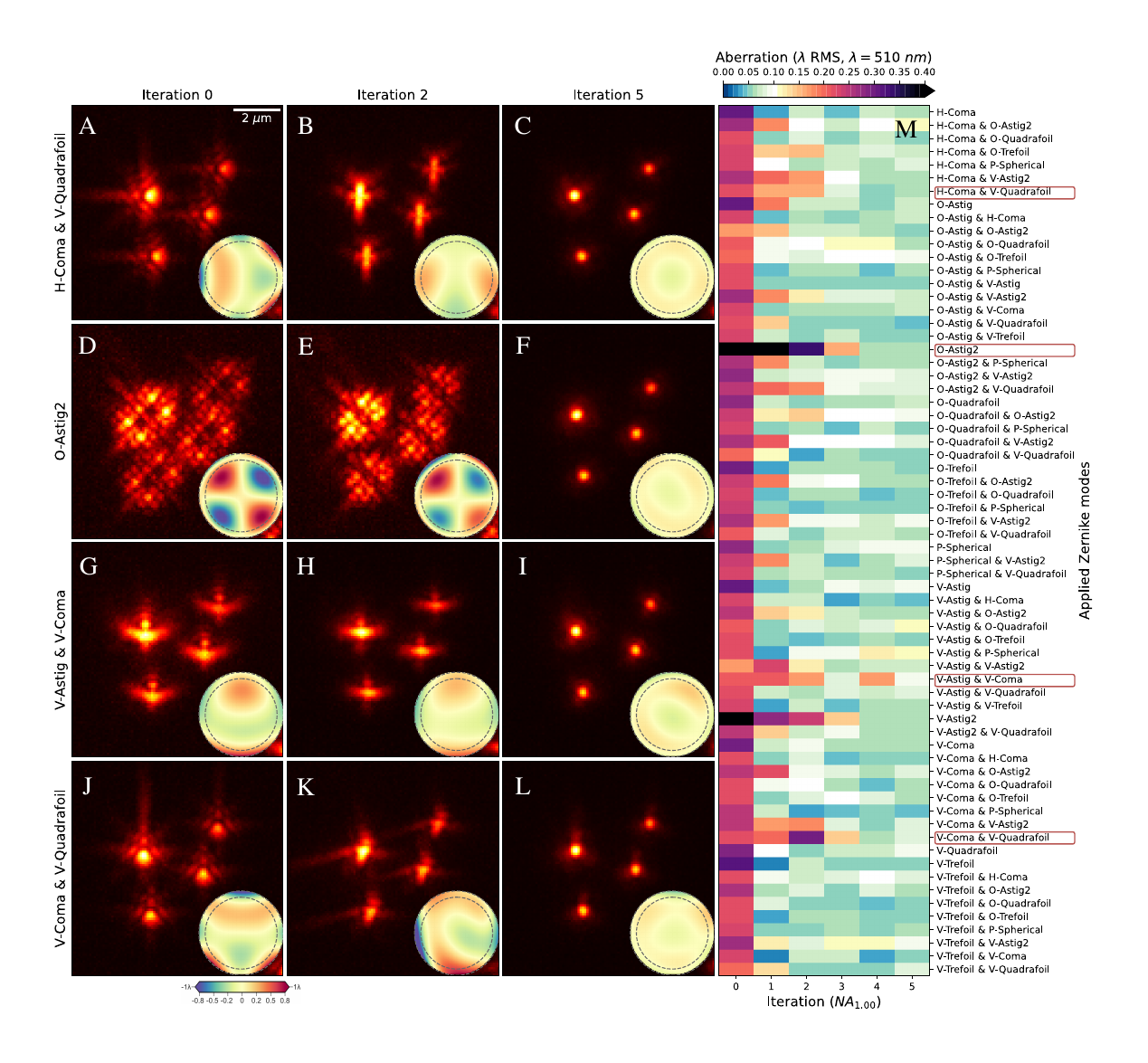}
    \caption{
    \textbf{Correction of beads with initial artificial aberrations}.
    Four examples, 
    H-Coma \& V-Quadrafoil $(Z^{m=1}_{n=3} + Z^{m=4}_{n=4})$, 
    O-Astig2 $(Z^{m=2}_{n=4})$, 
    V-Astig \& V-Coma $(Z^{m=2}_{n=2} + Z^{m=\text{-}1}_{n=3})$, 
    V-Coma \& V-Quadrafoil $(Z^{m=\text{-}1}_{n=3} + Z^{m=4}_{n=4})$, 
    where the initial aberration is artificially applied by the DM.
    \textit{Iteration 0} shows XY maximum projection of four beads with initial aberration imaged using LLS, upon which \model~makes predictions. 
    \textit{Iteration 2} shows the resulting field of beads after applying \model~prediction to the DM.
    \textit{Iteration 5} shows the results after applying the \model~prediction measured from \textit{Iteration 4}. 
    Insets show the \model~predicted wavefront over the $\text{NA}=1.0$ pupil with a dashed line at $NA=0.85$
    \textbf{M}. Heatmap of the residual aberration (as measured via phase retrieval on a well isolated bead) after application of \model~predictions starting with a single Zernike mode up to Mode 14 ($Z^{m=4}_{n=4}$) for up to 5 iterations.
    }
    \label{fig:beads-si}
\end{figure*}

\clearpage        
\begin{figure*}[!tp]
    \centering
    \includegraphics[width=\textwidth]{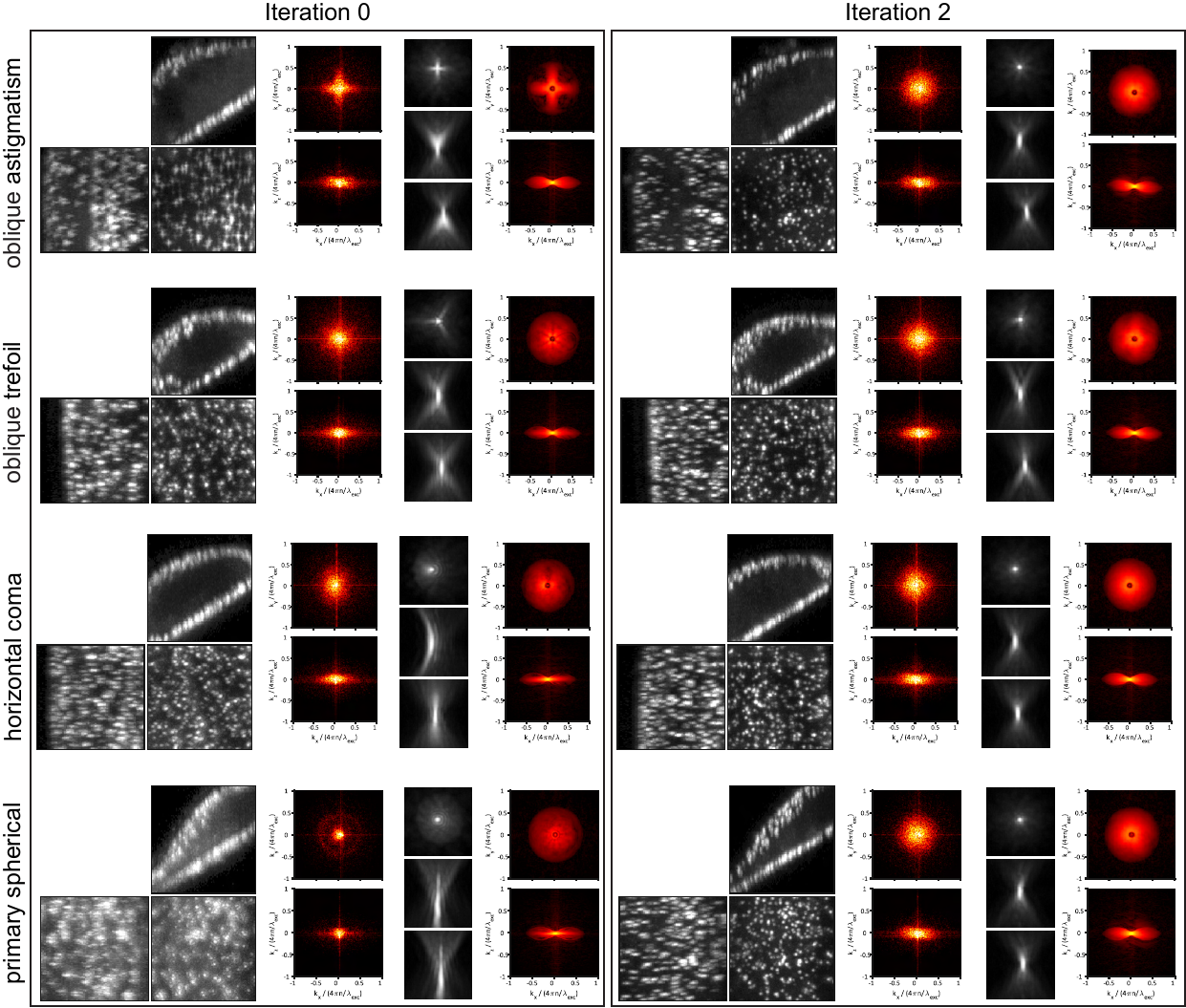}
    \caption{
    \textbf{Experimental evaluation of \model~on live SUM159 AP2 cells under single mode aberrations}.
    Four different single-mode aberrations (top to bottom) were introduced via the microscope DM to test \model~performance on live cells. Left panels show XY, XZ, YZ MIPs of a center-cropped region of cells; middle panels show the corresponding FFT of the cell volumes; and right panels show the widefield PSF and  OTF MIPs. Each box depicts the cells, cell volume FFTs, and widefield PSF/ OTF pairs with applied single mode aberrations (Iteration 0), and after two rounds of iterative \model~correction (Iteration 2).
    }
    \label{fig:Figure4_SI1}
\end{figure*}

\clearpage        
\begin{figure*}[!tp]
    \centering
    \includegraphics[width=\textwidth]{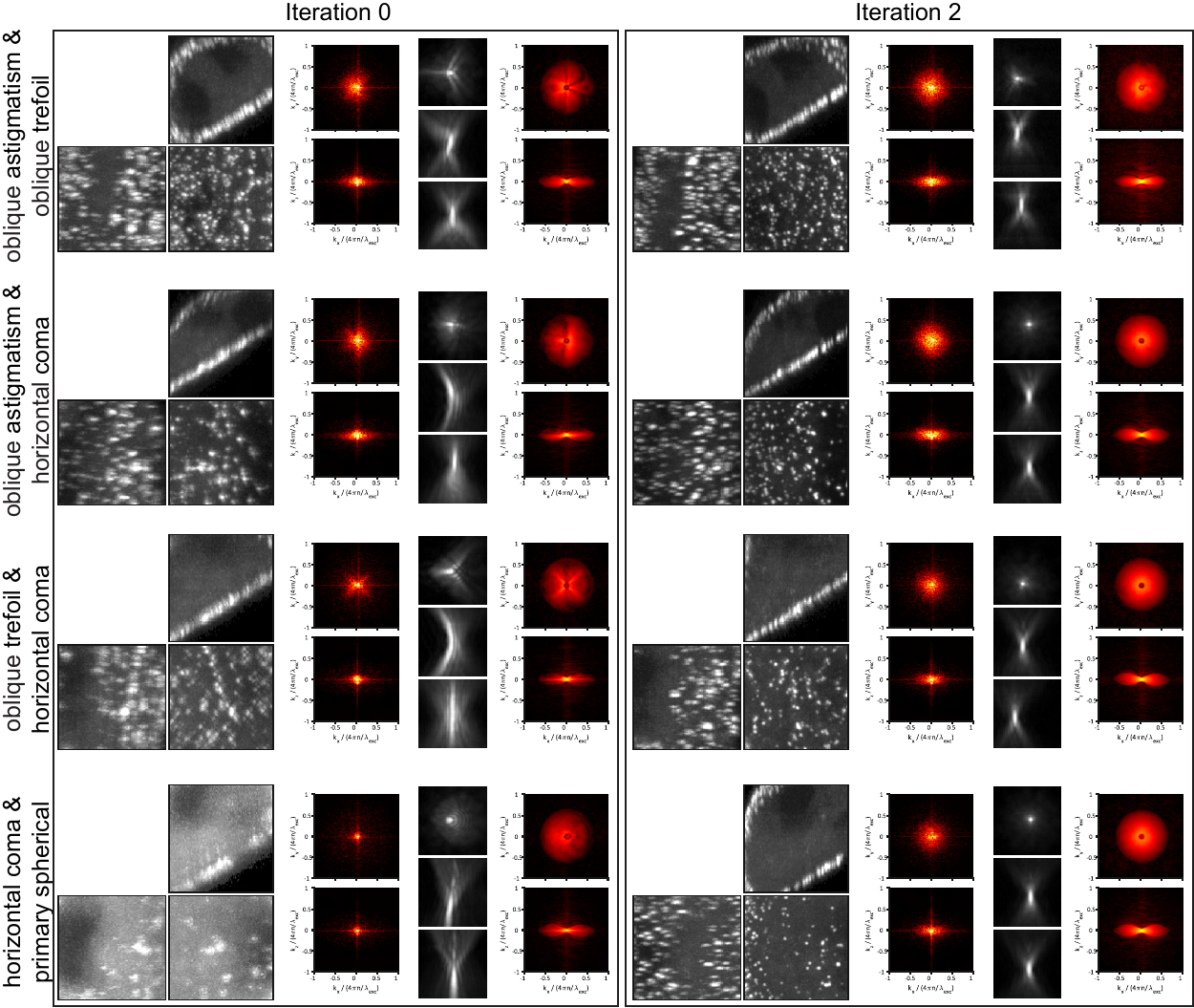}
    \caption{
    \textbf{Experimental evaluation of \model~on live SUM159 AP2 cells under multiple mode aberrations}.
    Four different two-mode aberration (top to bottom) were introduced via the microscope DM to test \model~performance on live cells. Left panels show XY, XZ, YZ MIPs of a center-cropped region of cells; middle panels show the corresponding FFT of the cell volumes; and right panels show the widefield PSF and OTF MIPs. Each box depicts the cells, cell volume FFTs, and widefield PSF/ OTF pairs with applied single mode aberrations (Iteration 0), and after two rounds of iterative \model~correction (Iteration 2)
    }
    \label{fig:Figure4_SI2}
\end{figure*}

\clearpage        
\begin{figure*}[!tp]
    \centering
    \includegraphics[width=0.95\textwidth]{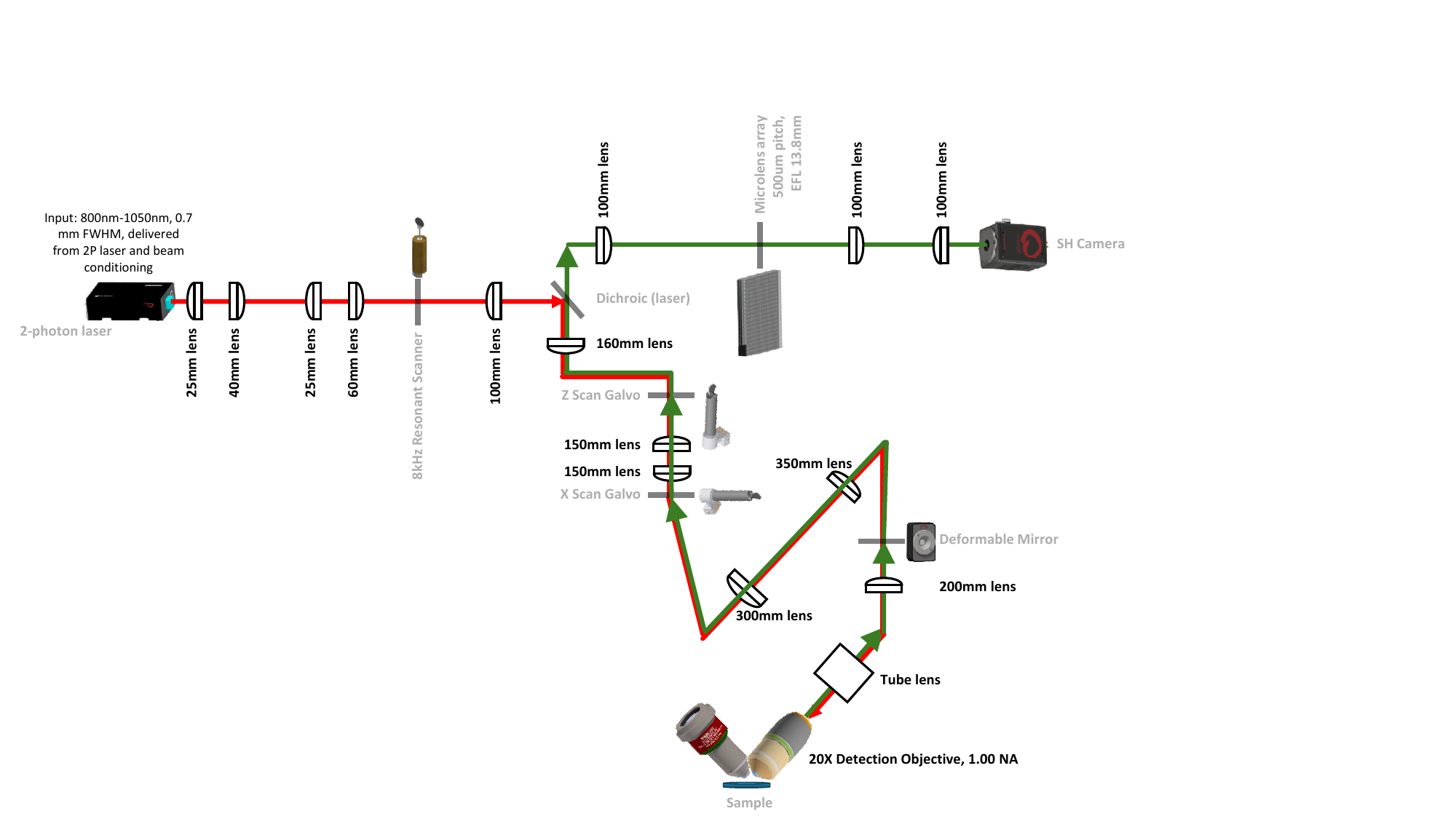}
    \caption{
        \textbf{AO-LLS microscope schematic illustrating the Shack-Hartmann wavefront detection mode}. Lenses are in 4f configuration. The X Scan Galvo, Z Scan Galvo, Annular Mask, and Deformable Mirror (DM) are at pupil conjugate planes. The Spatial Light Modulator (SLM), 4kHz wobble resonant scanner, and cameras are all at sample conjugate planes.
    }
    \label{fig:schematic_sh}
\end{figure*}

\clearpage
\begin{figure*}[!tp]
    \centering
    \includegraphics[width=0.8\linewidth]{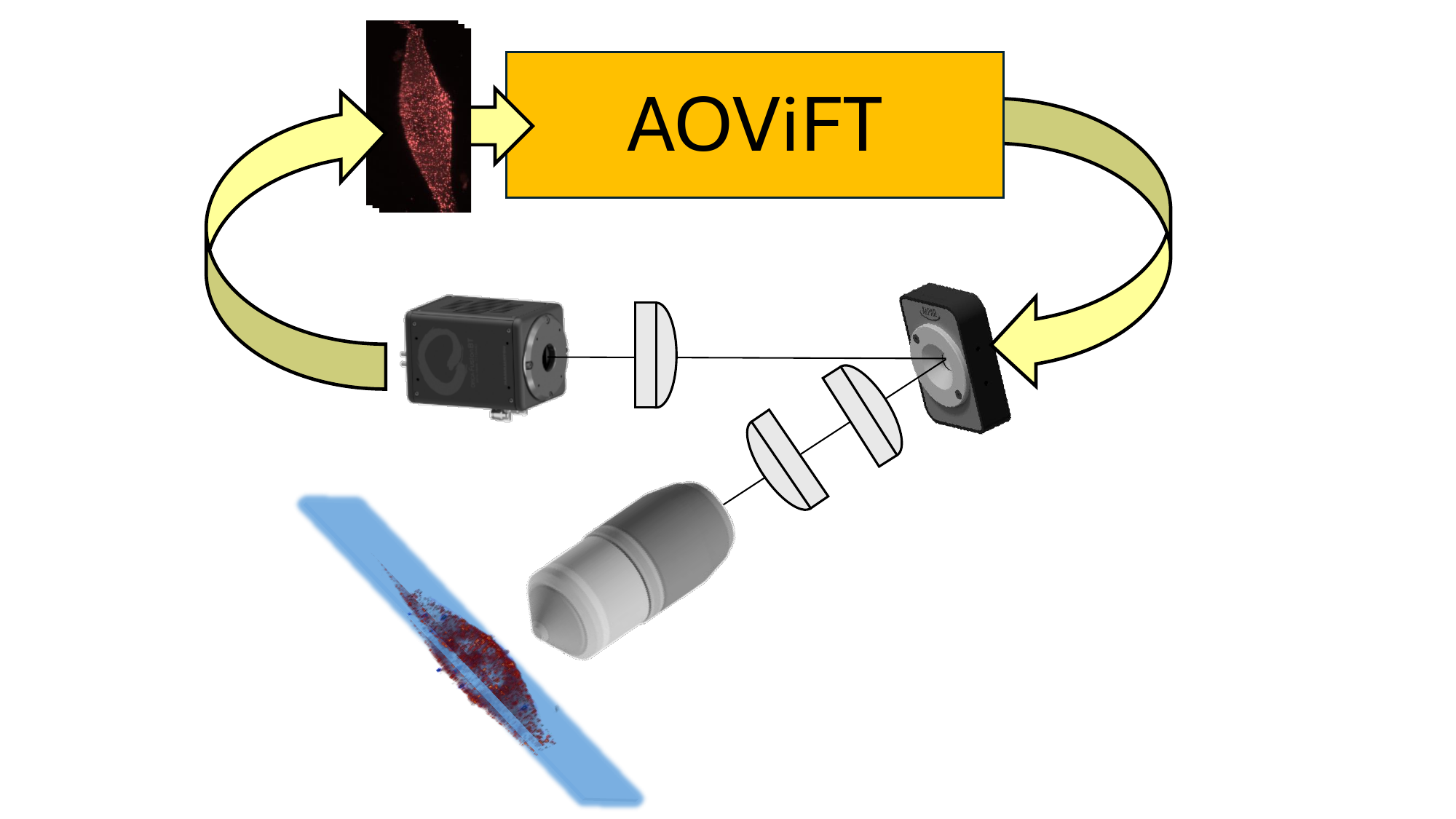}
    \caption{
        \textbf{\model~closed loop operation}. Light-sheet excitation (blue), illuminates the focal plane of the detection objective causing emission from fluorescent markers (red) in the specimen.  This emitted light is collected by the detection objective, relayed to a deformable mirror, and recorded by a camera.  The acquired image stack is fed into \model~which produces an update for the deformable mirror shape.  
    }
    \label{fig:schematic_cycle}
\end{figure*}

\clearpage        
\begin{figure*}[!tp]
    \centering
    \includegraphics[width=\textwidth]{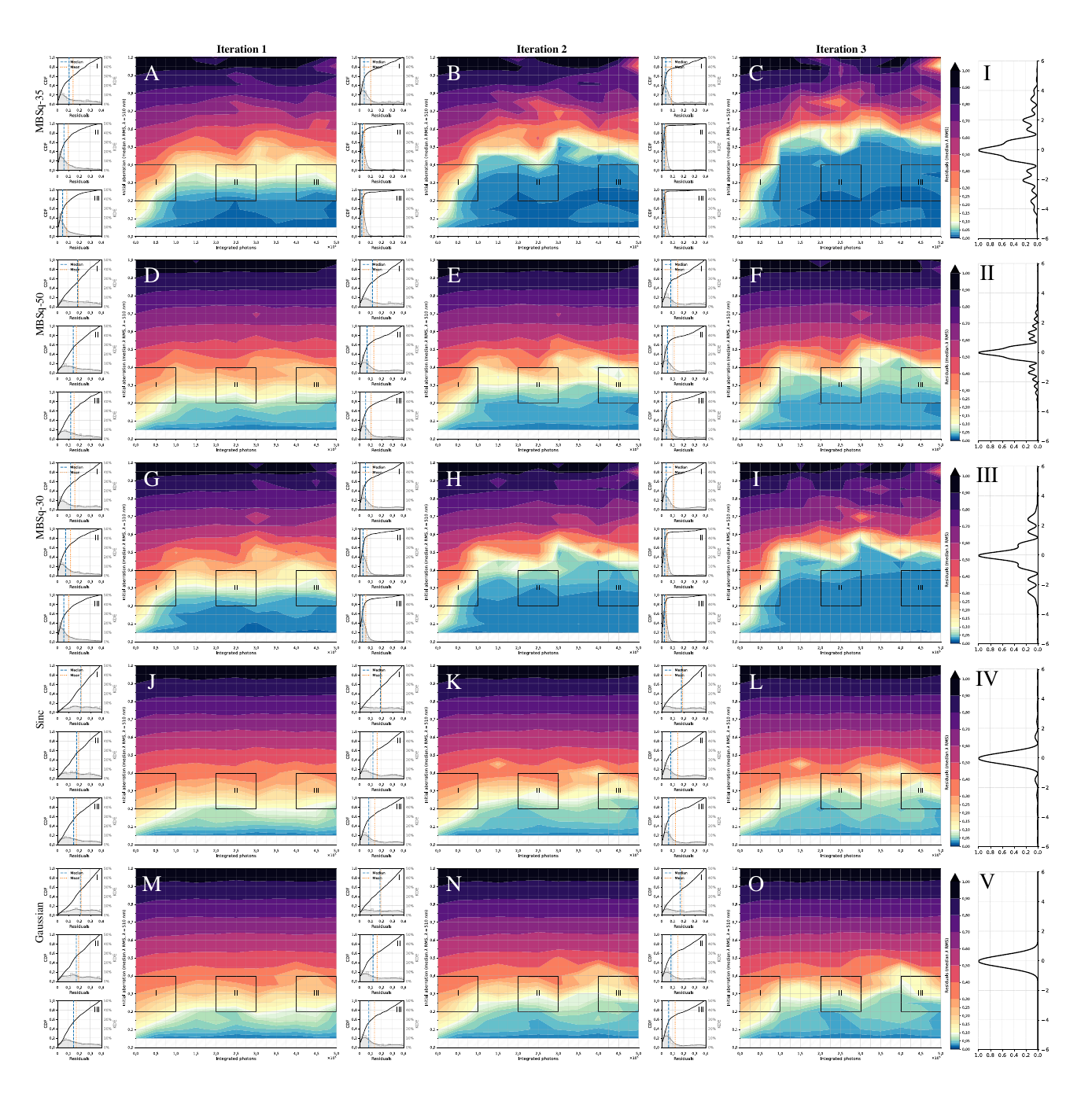}
    \caption{\textbf{Evaluation of our Small model (S) using several different light sheet types over 10K synthetic samples of a single bead with aberrations up to 1.0$\lambda$~RMS}. 
    \textbf{A--C}. MBSq-35  (the original LLS profile we used to simulate the training data). 
    \textbf{D--F}. MBSq-50.  
    \textbf{G--I}. MBSq-30.
    \textbf{J--L}. Sinc light sheet simulated by swept lateral standing wave.
    \textbf{M--O}. Gaussian light sheet.
    \textbf{I--V}. Excitation profiles. 
    See Supplementary Table~\ref{tab:lightsheets} for more details about the light sheets used for this test.
    }
    \label{fig:eval-mb-rms}
\end{figure*}

\clearpage        
\begin{figure*}[!tp]
    \centering
    \includegraphics[width=\textwidth]{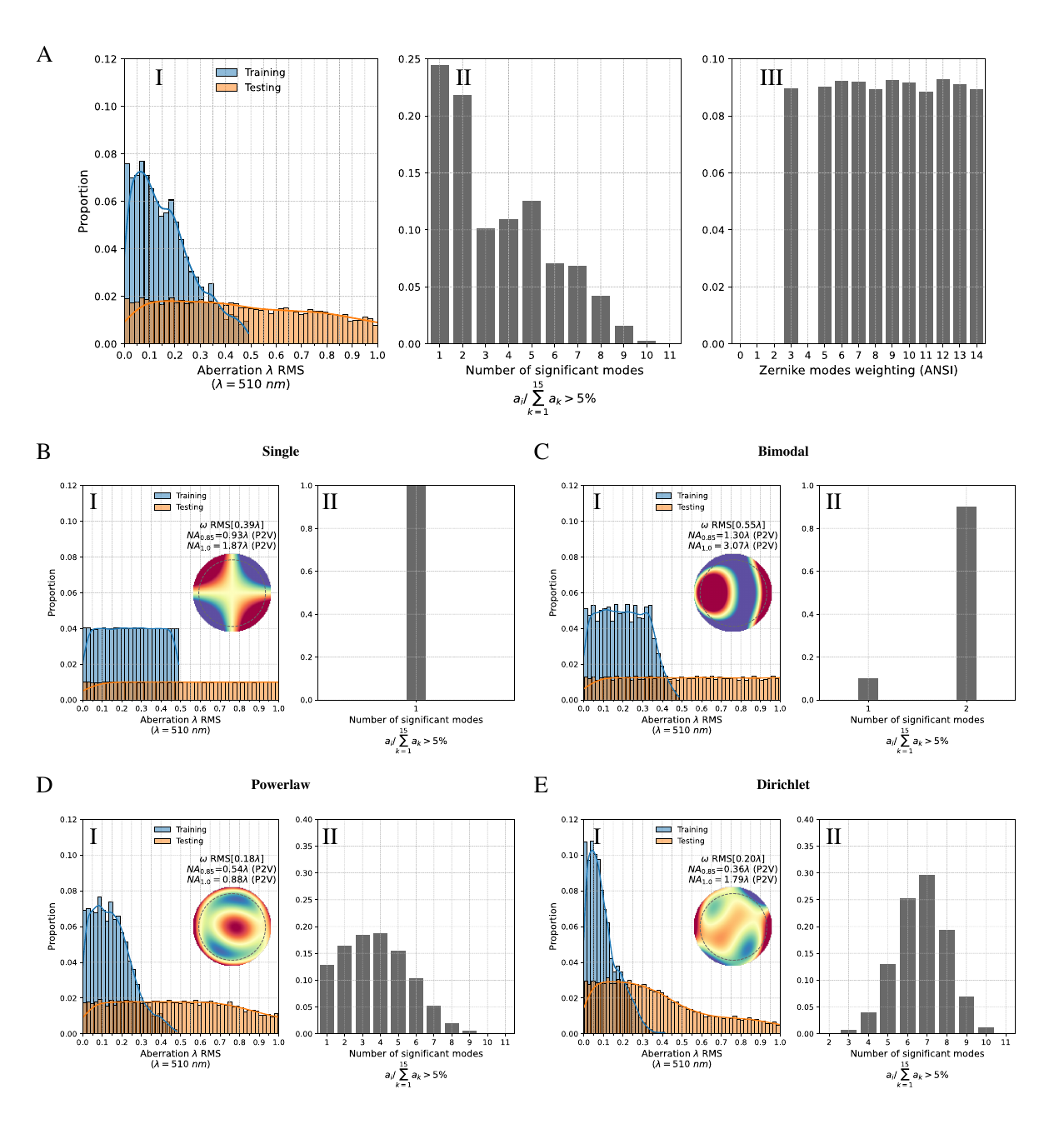}
    \caption{
    \textbf{Zernike distributions}. 
    \textbf{A}. Statistics for the overall training and testing with four different mode distributions: single, bimodal, powerlaw, and Dirichlet.
    \textbf{A(I)}. Magnitude histogram for training (blue), and testing (orange) distributions.
    \textbf{A(II)}. A histogram showing the number of modes that contribute more than 5\% of the total aberration for any given wavefornt (\ie dominant modes).  
    \textbf{A(III)}. A histogram illustrating the probability of choosing any of the first 15 Zernike modes. 
    \textbf{B}. Histograms for the single mode distribution along with example wavefronts. 
    \textbf{C}. Same, but for the bimodal distribution. 
    \textbf{D}. Same, powerlaw distribution. 
    \textbf{E}. Same, dirichlet distribution. 
    }
    \label{fig:zernike_distribution}
\end{figure*}

\clearpage        
\begin{figure*}[!tp]
    \centering
    \includegraphics[width=\textwidth]{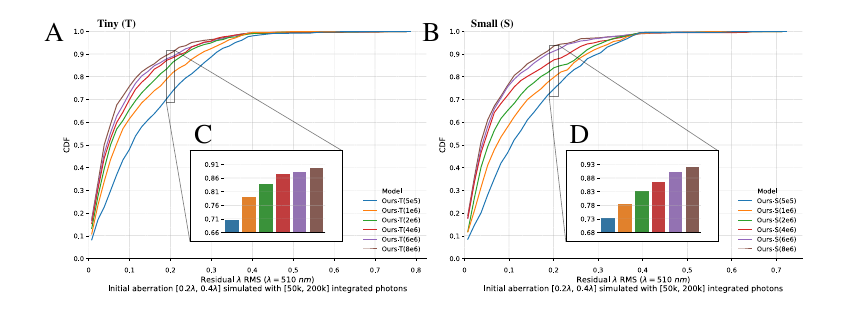}
    \caption{
    \textbf{Scaling performance as a function of training dataset size}. 
    Residual RMS for a series of models trained with an increasing number of training samples: 
    500K (blue), 1M (orange), 2M (green), 4M (red), 6M (purple), and 8M (brown). 
    To evaluate our models we show our residuals after a single correction over a test dataset of 10K samples with aberration ranging between 0.2$\lambda$~RMS and 0.4$\lambda$~RMS simulated using a single bead with 50K up to 200K integrated photons. 
    \textbf{A--B}. Cumulative distribution functions for our \textit{Tiny} model (\textbf{A}), and our \textit{Small} model (\textbf{B}).
    \textbf{C--D}. Inset plots showing the percentage of data below 0.2$\lambda$~RMS for each model. 
    }
    \label{fig:scaling-dataset-summary-rms}
\end{figure*}

\clearpage        
\begin{figure*}[!tp]
    \centering
    \includegraphics[width=.9\textwidth]{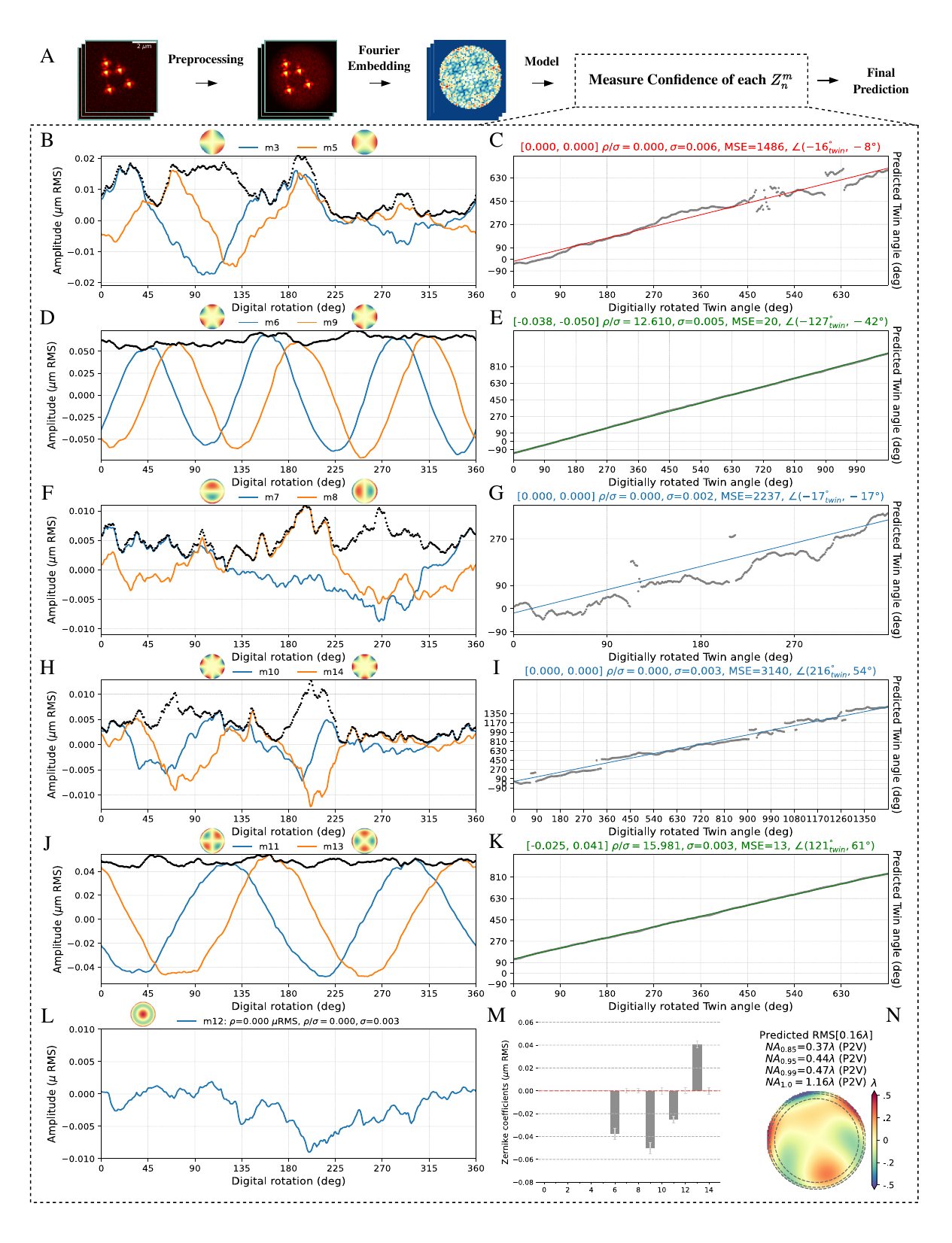}
    \caption{
    \textbf{Predictions on rotated embeddings gives measure of confidence}.
    \textbf{A}. Schematic of our prediction pipeline. 
    \textbf{B}. Predicted amplitudes of astigmatism for each digital rotation: ($Z^{m=\text{-}2}_{n=2}$, blue), ($Z^{m=2}_{n=2}$, orange), and the magnitude for both twin modes in black. 
    \textbf{C}. Regression fit between the digitally rotated angle and the predicted twin angle for astigmatism. 
    \textbf{D--E}. Trefoil.
    \textbf{F--G}. Coma.
    \textbf{H--I}. Quadrafoil. 
    \textbf{J--K}. Secondary astigmatism.
    \textbf{L}. Predicted amplitudes of primary spherical for each digital rotation.
    \textbf{M}. Final prediction for each Zernike mode.
    \textbf{N}. Predicted wavefront.
    }
    \label{fig:pipeline-confidence}
\end{figure*}

\clearpage        
\begin{figure*}[!tp]
    \centering
    \includegraphics[width=\textwidth]{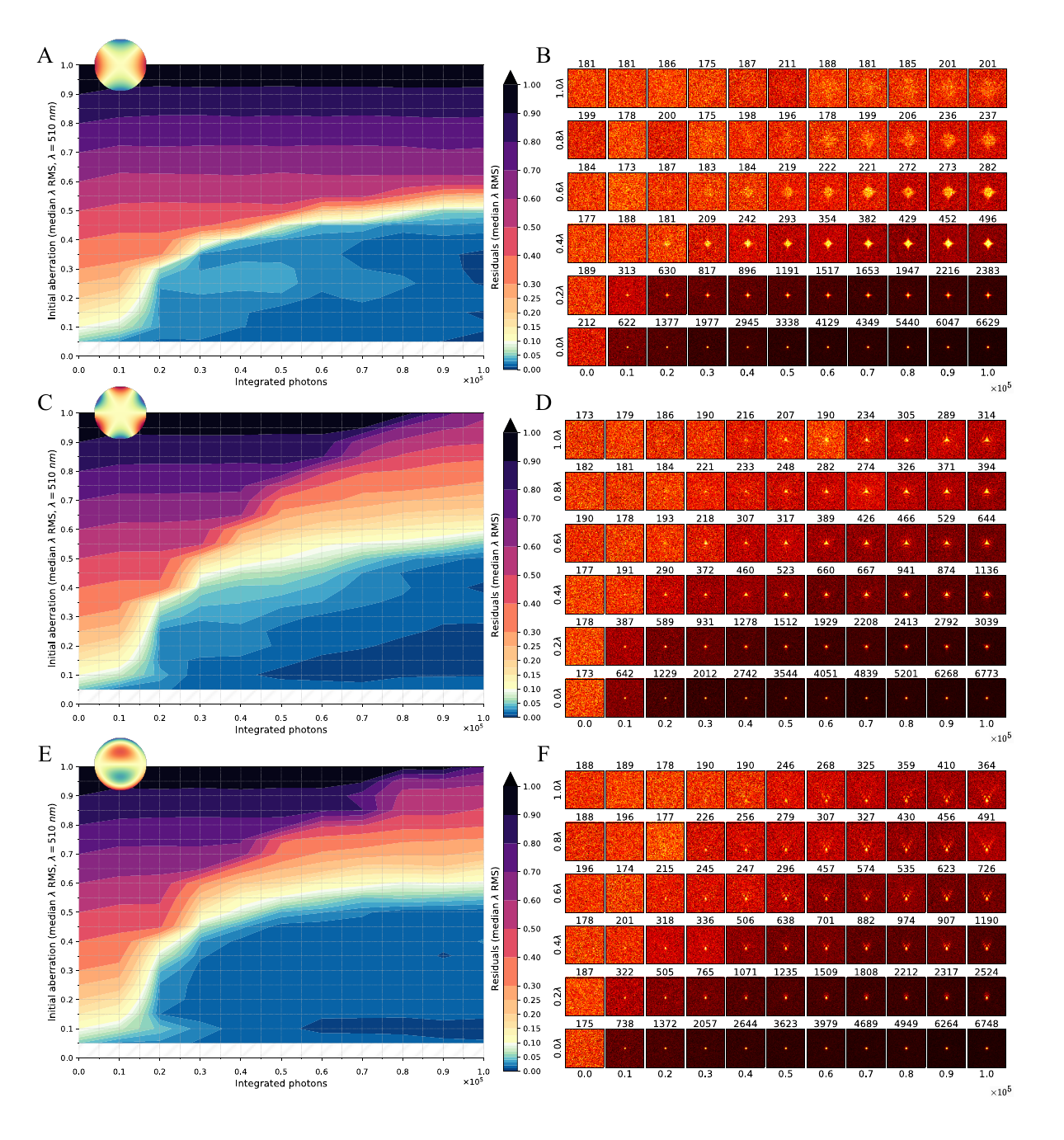}
    \caption{
    \textbf{Single mode evaluation (I)}.
    \textbf{A}. Residual $\lambda$~RMS after a single correction using our Small model (S) for vertical astigmatism $Z^{m=2}_{n=2}$.
    \textbf{B}.
        XY MIPs
        showing the initial aberration without correction for different amplitudes \wrt SNR, with the max counts
        highlighted above each PSF.
    \textbf{C--D}. Vertical trefoil $Z^{m=\text{-}3}_{n=3}$.
    \textbf{E--F}. Vertical coma $Z^{m=\text{-}1}_{n=3}$.
    }
    \label{fig:modes-eval-single-mode-5-7}
\end{figure*}

\clearpage        
\begin{figure*}[!tp]
    \centering
    \includegraphics[width=\textwidth]{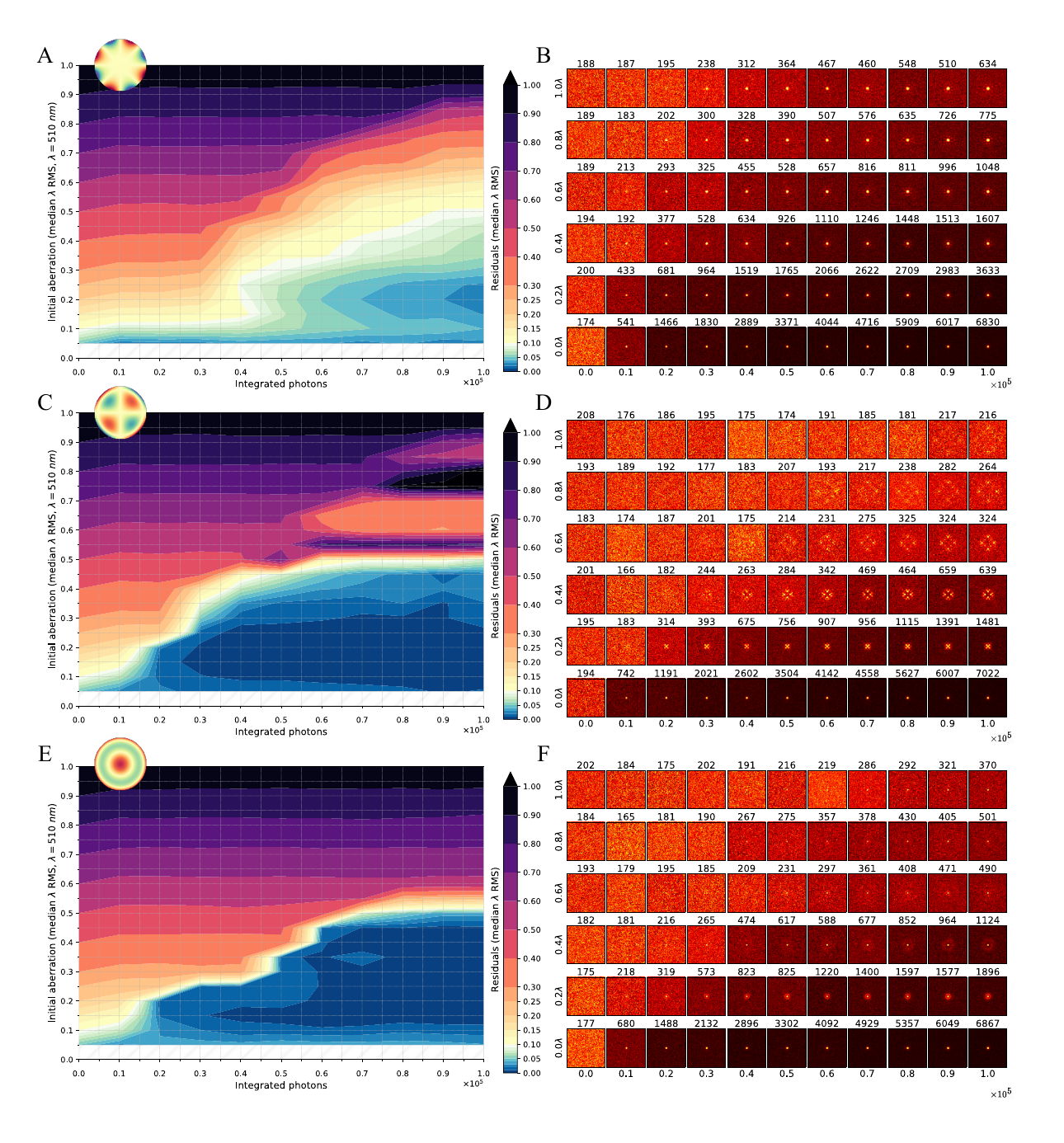}
    \caption{
    \textbf{Single mode evaluation (II)}.
    \textbf{A}. Residual $\lambda$~RMS after a single correction using our Small model (S) for oblique quadrafoil $Z^{m=\text{-}4}_{n=4}$.
 \textbf{B}.
        XY MIPs
        showing the intitial aberration without correction for different amplitudes \wrt SNR, with the max counts
        highlighted above each PSF.
    \textbf{C--D}. Oblique secondary astigmatism $Z^{m=\text{-}2}_{n=4}$.
    \textbf{E--F}. Primary spherical $Z^{m=0}_{n=4}$.
    }
    \label{fig:modes-eval-single-mode-10-12}
\end{figure*}

\clearpage        
\begin{figure*}[!tp]
    \centering
    \includegraphics[width=.9\textwidth]{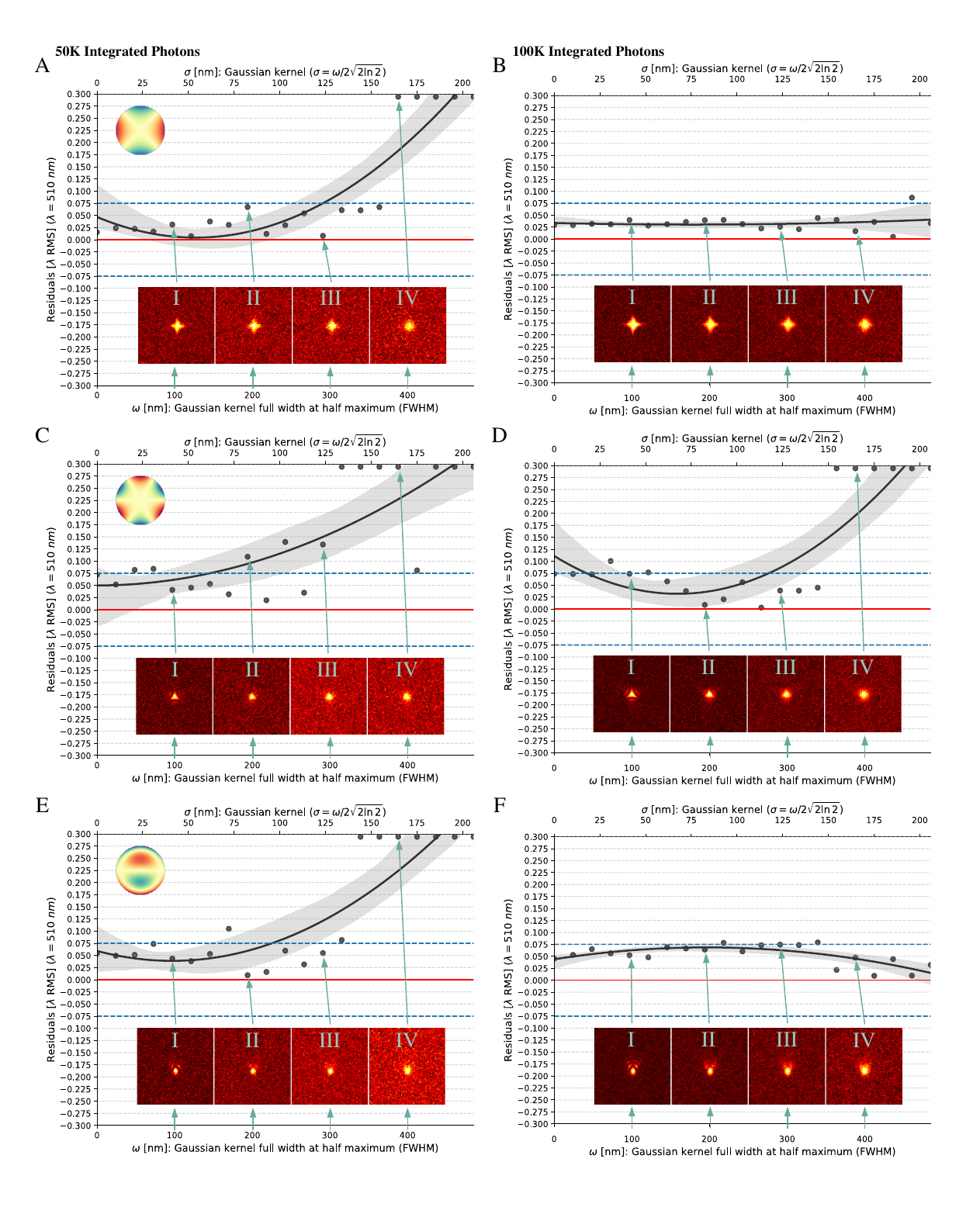}
    \caption{
    \textbf{Sensitivity to object size}.
     We use 0.3 $\lambda$~RMS to synthetically generate a single mode aberration, while increasing the size of the Gaussian kernel used to simulate the bead for each sample.  
    \textbf{A}. Residual $\lambda$~RMS after a single correction using our Small model (S) for vertical astigmatism $Z^{m=2}_{n=2}$ simulated with 50K integrated photons.
    \textbf{B}. Residual $\lambda$~RMS using PSFs simulated with 100K integrated photons.
    \textbf{C--D}. Residual $\lambda$~RMS for vertical trefoil $Z^{m=\text{-}3}_{n=3}$.
    \textbf{E--F}. Residual $\lambda$~RMS for vertical coma $Z^{m=\text{-}1}_{n=3}$.
    }
    \label{fig:modes-eval-bead-size-single-mode-5-7-rms}
\end{figure*}

\clearpage
\begin{table*}[!tp]

\caption{
\textbf{Comparisons with other ML-based AO methods}.
}
\begin{tabularx}{\textwidth}{lCCCc}
\textbf{Paper} & \makecell{\textbf{Speed}\\(time/isoplanatic patch)} & \textbf{Accuracy and Robustness} & \makecell{\textbf{Noninvasiveness}\\(photobleaching)}  & \makecell{\textbf{Code}\\\textbf{Available?}}\\
\toprule
\cite{saha2020practical} & \makecell{$\leq 1$s.} & \makecell{3D widefield PSFs.\\(Fig.~\ref{fig:comparisons-phase-retrieval-rms}--\ref{fig:comparisons-phase-retrieval-psf-rms-iterations})}  & \makecell{No extra imaging.} & \cmark \\
\midrule
\cite{rai2023deep}& \makecell{$\leq 5$s.} & \makecell{2D data.\\Samples that are sparse,\\cleared, fixed, and high SNR.}  & \makecell{1 extra (defocused) image.} & \xmark \\
\midrule
\cite{zhang2023deep} & \makecell{$\leq 1$s.} & \makecell{3D data.\\Single Molecule Light Microscopy\\data only with small aberration.\\Kalman filter needed for robustness.} & \makecell{No extra imaging.} & \cmark \\
\midrule
\cite{hu2023universal} & \makecell{Extra time needed\\to acquire M more extra images\\(M=2,4 or 2N, 4N).} & \makecell{2D data.\\Diverse sample types.\\Many manually-tuned parameters.\\ Best at 0.15--0.3 $\lambda$~RMS.\\Less accurate at 0.1--0.13 $\lambda$~RMS.}  & \makecell{Many extra images\\and motion correction\\recommended.} & \xmark \\
\midrule
\cite{kang2024coordinate} & \makecell{$\sim$5 minuets per isoplanatic\\patch of 32 megavoxels\\(200$\times$400$\times$400).} & \makecell{3D data.\\Local thresholding required\\to remove hallucinations.\\Tested with beads and\\sparse neuronal structures.\\Aberration up to 0.3 $\lambda$~RMS.\\Many manually-tuned parameters.} & \makecell{No extra imaging.} & \cmark \\
\midrule
Ours &  \makecell{$\leq 3$s (Table.~\ref{tab:performance_SH_vs_model}).} & \makecell{3D data with puncta.\\Aberration up to 1.0 $\lambda$~RMS\\(Appendix~\ref{sup:insilico}, Fig.~\ref{fig:eval-mb-rms}).\\ Confidence scores\\(Appendix~\ref{sup:confidence}).} & \makecell{No extra imaging.} & \cmark \\
\end{tabularx}
\label{tab:ao_comparisons}
\end{table*}

\clearpage
\begin{table*}[!tp]

\caption{
\textbf{Hyperparameters for our model variants.}
A breakdown of the hyperparameters used for each stage i, 
where $p_i$ is the patch size used to tile in the input tensor, 
$n_i$ is the number of transformer layers,
$h_i$ is the number of heads for each transformer layer,
$\epsilon_i$ is the embedding size, 
and $x_i$ is the MLP size. 
Furthermore, we show the total number of trainable parameters based on the scheme used to derive each variant with the total number of transformer layers used across all stages ($n$) and the number of heads for each transformer layer ($h$).
}
\begin{tabularx}{\textwidth}{r|CCCCC|CCCCC|CC|C}
{} & \multicolumn{5}{c}{\textbf{Stage $1$}} & \multicolumn{5}{c|}{\textbf{Stage $2$}} & \multicolumn{2}{c|}{\textbf{Scheme}} & \textbf{Params} \\
\textbf{Name} & \makecell{Patch\\$p_1$} & \makecell{Layers\\$n_1$} & \makecell{Heads\\$h_1$} & \makecell{EMB\\$\epsilon_1$} & \makecell{MLP\\$x_1$} & \makecell{Patch\\$p_2$} & \makecell{Layers\\$n_2$} & \makecell{Heads\\$h_2$} & \makecell{EMB\\$\epsilon_2$} & \makecell{MLP\\$x_2$} & \makecell{Layers\\$n$} & \makecell{Heads\\$h$} & millions \\
\toprule
\textbf{Tiny  (T)} & 32 & 3 & 6 & 1024 & 4096 & 16 & 3 & 6 & 256 & 1024 & 6 & 6 & 33.9 \\
\textbf{Small (S)} & 32 & 4 & 8 & 1024 & 4096 & 16 & 4 & 8 & 256 & 1024 & 8 & 8 & 47.4 \\
\textbf{Base  (B)} & 32 & 6 & 12 & 1024 & 4096 & 16 & 6 & 12 & 256 & 1024  & 12 & 12 & 78.3 \\
\textbf{Large (L)} & 32 & 12 & 16 & 1024 & 4096 & 16 & 12 & 16 & 256 & 1024 & 24 & 16 & 171.3 \\
\textbf{Huge  (H)} & 32 & 16 & 16 & 1024 & 4096 & 16 & 16 & 16 & 256 & 1024 &  32 & 16 & 227.9 \\
\end{tabularx}
\label{tab:variants}
\end{table*}

\clearpage
\begin{table*}[htp!]

\caption{
\textbf{Evaluation benchmark}.
}
\begin{tabularx}{\textwidth}{ccRRRRRRRRRR}
{} & {} & \multicolumn{2}{c}{\multirow{2}{*}{\textbf{Testing $\lambda$ RMS}}}   & \multirow{2}{*}{\textbf{Loss}}         & \multirow{2}{*}{\textbf{Training}}  & \multirow{2}{*}{\textbf{Training}}   & \multirow{2}{*}{\textbf{Memory}}   & \multirow{2}{*}{\textbf{Throughput}}  & \multirow{2}{*}{\textbf{Latency}}     & \multirow{2}{*}{\textbf{Inference}}   & \multirow{2}{*}{\textbf{Parameters}} \\
{} & {} & \textbf{(one shot)} & \textbf{(two shot)$\downarrow$} & \textbf{($\mu$m RMS)$\downarrow$}   & \textbf{hours$\downarrow$}   & \textbf{EFLOPs$\downarrow$}      & \textbf{(GB)$\downarrow$}     & \textbf{(img/s)$\uparrow$}  & \textbf{(ms/img)$\downarrow$}    & \textbf{GFLOPs$\downarrow$}       & \textbf{(millions)$\downarrow$}  \\
\toprule
\parbox[t]{2mm}{\multirow{4}{*}{\rotatebox[origin=c]{90}{\textbf{ConvNext}}}} & \textbf{T} &  0.07 & 0.04 &          85.6$e^{-7}$ &                    37.5 &            12.9 &                     6.8 &                846 &                18.7 &   4.3 &           27.9  \\
{} & \textbf{S} &  0.04 & 0.02 &          18.8$e^{-7}$ &                    73.8 &            25.0 &                    9.4 &                648 &                32.2 &   8.4 &           49.6  \\
{} & \textbf{B} &  0.03 & 0.02 &          8.7$e^{-7}$ &                    92.0 &            44.6 &                    12.6 &                519 &                32.8 &   14.9 &            87.7  \\
{} & \textbf{L} &  0.04 & 0.02 &          7.4$e^{-7}$ &                    146.6 &            99.7 &                    19.1 &                369 &                33.3 &   33.5 &           196.4  \\
\midrule
\parbox[t]{2mm}{\multirow{2}{*}{\rotatebox[origin=c]{90}{\textbf{ViT/16}}}} & \textbf{S}   &   0.03 & 0.02 &          28.4$e^{-7}$ &                    27.3 &            8.4 &                    2.3 &                  1326 &                15.7 &   2.8 &           71.2  \\
{} & \textbf{B}   &   0.03 & 0.02 &          4.7$e^{-7}$ &                    75.8 &            33.2 &                    5.5 &                 498 &                17.0 &   11.1 &           397.6  \\
\midrule
\parbox[t]{2mm}{\multirow{3}{*}{\rotatebox[origin=c]{90}{\textbf{ViT/32}}}} & \textbf{S}   &  0.04 & 0.02 &          26.8$e^{-7}$ &                    15.8 &            2.1 &                    0.9 &               2345 &                16.0 &   0.7 &           71.5  \\
{} & \textbf{B}   &  0.04 & 0.02 &          5.9$e^{-7}$ &                    28.7 &            8.4 &                    2.6 &               1557 &                16.1 &   2.8 &           398.2  \\
{} & \textbf{L}   &  0.04 & 0.02 &          2.8$e^{-7}$ &                    103.9 &            29.4 &                    9.3 &                488 &                30.6 &   9.8 &            1815.0  \\
\midrule
\parbox[t]{2mm}{\multirow{5}{*}{\rotatebox[origin=c]{90}{\textbf{Ours}}}} & \textbf{T}     &  0.04 & 0.02 &          36.3$e^{-7}$ &                    10.9 &             4.7 &                     1.4 &               2976 &                11.1 &    1.6 &           33.9  \\
{} & \textbf{S}     &  0.04 & 0.02 &          22.9$e^{-7}$ &                    13.9 &            6.3 &                    1.7 &               2522 &                13.3 &    2.1 &           47.4  \\
{} & \textbf{B}     &  0.04 & 0.02 &          11.7$e^{-7}$ &                    19.5 &            9.3 &                    2.2 &               2189 &                17.2 &   3.1 &           78.3  \\
{} & \textbf{L}     &  0.04 & 0.02 &          3.3$e^{-7}$ &                    37.0 &            18.4 &                    3.6 &                1233 &                29.0 &   6.1 &           171.3 \\
 {} & \textbf{H}     &  0.04 & 0.02 &          2.7$e^{-7}$ &                    48.2 &            24.4 &                      4.6 &                1003 &                37.7 &   8.2 &             227.9 \\
\end{tabularx}
\label{tab:benchmark}
\end{table*}

\clearpage
\begin{table*}[!tp]

\caption{\textbf{Residual wavefront measured using phase retrieval with widefield on an isolated bead for Fig.~\ref{fig:cells}}.}
\begin{tabularx}{\textwidth}{LCCC}
\textbf{Iteration} &    \textbf{NA} &   \makecell{\textbf{Residual WF}\\$\lambda$ RMS} & \makecell{\textbf{Residual WF}\\$\lambda$ P2V} \\
\toprule
\multicolumn{4}{c}{\textbf{Fig.~\ref{fig:cells}a}}\\
\midrule
0               &  1.00 &  0.450701 &  2.940459 \\
0               &  0.95 &  0.380205 &  2.425910 \\
0               &  0.85 &  0.292638 &  1.435171 \\
\midrule
1               &  1.00 &  0.155836 &  1.032606 \\
1               &  0.95 &  0.135578 &  0.927065 \\
1               &  0.85 &  0.096931 &  0.705702 \\
\midrule
2               &  1.00 &  0.089000 &  0.479712 \\
2               &  0.95 &  0.083749 &  0.370095 \\
2               &  0.85 &  0.077435 &  0.336847 \\
\midrule
\multicolumn{4}{c}{\textbf{Fig.~\ref{fig:cells}b}}\\
\midrule
0               &  1.00 &  0.536977 &  3.053908 \\
0               &  0.95 &  0.465103 &  2.359822 \\
0               &  0.85 &  0.425599 &  1.941650 \\
\midrule
1               &  1.00 &  0.071351 &  0.459885 \\
1               &  0.95 &  0.061207 &  0.330569 \\
1               &  0.85 &  0.054622 &  0.244614 \\
\midrule
2               &  1.00 &  0.068686 &  0.459620 \\
2               &  0.95 &  0.059220 &  0.362715 \\
2               &  0.85 &  0.048254 &  0.218788 \\
\end{tabularx}
\label{tab:cells_wavefronts}
\end{table*}

\clearpage
\begin{table*}[!tp]

\caption{
\textbf{Performance of SH and \model-S running on a single node with either one or four A100 GPUs}.
Benchmark for three modes of operations: default mode, smart ROI detection mode where a ROI containing diffraction-limited puncta are cropped from a scanned volume, and tiling mode. 
\model-S measurement is done on the entire 3D volume (z $\times$ y $\times$ x $\mu\text{m}^3$), whereas SH measurement is done on 3 (y $\times$ x $\mu\text{m}^2$) planes.
Total time includes initialization and setup time (\eg loading model onto each GPU), which may take between 5 and 10 seconds but is only needed at the very beginning of an imaging experiment. Since aberration measurement by SH is done serially, the SH [tile] values reported here represent the average measurement time for 288 ROIs based on the SH [default] measurement time for a single ROI. 
We also show a breakdown of the time needed for preprocessing \& embedding, smart ROI detection, and model inference time. 
}
\begin{tabularx}{\textwidth}{lCCCCCC}
\textbf{Method} & \makecell{\textbf{Predictions}\\w/ 361 rotations} & \makecell{\textbf{Preprocessing}\\\textbf{Time}} & \makecell{\textbf{ROI Detection}\\\textbf{Time}} & \makecell{\textbf{Inference}\\\textbf{Time}} & \makecell{\textbf{Total}\\\textbf{Time}} & \makecell{\textbf{GPUs}\\(A100)} \\
\toprule
\multicolumn{3}{l}{\textbf{Beads} 12.8 $\times$ 9.3 $\times$ 9.3 $\mu\text{m}^3$} & \multicolumn{4}{r}{}\\
\midrule
\model-S [default] & 1 & 2s & N/A & 3s & 8.6s & 1 \\
\model-S [default] & 361 & 2s & N/A & 3s & 9.2s & 1 \\
\model-S [default] & 361 & 2s & N/A & 9s & 16.9s & 4 \\
\midrule
\multicolumn{3}{l}{\textbf{Cells}  25.6 $\times$ 15.5 $\times$ 55.9 $\mu\text{m}^3$} & \multicolumn{4}{r}{(ROI = 12.8, 6.2, 6.2 $\mu\text{m}^3$)}\\
\model-S [default] & 1  & 3s & N/A & 3s & 8.9s  & 1 \\
\model-S [default] & 361  & 3s & N/A & 3s & 9.3s  & 1 \\
\model-S [default] & 361  & 3s & N/A & 9s & 18.1s  & 4 \\
\model-S [roi-1] & 361  & 3s & 3s & 9s & 25.3s & 4 \\
\model-S [roi-10] & 3610 & 6s & 9s & 15s & 40.6s & 4 \\
\midrule
\multicolumn{3}{l}{\textbf{Zebrafish small FOV} 12.8 $\times$ 12.4 $\times$ 12.4 $\mu\text{m}^3$} & \multicolumn{4}{r}{(ROI = 12.8, 6.2, 6.2 $\mu\text{m}^3$)}\\
\midrule
SH [default] & 1 & N/A & N/A & N/A &  5.1s & N/A \\
\model-S [default] & 1  & 3s & N/A & 3s &  8.8s & 1 \\
\model-S [default] & 361  & 3s & N/A & 3s &  9.2s & 1 \\
\model-S [default] & 361  & 3s & N/A & 9s &  17.0s & 4 \\
\model-S [roi-1] & 361 & 3s & 3s & 9s & 23.7s  & 4 \\
\model-S [roi-10] & 3610 & 6s  & 9s & 15s & 38.9s & 4 \\
\midrule
\multicolumn{3}{l}{\textbf{Zebrafish large FOV} 12.8 $\times$ 49.7 $\times$ 223.5 $\mu\text{m}^3$} & \multicolumn{4}{r}{(TILE = 12.8, 9.3, 9.3 $\mu\text{m}^3$)}\\
\midrule
SH [tile]  & 288 tiles & N/A & N/A & N/A &  1,468.8s & N/A \\
\model-S [tile]  & \makecell{103,968\\(288 tiles  $\times$  361)} & 30.6s & N/A & 91.4s &  139.8s & 4 \\
\end{tabularx}
\label{tab:performance_SH_vs_model}
\end{table*}

\clearpage
\begin{table*}[htp!]

\caption{
\textbf{Light sheet specifications}. \textit{MBSq-35} LLS excitation profile was used for simulating training data.
}
\begin{tabularx}{\textwidth}{LlCCCCCCC}
\textbf{Name} & \textbf{Lattice type} & \textbf{$\text{NA}_{lattice}$} & \textbf{$\text{NA}_{exc}$} & \textbf{$\text{NA}_{sinc}$} & \textbf{$\text{NA}_{annulus}$} & \textbf{$\sigma_{\text{NA}}$} & \textbf{FWHM} & \textbf{Crop} \\
\toprule
\textbf{MBSq-30}     & Multi-Bessel square LLS                    & 0.30 & - & - & 0.375/0.225 & 0.10 & 48.5 & - \\
\textbf{MBSq-35}     & Multi-Bessel square LLS                    & 0.35 & - & - & 0.40/0.30 & 0.10 & - & - \\
\textbf{MBSq-50}     & Multi-Bessel square LLS                    & 0.50 & - & - & 0.40/0.30 & 0.10 & - & - \\
\textbf{Sinc}        & Simulated by swept lateral standing wave   & - & 0.32 & 0.24 & 0.40/0.20 & - & 51.5 & - \\
\textbf{Gaussian}    & Simulated by swept lateral standing wave   & - & 0.21 & - & 0.40/0.20 & 0.21 & 51.0 & 0.1 \\
\end{tabularx}
\label{tab:lightsheets}
\end{table*}

\clearpage
\begin{table*}[htp!]

\caption{
\textbf{Imaging configuration.} All data was collected using \textit{MBSq-35} light sheet described in Supplementary Table~\ref{tab:lightsheets}.
}
\begin{tabularx}{\textwidth}{lCCCC}
\textbf{Figure | Scan type} & \makecell{\textbf{Voxel size}\\(obj x, obj y, obj z)} & \makecell{\textbf{Volume size [voxels]}\\(obj x, obj y, obj z)} & \makecell{\textbf{Excitation}\\\textbf{wavelength}} & \makecell{\textbf{Camera}\\\textbf{Exposure}} \\
\toprule
\textbf{Fig.~\ref{fig:cells}} | Widefield Scan & 97, 97, 100 nm & 96, 96, 128  & 488 nm  & 50 ms \\
\textbf{Fig.~\ref{fig:cells}} | Cell Scan & 97, 97, 200 nm & 160, 576, 128  & 488 nm, 560 nm  & 50 ms \\
\midrule
\textbf{Fig.~\ref{fig:fish_online}a} | No AO Full Chip Scan & 97, 97, 200 nm & 2304, 2304, 64  & 488 nm , 560 nm  & 50 ms \\
\midrule
\textbf{Fig.~\ref{fig:fish_online}b} | No AO Scan 1 & 97, 97, 200 nm & 128, 128, 64  & 488 nm, 560 nm  & 50 ms \\
\textbf{Fig.~\ref{fig:fish_online}b} | DSH 1 & 0.1, 6.2, 6.3 um & 125, 3, 3  & 920 nm  & 500 ms \\
\textbf{Fig.~\ref{fig:fish_online}b} | DSH 2 & 0.1, 6.2, 6.3 um & 125, 3, 3  & 920 nm  & 400 ms \\
\textbf{Fig.~\ref{fig:fish_online}b} | DSH 3 & 0.1, 6.2, 6.3 um & 125, 3, 3  & 920 nm  & 400 ms \\
\textbf{Fig.~\ref{fig:fish_online}b} | SH AO Scan & 97, 97, 200 nm & 128, 128, 64  & 488 nm , 560 nm  & 50 ms \\
\textbf{Fig.~\ref{fig:fish_online}b} | No AO Scan 2 & 97, 97, 200 nm & 128, 128, 64  & 488 nm , 560 nm  & 50 ms \\
\textbf{Fig.~\ref{fig:fish_online}b} | AOViFT Iter 1 Scan & 97, 97, 200 nm & 128, 128, 64  & 488 nm , 560 nm  & 50 ms \\
\textbf{Fig.~\ref{fig:fish_online}b} | AOViFT Iter 2 Scan & 97, 97, 200 nm & 128, 128, 64  & 488 nm , 560 nm  & 50 ms \\
\midrule
\textbf{Fig.~\ref{fig:fish_online}c} | No AO Full Chip Scan & 97, 97, 200 nm & 2304, 2304, 64 & 488 nm, 560 nm & 50 ms, 50 ms \\
\textbf{Fig.~\ref{fig:fish_online}c} | No AO Scan 1 & 97, 97, 200 nm & 128, 128, 64  & 488 nm, 560 nm & 50 ms, 50 ms \\
\textbf{Fig.~\ref{fig:fish_online}c} | DSH 1 & 0.1, 6.2, 6.3 um & 125, 3, 3  & 920 nm & 500 ms \\
\textbf{Fig.~\ref{fig:fish_online}c} | DSH 2 & 0.1, 6.2, 6.3 um & 125, 3, 3  & 920 nm  & 500 ms \\
\textbf{Fig.~\ref{fig:fish_online}c} | DSH 3 & 0.1, 6.2, 6.3 um & 125, 3, 3  & 920 nm  & 500 ms \\
\textbf{Fig.~\ref{fig:fish_online}c} | SH AO Scan & 97, 97, 200 nm & 128, 128, 64  & 488 nm, 560 nm & 50 ms, 50 ms \\
\textbf{Fig.~\ref{fig:fish_online}c} | No AO Scan 2 & 97, 97, 200 nm & 128, 128, 64  & 488 nm, 560 nm & 50 ms, 50 ms \\
\textbf{Fig.~\ref{fig:fish_online}c} | AOViFT Iter 1 Scan & 97, 97, 200 nm & 128, 128, 64  & 488 nm, 560 nm & 50 ms, 50 ms \\
\textbf{Fig.~\ref{fig:fish_online}c} | AOViFT Iter 2 Scan & 97, 97, 200 nm & 128, 128, 64  & 488 nm, 560 nm & 50 ms, 50 ms \\
\textbf{Fig.~\ref{fig:fish_online}c} | AOViFT Full Chip Scan & 97, 97, 200 nm & 2304, 2304, 64  & 488 nm, 560 nm & 50 ms, 50 ms \\
\midrule
\textbf{Fig.~\ref{fig:fish_offline}b} | AOViFT Full Chip Scan & 97, 97, 200 nm & 2304, 2304, 64  & 488 nm & 50 ms \\
\midrule
\textbf{Fig.~\ref{fig:fish_offline}d} | AOViFT Full Chip Scan & 97, 97, 200 nm & 2304, 2304, 64  & 560 nm & 50 ms \\
\midrule
\textbf{Fig.~\ref{fig:Figure4_SI1}--\ref{fig:Figure4_SI2}} | No AO Full Chip Scan & 97, 97, 200 nm & 2304, 2304, 64 & 488 nm, 560 nm & 50 ms, 50 ms \\
\textbf{Fig.~\ref{fig:Figure4_SI1}--\ref{fig:Figure4_SI2}} | No AO Scan 1 & 97, 97, 200 nm & 256, 256, 64  & 488 nm, 560 nm & 50 ms, 50 ms \\
\textbf{Fig.~\ref{fig:Figure4_SI1}--\ref{fig:Figure4_SI2}} | DSH 1 & 0.1, 6.2, 6.2 um & 125, 3, 3  & 920 nm & 400 ms \\
\textbf{Fig.~\ref{fig:Figure4_SI1}--\ref{fig:Figure4_SI2}} | DSH 2 & 0.1, 6.2, 6.2 um & 125, 3, 3  & 920 nm  & 400 ms \\
\textbf{Fig.~\ref{fig:Figure4_SI1}--\ref{fig:Figure4_SI2}} | DSH 3 & 0.1, 6.2, 6.2 um & 125, 3, 3  & 920 nm  & 400 ms \\
\textbf{Fig.~\ref{fig:Figure4_SI1}--\ref{fig:Figure4_SI2}} | SH AO Scan & 97, 97, 200 nm & 256, 256, 64  & 488 nm, 560 nm & 50 ms, 50 ms \\
\textbf{Fig.~\ref{fig:Figure4_SI1}--\ref{fig:Figure4_SI2}} | No AO Scan 2 & 97, 97, 200 nm & 256, 256, 64  & 488 nm, 560 nm & 50 ms, 50 ms \\
\textbf{Fig.~\ref{fig:Figure4_SI1}--\ref{fig:Figure4_SI2}} | AOViFT Iter 1 Scan & 97, 97, 200 nm & 256, 256, 64  & 488 nm, 560 nm & 50 ms, 50 ms \\
\textbf{Fig.~\ref{fig:Figure4_SI1}--\ref{fig:Figure4_SI2}} | AOViFT Iter 2 Scan & 97, 97, 200 nm & 256, 256, 64  & 488 nm, 560 nm & 50 ms, 50 ms \\
\end{tabularx}
\label{tab:imaging_configuration}
\end{table*}

\clearpage
\begin{table*}[htp!]

\caption{
\textbf{Training configuration}.
}
\begin{tabularx}{\textwidth}{L|C}
\toprule
\multicolumn{2}{c}{\textbf{Optical configuration}} \\
\midrule
Lattice type                & \textit{MBSq-35} (Supplementary Table~\ref{tab:lightsheets}) \\
$\text{NA}_{exc}$           & 0.35 \\
$\text{NA}_{annulus}$       & 0.4/0.3 \\
$\text{NA}_{detection}$     & 1.0 \\
$\sigma_{\text{NA}}$        & 0.10 \\
Refractive index            & 1.33 \\
Excitation wavelength       & 488 nm \\
Detection wavelength ($\lambda$)      & 510 nm \\
Voxel size                  & $200^Z, 125^Y, 125^X$ nm \\
\midrule
\multicolumn{2}{c}{\textbf{Training Dataset}} \\
\toprule
Dataset size                & 2M synthetic samples \\
Number of beads             & $b \in \{1 \to 5\}$ \\
Bead size                   & $\text{FWHM} \in \{100, 200, 300, 400\}nm$ \\
Mode distribution(s)        & $\mathcal{D} \in$ \{Single, Bimodal, Powerlaw, Dirichlet\} \\ 
Mode weighting ($Z_n^m$)    & Uniform $n \in \{2\to4\}$ \& linear decay $n \in \{4\to10\}$ \wrt $m$ \\ 
Magnitude range             & $a \in \{0\to0.3\}\mu$m $\Longleftrightarrow$ $a \in \{0\to5\}\lambda$ \\
Image size                  & $64^Z, 64^Y, 64^X$ voxels $\Longleftrightarrow$ $12.8^Z, 8^Y, 8^X$ $\mu$m \\
\midrule
\multicolumn{2}{c}{\textbf{Hyperparameters}} \\
\toprule
Batch size                  & 4096      \\
Loss                        & MSE       \\
Optimizer                   & AdamW~\cite{loshchilov2018decoupled} \\
Layerwise decay             & LAMB~\cite{you2020large} \\ 
Momentum $\beta_1$          & 0.9        \\
Momentum $\beta_2$          & 0.99       \\
Initial learning rate (LR)  & 0.0001     \\
Weight decay                & 0.001      \\
\midrule
\multicolumn{2}{c}{\textbf{Training schedulers}} \\
\toprule
Total epochs                & 500           \\
Warmup epochs               & 25            \\
Warmup scheduler            & Linear        \\
Learning rate decay         & Cosine~\cite{loshchilov2016sgdr} \\
\bottomrule
\end{tabularx}
\label{tab:training_config}
\end{table*} 


\end{appendices}

\end{document}